\documentclass[12pt,runningheads]{article}
\usepackage[utf8]{inputenc}
\usepackage{array,multicol,mathpazo}
\usepackage[normalem]{ulem}
\usepackage{fullpage}
\usepackage{soul}
\usepackage{floatrow}
\usepackage[top=1in, bottom=1in, left=1in, right=1in]{geometry}
\usepackage{setspace}
   
\fontsize{12}{0}
\usepackage{amsmath}
\usepackage{amssymb}
\usepackage{bm}
\usepackage{amsthm}
\usepackage{mathtools} 
\usepackage{exscale}
\usepackage[mathscr]{eucal}
\usepackage{bm}
\usepackage{eqlist} 
\usepackage[final]{graphicx}
\usepackage[dvipsnames]{color}
\usepackage{verbatim}
\usepackage[square,sort,comma]{natbib}
\usepackage{natbib}
\usepackage{caption}
\usepackage{enumerate}
\usepackage{subcaption}
\usepackage{algpseudocode}
\usepackage{rotating}
\usepackage{etoolbox}
\usepackage{color}
\usepackage[linesnumbered,ruled]{algorithm2e}
\usepackage[thinc]{esdiff}
\usepackage{multirow}
\usepackage[shortlabels]{enumitem}
\PassOptionsToPackage{hyphens}{url}\usepackage{hyperref}

\usepackage{graphicx}
\usepackage{array}
\newcolumntype{N}{>{\centering\arraybackslash}m{.5in}}
\newcolumntype{G}{>{\centering\arraybackslash}m{2in}}

\makeatletter
\patchcmd{\env@cases}{1.2}{0.6}{}{}
\makeatother

\DeclareGraphicsExtensions{.pdf, .jpg}

\theoremstyle{definition}

\newtheorem{condition}{Condition}

\newtheorem{proposition.a}{Proposition A\ignorespaces}
\newtheorem{proposition.s}{Proposition S\ignorespaces}

\newtheorem{remark.s}{Remark S\ignorespaces}
\newtheorem{thm}{Theorem}
\newtheorem{thm.s}{Theorem S\ignorespaces}

\newtheorem{cor.s}{Corollary S\ignorespaces}

\newtheorem{lem.s}{Lemma S\ignorespaces}

\newcommand{\x}{\bm{x}}
\newcommand{\bma}{\bm{a}}
\newcommand{\z}{\bm{z}}

\newcommand{\w}{\bm{w}}

\newcommand{\y}{\bm{y}}
\newcommand{\X}{\bm{X}}

\newcommand{\h}{{h}}
\newcommand{\Z}{\bm{Z}}
\newcommand{\I}{\bm{I}}
\newcommand{\W}{\bm{W}}

\newcommand{\bmeps}{\bm{\epsilon}}
\newcommand{\bmbeta}{\bm{\beta}}
\newcommand{\bmalpha}{\bm{\alpha}}

\newcommand{\tE}{\mbox{E}}
\newcommand{\var}{\mbox{Var}}

\newcommand{\tr}{\mbox{tr}}
\newcommand{\bbR}{\mathbb{R}}

\newcommand{\bmI}{\mbox{I}}
\newcommand{\bmSigma}{\bm{\Sigma}}

\newcommand{\bmPhi}{\bm{\Phi}}

\makeatletter
\newcommand*\rel@kern[1]{\kern#1\dimexpr\macc@kerna}
\newcommand*\widebar[1]{%
  \begingroup
  \def\mathaccent##1##2{%
    \rel@kern{0.8}%
    \overline{\rel@kern{-0.8}\macc@nucleus\rel@kern{0.2}}%
    \rel@kern{-0.2}%
  }%
  \macc@depth\@ne
  \let\math@bgroup\@empty \let\math@egroup\macc@set@skewchar
  \mathsurround\z@ \frozen@everymath{\mathgroup\macc@group\relax}%
  \macc@set@skewchar\relax
  \let\mathaccentV\macc@nested@a
  \macc@nested@a\relax111{#1}%
  \endgroup
}
\makeatother
\usepackage{newfloat}
\usepackage[labelfont=bf]{caption}
\usepackage{tabularx}
\usepackage{lipsum}
\def\bxz{\color{black}}
\usepackage{xcolor}

\title{\LARGE  
Estimating trans-ancestry genetic correlation with unbalanced data resources} 
\author{Bingxin Zhao\footnote{Department of Statistics, Purdue University. Email: bingxin@purdue.edu}, Xiaochen Yang\footnote{Department of Statistics, Purdue University. Email: yang1641@purdue.edu}, and Hongtu Zhu\footnote{Department of Biostatistics, University of North Carolina at Chapel Hill. Email: htzhu@email.unc.edu}\\~\\
}
\begin{document}
\maketitle
\date{\vspace{-2ex}}
\date{}
\abstract{
The aim of this paper is to  propose a novel estimation method of 
using genetic-predicted observations to estimate trans-ancestry genetic correlations, which  describes how genetic architecture of complex traits varies among populations, in 
genome-wide association studies (GWAS).   
 Our new estimator corrects for  prediction errors caused by high-dimensional weak GWAS signals, while addressing 
 the heterogeneity of GWAS data across ethnicities, such as linkage disequilibrium (LD) differences, which  can lead to biased findings in homogeneity-agnostic analyses.
 Moreover,  our estimator 
 only requires one population to have a large GWAS sample size, and the second population can only have a much smaller number of participants (for example, hundreds). It is designed to specifically  address the unbalanced data resources such that the GWAS sample size for European populations is usually larger than that of non-European ancestry groups.  
 Extensive simulations and real data analyses of $30$ complex traits in the UK Biobank study show that our method is capable of providing reliable estimates of a wide range of complex traits. Our results  provide deep  insights into the transferability of population-specific genetic findings.
 \\

\noindent \textbf{Keywords.}  Data heterogeneity; 
 GWAS; High-dimensional prediction;  
 Trans-ancestry genetic correlation; UK Biobank. 
}

\section{Introduction}\label{sec1}

{The distribution of heritable complex traits and diseases is typically population-specific. For example, Hispanics and Blacks  are more likely to suffer from diseases that harm white matter in the brain, such as strokes \citep{gardener2020race}  or Alzheimer's disease \citep{chen2018racial}.  
Such phenotypic differentiation across populations may result from differences in the underlying genetic, environmental, and demographic factors, such as allele frequency, linkage disequilibrium (LD), genetic effects, and life style. 
Particularly, it would be critically important to understand the trans-ancestry genetic effects on phenotypic variation for  downstream analyses of disease mechanisms and drug discovery \citep{mahajan2014genome}.  
In popular genome-wide association studies (GWAS) \citep{van2019genetic}, 
the degree of genetic similarity between two populations can be measured by trans-ancestry genetic correlation. 
Briefly, genetic correlation can be quantified as the correlation between two sets of genetic effect sizes across the genome. 
A higher trans-ancestry genetic correlation indicates better generalizability and transferability of population-specific genetic findings. 
} 

Estimating trans-ancestry genetic correlations in GWAS faces two major challenges, even though  
numerous GWAS-based methods have been proposed for genetic correlation estimation  within the same population by  using either individual-level data or GWAS summary statistics \citep{lee2012estimation,loh2015contrasting,bulik2015atlas,lu2017powerful,guo2019optimal,wang2021estimation,ning2020high,speed2019sumher,zhao2021genetic}. 
The first challenge is the heterogeneity of GWAS across populations, such as LD differences, which can cause   
  biased findings in homogeneity-agnostic analysis  \citep{zhang2021estimating}. 
However,  
most of the existing estimation methods for genetic correlation  assume that the LD patterns among the genetic variants are the same for both GWAS or that the genetic variants are independent.   
Furthermore, because of the heterogeneity caused by LD, the definition of trans-ancestry genetic correlation is even unclear \citep{wang2021estimation}. 
The second challenge is the unbalanced distribution of GWAS data among global populations. Specifically, most GWAS data are collected from European populations, whereas most non-European ancestry groups only  have limited data available.
Specifically, about $79\%$ of all GWAS participants are of European descent, and the fraction of non-Europeans in GWAS has stagnated or declined since late 2014 \citep{martin2019clinical}. Moreover, when the non-European GWAS have small sample sizes, 
summary statistics-based approaches, such as the Popcorn \citep{brown2016transethnic}, may have poor performance for trans-ancestry genetic correlation estimation \citep{ni2018estimation,zhang2021estimating}.


This paper aims to address these two challenges in trans-ancestry genetic correlation estimation by developing a LD difference-aware method that is applicable to GWAS with a small number of subjects. 
The proposed method is based on constructing genetic-predicted traits in non-European GWAS, in which the genetic effects are learned from large-scale European GWAS. 
As a result, this estimator only needs the European population to have a large GWAS sample size, and the non-European population can only have a much smaller number of individuals (for example, hundreds).
We estimate and correct the prediction-induced bias in high-dimensional genetic variants data 
and alleviate the negative influences of mismatched LD. 
Additionally, we examine the popular reference panel-based approaches \citep{pasaniuc2017dissecting} in trans-ancestry analysis.
We develop a pipeline to implement our estimator on real genotype data from the UK Biobank \citep{bycroft2018uk}. 
We use extensive simulations and real data analyses to show that our method can provide reliable estimators for complex traits from different   domains. 

This paper proceeds as follows.
In Section~\ref{sec2}, we introduce the model setups and definition for trans-ancestry genetic correlation. 
In Section~\ref{sec3}, we develop our estimator and study the influence of LD heterogeneity.  
Section~\ref{sec4} analyzes reference panel-based approaches in trans-ancestry analysis. 
Section~\ref{sec5} provides numerical details, 
including the simulation results, implementation of the estimator in real GWAS data, and real data analysis. We discuss a few future topics in Section~\ref{sec6}. 
Most of the technical details are provided in the supplementary file. 
\section{Modeling framework}\label{sec2}
\subsection{Model setups and assumptions}\label{sec2.1}
Consider two independent GWAS that are conducted on individuals from two different ancestry groups (e.g., European and Asian) with the same $p$ genetic variants, most of which are single nucleotide polymorphisms (SNPs): 
\begin{itemize}
\item Population-I GWAS: $(\X,\y)$ with $\X=(\x_{1},\ldots,\x_{p}) \in \bbR^{n\times p}$ and $\y \in \bbR^{n \times 1}$; 
\item Population-II GWAS: $(\Z,\y_{z})$ with $\Z=(\z_{1},\ldots,\z_{p}) \in \bbR^{n_{z}\times p}$ and $\y_{z} \in \bbR^{n_{z} \times 1}$, 
\end{itemize}
where $\y$ and $\y_{z}$ are continuous complex traits measured in the two GWAS with sample sizes $n$ and $n_z$, respectively. In practice, they may represent either the same trait in two different populations, such as height, or two different but genetically related traits, such as regional brain volume and intelligence \citep{zhao2019genome}.
The linear additive polygenic models are assumed between complex traits and genetic variants   \citep{jiang2016high}  as follows: 
\begin{flalign}
\y= \X\bmbeta+\bmeps  \quad \text{and}  \quad
\y_{z}=\Z\bmalpha+\bmeps_{z},  
\label{equ2.1}
\end{flalign}
where $\bmbeta^T=(\beta_1,\ldots, \beta_{p})^{T}$ and $\bmalpha^T=(\alpha_1,\ldots, \alpha_{p})^{T}$
are population-speciﬁc genetic effects and $\bmeps$ and $\bmeps_{z}$ represent population-speciﬁc random error vectors. Then,  
the genetic heritability of $\y$ and that of $\y_z$ are, respectively,  given by  
\begin{flalign}
\h^2_{\beta}=\frac{\bmbeta^T\X^T\X\bmbeta}{\bmbeta^T\X^T\X\bmbeta+\bmeps^T\bmeps} \quad \text{and}  \quad
\h^2_{\alpha}=\frac{\bmalpha^T\Z^T\Z\bmalpha}{\bmalpha^T\Z^T\Z\bmalpha+\bmeps_{z}^T\bmeps_{z}}.  
\label{equ1.2.3}
\end{flalign}
The $\h^2_{\beta}$ (or $\h^2_{\alpha}$) measures the proportion of variation in  $\y$ (or $\y_z$) that can be explained by additive genetic effects across the genome. 

We introduce some assumptions on complex traits and genetic variants in order to quantify the effect of LD heterogeneity on trans-ancestry genetic correlations.

\paragraph{SNP data} We summarize the assumptions on SNP data $\X$ and $\Z$ in Condition~\ref{con1}.
\begin{condition}
\label{con1}
\begin{enumerate}[(a).]
\item We assume $\X={\X_0}\bmSigma_{X}^{1/2}$ and $\Z={\Z_0}\bmSigma_{Z}^{1/2}$. Entries of $\X_0$ and $\Z_0$ are real-value i.i.d. random variables with mean zero, variance one,  and a finite $4$th order moment.
The $\bmSigma_X$ and $\bmSigma_Z$ are $p\times p$ population level deterministic positive definite matrices with uniformly bounded eigenvalues. 
Specifically, we have 
$0<c\le \lambda_{min}(\bmSigma_{X}) \le \lambda_{max}(\bmSigma_{X})\le C$ for all $p$ and some constants $c,C$, where  
$\lambda_{min}(\cdot)$ and $\lambda_{max}(\cdot)$ are the smallest and largest eigenvalues of a matrix, respectively. The $\bmSigma_{Z}$ satisfies similar conditions. 
For simplicity, we assume $\bmSigma_{X_{ii}}=\bmSigma_{Z_{ii}}=1$ for $i=1, \ldots,p$, or equivalently, $\X$ and $\Z$ have been column-standardized. 
\item 
Let $F^{\bmSigma_X}_p(x)=p^{-1}\cdot\sum^{p}_{i=1}\bmI(\lambda_i(\bmSigma_X)\le x )$ denote the  empirical spectral distributions (ESD) of $\bmSigma_X$, where $\bmI(\cdot)$ is the indicator function,   $\lambda_i(\cdot)$ is the $i$th eigenvalue of a matrix and $x \in \bbR$. 
As $p \to \infty$, 
  the sequence of ESDs $\{F^{\bmSigma_X}_p(x)\}_{p>1}$ converges weakly to the limiting spectral distribution (LSD) of $\bmSigma_X$, denoted as $H_X(x)$.
Similarly,   the LSDs of $\bmSigma_Z$, $\bmSigma_X^{1/2}\bmSigma_Z^{1/2}$, $\bmSigma_X\bmSigma_Z$, and 
$\bmSigma_X^2\bmSigma_Z$   exist and are denoted as $H_Z(x)$, $H_{X^{1/2}Z^{1/2}}(x)$, $H_{XZ}(x)$,  $H_{X^2Z}(x)$, and $H_{Z^2X}(x)$, respectively. 
\item 
As $\min(n,n_z) \to \infty$, we assume $p/n \to \omega$ and $p/n_z \to \omega_z$ for $\omega$ and $\omega_z \in (0,\infty)$. 
\end{enumerate}
\end{condition}
Conditions~\ref{con1}~(a) and (b) are frequently used in the  application  of random matrix theory for  high-dimensional data \citep{ledoit2011eigenvectors,dobriban2018high}. 
Moreover, $\bmSigma_{X}$ and $\bmSigma_{Z}$ can be different,  representing  different patterns of LD  in diverse populations.  
In Condition~\ref{con1}~(c),  it is natural to assume that the GWAS sample sizes $n$ and $n_z$ and the number of genetic variants $p$ are proportional to each other \citep{jiang2016high}.   Moreover, we allow a flexible range for $\omega$ and $\omega_z$, where $\omega$ can be close to one and $\omega_z$ can be much larger. 
For GWAS, unimputed genotype data typically have about half a million genetic variants and genotype imputation can increase the number to several millions.
In contrast,  biobank-scale European GWAS often have large  sample sizes (e.g., over 1 million for certain traits), whereas non-European GWAS typically have much smaller sample sizes (e.g., several thousands). 
The framework and methods developed in this paper can also be used to perform within-population genetic correlation analyses between two different traits, while controlling for LD heterogeneity among different datasets.  

\paragraph{Genetic effects and random errors}
Let $F(0,V)$ denote a generic distribution with mean zero, (co)variance $V$, and finite $4$th order moments.
We introduce the following conditions on genetic effects and random errors.
\begin{condition}\label{con2}
\begin{enumerate}[(a).]
\item
Let $\bmPhi_{\beta \beta}$, $\bmPhi_{\alpha \alpha}$, and $\bmPhi_{\beta\alpha}$ be diagonal matrices, in which $\bmPhi_{\beta \beta}=
\mbox{Diag}(\phi^2_{\beta_{1} },\ldots,\phi^2_{\beta_{i} },\ldots,\phi^2_{\beta_{p}})$,
$\bmPhi_{\alpha \alpha}=
\mbox{Diag}(\phi^2_{\alpha_{1}},\ldots,$  $\phi^2_{  \alpha_{i}},\ldots,\phi^2_{\alpha_{p}})$,  and 
$\bmPhi_{\beta\alpha}=
\mbox{Diag}$
$(\phi_{\beta_1\alpha_{1}},\ldots,\phi_{\beta_i\alpha_{i}},\ldots,\phi_{\beta_p\alpha_{p}})$ with all diagonal elements in $\in [0,\infty)$.
The joint distribution of $\bmbeta$ and $\bmalpha$ is given by   
\begin{flalign*}
\begin{pmatrix} 
\bmbeta\\
\bmalpha
\end{pmatrix}
\stackrel{}{\sim} F 
\left (
\begin{pmatrix} 
{\bf 0}\\
{\bf 0} 
\end{pmatrix},
p^{-1} \cdot
\begin{pmatrix} 
\bmPhi_{\beta \beta} & \bmPhi_{\beta\alpha} \\
\bmPhi_{\beta\alpha}^T & \bmPhi_{\alpha \alpha},
\end{pmatrix}
\right ).  
\end{flalign*}
 In addition, we have $\phi_{\beta_i\alpha_{i}}=0$ if either $\phi^2_{\beta_{i}}=0$ or $\phi^2_{\alpha_{i}}=0$.
Let $m_{\beta}$,  $m_{\alpha}$, and $m_{\beta\alpha}$ denote the number of positive entries in the $\bmPhi_{\beta \beta}$, $\bmPhi_{\alpha \alpha}$, and $\bmPhi_{\beta\alpha}$, respectively.  
As $p \to \infty$, we assume  $m_{\beta}/p \to \kappa_\beta\in (0, 1]$,  $m_{\alpha}/p \to \kappa_\alpha\in (0,1]$, $m_{\beta \alpha} / p \to \delta_{\beta \alpha} \in (0, 1]$, and  $m_{\beta\alpha}/\sqrt{m_{\beta}m_{\alpha}}\to \kappa_{\beta\alpha}\in (0,1]$.
For random errors,   $\epsilon_{j}$s in $\bmeps$ and $\epsilon_{z_{j}}$s in $\bmeps_z$ are independent random variables and  have distributions
\begin{flalign*}
\epsilon_{j} \stackrel{iid}{\sim} F(0,\sigma^2_{\epsilon}),\quad j=1, \ldots, n; \quad \mbox{and}\quad
\epsilon_{z_j} \stackrel{iid}{\sim} F(0,\sigma^2_{\epsilon_{z}}), \quad
j=1,\ldots,n_{\z}. 
\end{flalign*}
\item We assume 
$\tr(\bmSigma_X\bmSigma_Z\bmPhi_{\beta\alpha})=\phi_{\beta\alpha}\cdot \tr(\bmSigma_X\bmSigma_Z) \cdot(1+o_p(1))$, $\tr( \bmSigma_Z \bmSigma_X \bmPhi_{\beta \alpha} \bmSigma_Z \bmSigma_X \bmPhi_{\beta \alpha} ) = \phi_{\beta \alpha}^2 \cdot \tr(\bmSigma_Z \bmSigma_X \bmSigma_Z \bmSigma_X) \cdot (1 + o_p(1))$, 
and  
$\tr(\widehat{\bmSigma}_X\bmSigma_Z\widehat{\bmSigma}_X\bmPhi_{\beta\beta})=$ $\phi_{\beta}^2\cdot \tr(\widehat{\bmSigma}_X^2\bmSigma_Z)  \cdot(1+o_p(1))$
, where $\widehat{\bmSigma}_X=n^{-1} \X^T\X$, $\phi_{\beta\alpha}=\tr(\bmPhi_{\beta\alpha})/p=\sum_{i=1}^{p}\phi_{\beta_i\alpha_{i}}/p$, and
$\phi_{\beta}^2=\tr(\bmPhi_{\beta\beta})/p$ $=\sum_{i=1}^{p}\phi^2_{\beta_{i}}/p$.  
\end{enumerate}
\end{condition}
Condition~\ref{con2}~(a) details a random effect model, in which genetic effects are  independent and may vary in scale  and an arbitrary proportion of them is allowed to be zero. 
Moreover, without further restrictions on their sparsity,  $m_\beta$, $m_\alpha$, and $m_{\alpha\beta}$  are   proportional to the number of all genetic variants $p$.  
Our random effect model  weakens the classical i.i.d random effect assumption in GWAS, which typically assumes  $\bmPhi_{\beta \beta}=\mbox{Diag}[\phi^2_{\beta} \cdot \bm{I}_{m_\beta},{\bm 0_{p-m_\beta}} ]$ for some constant genetic effect $\phi^2_{\beta}$ \citep{jiang2016high,yang2011gcta,bulik2015atlas}. 
Condition~\ref{con2}~(b) provides the additional relationships between the LD structures and genetic effects required for our non-i.i.d random effect model. 
The $\phi_{\beta}^2$ is the average per-variant genetic effect and $\phi_{\beta\alpha}$ is the average per-variant contribution to the genetic correlation between $\y$ and $\y_z$.  
Intuitively, we need the entries of $\bmSigma_X\bmSigma_Z$ and $\widehat{\bmSigma}_X^2\bmSigma_Z$ to be balanced across the genome when we have non-i.i.d genetic effects. The fact that the real LD is accompanied by a block-diagonal structure may support this assumption. 
The i.i.d random effect model is a special case satisfying the Condition~\ref{con2}~(b). 
For example, let $\bmPhi_{\beta \alpha}=\phi_{\beta\alpha} \cdot \bm{I}_p$,
we have $\tr(\bmSigma_X\bmSigma_Z\bmPhi_{\beta\alpha})=\phi_{\beta\alpha}\cdot\tr(\bmSigma_X\bmSigma_Z) \cdot(1+o_p(1))$. Our condition also provide insights into the robustness of i.i.d random effect models in GWAS. 
\subsection{Heritability and trans-ancestry genetic correlation}\label{sec2.2}
According to Conditions~\ref{con1}~and~\ref{con2}, we have the following results for heritability and trans-ancestry genetic correlation. 
\paragraph{Heritability}
The heritability $h^2_{\beta}$ defined in  (\ref{equ1.2.3}) can be approximated as
\begin{flalign*} 
\h^2_{\beta}=\frac{\Vert\bmbeta\Vert_{\bmSigma_X}}{\Vert\bmbeta\Vert_{\bmSigma_X}+\sigma^2_{\epsilon}}+o_p(1) =\frac{\tr(\bmSigma_X\bmPhi_{\beta \beta})/p}{\tr(\bmSigma_X\bmPhi_{\beta \beta})/p+\sigma^2_{\epsilon}}+o_p(1) 
=\frac{\tr(\bmPhi_{\beta \beta})/p}{\tr(\bmPhi_{\beta \beta})/p+\sigma^2_{\epsilon}}+o_p(1),  
\end{flalign*}
where $\Vert\bma\Vert_{\bmSigma}=\bma^T\bmSigma\bma$ for a generic $p\times 1$ vector $\bma$ and a generic $p\times p$ matrix $\bmSigma$.
In addition, we have $\h^2_{\alpha}=[\tr(\bmPhi_{\alpha\alpha})/p]/[\tr(\bmPhi_{\alpha\alpha})/p+\sigma^2_{\epsilon_z}]+o_p(1)$.   
Our heritability is based on standardized genotypes, in which {the effects of allele frequency on phenotypes have been incorporated in the genetic effects} \citep{yang2011gcta}. Similar heritability definitions have been introduced  for the special case $\bmSigma_X=\bmSigma_Z=\I_p$ in the literature  \citep{jiang2016high,guo2019optimal}. 
 
\paragraph{Trans-ancestry genetic correlation} 
We consider two popular definitions and highlight their differences and connections, even though 
there are {\bxz several} different definitions of genetic correlation due to the LD heterogeneity in the two GWAS  \citep{brown2016transethnic}. The first one is 
the Pearson correlation of population-specific genetic effect vectors given by 
\begin{flalign*}
\varphi_{\beta\alpha}=\frac{\bmbeta^T\bmalpha}{\Vert\bmbeta\Vert\cdot\Vert\bmalpha\Vert}
=\frac{\tr(\bmPhi_{\beta\alpha})}{\{\tr(\bmPhi_{\beta\beta})\cdot
\tr(\bmPhi_{\alpha\alpha})\}^{1/2}}+o_p(1)
, 
\end{flalign*}
where  $\Vert\bma\Vert^2=\bma^T\bma$  for a generic $p\times 1$ vector $\bma$.
The $\varphi_{\beta\alpha}$, referred 
 as the ``genetic-effect correlation" in \cite{brown2016transethnic}, has been widely used  in within-population genetic correlation analysis \citep{guo2019optimal,lu2017powerful,bulik2015atlas}. 
However, a major issue is that  $\varphi_{\beta\alpha}$ does not   account  for the indirect correlation of genetic effects due to the LD among causal variants \citep{wang2021estimation,zhao2019cross}. 
Incorporating the LDs of both GWAS leads to the second   one as follows:  
\begin{flalign*}
\varphi^*_{\beta\alpha}
=\frac{\bmbeta^T\bmSigma_X^{1/2}\bmSigma_Z^{1/2}\bmalpha}{\Vert\bmbeta\Vert_{\bmSigma_X}\cdot\Vert\bmalpha\Vert_{\bmSigma_Z}}
 =\frac{\tr(\bmSigma_X^{1/2}\bmSigma_Z^{1/2}\bmPhi_{\beta\alpha})}{\{\tr(\bmSigma_X\bmPhi_{\beta\beta})\cdot
 \tr(\bmSigma_Z\bmPhi_{\alpha\alpha})\}^{1/2}}+o_p(1). 
\end{flalign*}

We discuss the connections  between $\varphi^*_{\beta\alpha}$ and $\varphi_{\beta\alpha}$. 
First, we consider the  balanced case satisfying  $\bmPhi_{\beta\alpha}=\phi_{\beta\alpha} \cdot \I_p$,
$\bmPhi_{\beta\beta}=\phi_{\beta}^2 \cdot \I_p$, and
$\bmPhi_{\beta\alpha}=\phi_{\alpha}^2 \cdot \I_p$. 
In this case, 
  we have 
\begin{flalign*}
\varphi^*_{\beta\alpha}
=\varphi_{\beta\alpha}\cdot b_1(\bmSigma_X^{1/2}\bmSigma_Z^{1/2})+o_p(1), 
\end{flalign*}
where {\bxz $b_1(\bmSigma_X^{1/2}\bmSigma_Z^{1/2})=\int_{\bbR}t dH_{X^{1/2}Z^{1/2}}(t)=\tE_{H_{X^{1/2}Z^{1/2}}}(t)$} is the first moment of the LSD of $\bmSigma_X^{1/2}\bmSigma_Z^{1/2}$. 
Thus, the ratio of $\varphi^*_{\beta\alpha}$ over $\varphi_{\beta\alpha}$ can be approximated by $b_1(\bmSigma_X^{1/2}\bmSigma_Z^{1/2})$. 
In Section~\ref{sec5}, we will  introduce a consistent estimator of $b_1(\bmSigma_X^{1/2}\bmSigma_Z^{1/2})$ in real GWAS data. 
{\bxz 
Then $\varphi^*_{\beta\alpha}$ can be obtained from $\varphi_{\beta\alpha}$ by applying this consistent estimator of $b_1(\bmSigma_X^{1/2}\bmSigma_Z^{1/2})$. }
Second, we can directly estimate   $\varphi^*_{\beta\alpha}$ through decorrelating $\X$ and $\Z$ into $\X_0=\X \bmSigma_X^{-1/2}$ and $\Z_0=\Z \bmSigma_Z^{-1/2}$. Therefore, models in (\ref{equ2.1}) reduce to  
\begin{flalign*}
\y= \X_0\widetilde{\bmbeta}+\bmeps=\X_0\bmSigma_X^{1/2}\bmbeta+\bmeps  \quad \text{and}  \quad
\y_{z}=\Z_0\widetilde{\bmalpha}+\bmeps_{z}=\Z_0\bmSigma_Z^{1/2}\bmalpha+\bmeps_{z}, 
\end{flalign*}
where  $\widetilde{\bmbeta}$ and $\widetilde{\bmalpha}$ are the corresponding genetic effects. 
Therefore, $\varphi^*_{\beta\alpha}$ is equal to the Pearson correlation of genetic effects of decorrelated SNP data $\X_0$ and $\Z_0$.  
In practice, SNP data decorrelation can be performed either  within each predetermined independent LD block as in  \citep{berisa2016approximately} or with a given window size as in \cite{bulik2015atlas}. 
As $\varphi_{\beta\alpha}$ and $\varphi^*_{\beta\alpha}$ are closely connected, we focus on estimating $\varphi_{\beta\alpha}$ from now on. 

\section{Estimation using genetic-predicted traits}\label{sec3}
In this section, we propose a consistent estimator of $\varphi_{\beta\alpha}$ and   investigate the effects of LD heterogeneity on trans-ancestry analysis. 

\subsection{Consistent estimators of trans-ancestry genetic correlation}\label{sec3.1}

Our estimator is built on the popular  GWAS marginal summary association statistics  \citep{pasaniuc2017dissecting} generated from Population-I GWAS and genetic-predicted traits   for all subjects in the Population-II GWAS. 
   We then estimate $\varphi_{\beta\alpha}$ by using either predicted or observed values on the same set of individuals in Population-II GWAS after correcting for the prediction error and LD differences.  
 We use $\widehat{\bmbeta}=n^{-1}\X^T\y$ to denote 
 Population-I GWAS summary statistics for $\y$ and calculate   the genetic-predicted values on the Population-II GWAS according to  $\widehat{\y}_\beta=\Z\widehat{\bmbeta}$. 
  The popular $\widehat{\y}_\beta$   are typically referred to as the cross-population polygenic risk scores \citep{duncan2019analysis}, the genetic endowments linked to this trait \citep{barth2020genetic}, or the genetically determined trait \citep{codd2021polygenic}.  
Based on $\widehat{\y}_\beta$, a popular estimator of trans-ancestry genetic correlation   is  given by  $G_{\beta\alpha}=\y_{z}^T\widehat{\y}_\beta/(\big\Vert\y_{z}\big\Vert\cdot\big\Vert\widehat{\y}_\beta\big\Vert)$.  
The $G_{\beta\alpha}$ has been widely   reported in the literature \citep{pirruccello2021genetic}, but its asymptotic property is largely unknown. We investigate the  asymptotic limit of $G_{\beta\alpha}$ in the following theorem.

\begin{thm}\label{thm1}
Under polygenic model~(\ref{equ2.1}) and Conditions~\ref{con1} and~\ref{con2}, 
as $\mbox{min}(n$, $n_z$, $m_{\beta\alpha}$, $p)\rightarrow\infty$,
for any $\omega, \omega_z \in (0,\infty)$, $\h_{\beta}^2, \h_{\alpha}^2\in (0,1]$, and $\varphi_{\beta\alpha} \in [-1,1]$,  
we have 
\begin{flalign*}
&G_{\beta\alpha}=\varphi_{\beta\alpha}\cdot \h_{\alpha} \cdot
\Big[\frac{b_1(\bmSigma_X^2\bmSigma_Z)}{b_1^2(\bmSigma_X\bmSigma_Z)}
+ \frac{\omega}{\h^2_{\beta} \cdot b_1(\bmSigma_X\bmSigma_Z) }\Big]^{-1/2}+o_p(1), 
\end{flalign*}
where 
$b_1(\bmSigma_X^2\bmSigma_Z)=\int_{\bbR}t dH_{X^2Z}(t)=\tE_{H_{X^{2}Z}}(t)$ and $b_1(\bmSigma_X\bmSigma_Z)=\int_{\bbR}t dH_{XZ}(t)=\tE_{H_{XZ}}(t)$. 
\end{thm}


Theorem~\ref{thm1} shows that the $G_{\beta\alpha}$ is a shrinkage estimator of $\varphi_{\beta\alpha}$  due to substantial prediction errors and LD differences. 
Intuitively, the shrinkage is largely caused by using genetic-predicted values rather than real observed observations in estimating the correlation. 
Even in within-population analyses, it is widely observed that GWAS show substantial discrepancies between the prediction accuracy and heritability for numerous complex traits \citep{daetwyler2008accuracy}. 
Similarly, utilizing the predicted values to access genetic correlations between two traits may lead to seriously underestimated results. \cite{zhao2021genetic} quantifies the potential bias for within-population analysis, which can be viewed as a special case of our results under independent genetic variant  (that is, $\bmSigma_X=\bmSigma_Z=\I_p$) and i.i.d random effect model assumptions. Under more general settings, we use  new theoretical techniques from random matrix theory \citep{bai2010spectral}  to show that the shrinkage of $\varphi_{\beta\alpha}$ in trans-ancestry analysis is jointly determined by the heritability metrics of both populations, the sample size of Population-I GWAS, and the first moments of the LSDs of $\bmSigma_X\bmSigma_Z$ and $\bmSigma_X^2\bmSigma_Z$. 
These results inspire us to propose a consistent estimator of $\varphi_{\beta\alpha}$.


\paragraph{Consistent estimator of $\varphi_{\beta\alpha}$.}
It follows from Theorem~\ref{thm1} that we have 
\begin{flalign*}
& G_{\beta\alpha}^M=G_{\beta\alpha}\cdot \Big[\frac{b_1(\bmSigma_X^2\bmSigma_Z)}{\h^2_{\alpha} \cdot b_1^2(\bmSigma_X\bmSigma_Z)}
+ \frac{\omega}{\h^2_{\beta}\h^2_{\alpha} \cdot b_1(\bmSigma_X\bmSigma_Z) }\Big]^{1/2}
=\varphi_{\beta\alpha}+o_p(1),
\end{flalign*}
which is a consistent estimator of $\varphi_{\beta\alpha}$.
For most complex traits, reliable estimates of $\h^2_{\beta}$ and $\h^2_{\alpha}$ exist \citep{yang2011gcta,jiang2016high,hou2019accurate,speed2019sumher}. 
The major challenge to approximate $G_{\beta\alpha}^M$
is to estimate  $b_1(\bmSigma_X\bmSigma_Z)$ and $b_1(\bmSigma_X^2\bmSigma_Z)$, in which the dimensions of $\bmSigma_X$ and $\bmSigma_Z$ are very large. Details about our implementation will be provided in Section~\ref{sec5.2}.

{\bxz In addition, similar to Condition~\ref{con2}~(b), we assume that 
$\tr(\bmSigma_Z \bmPhi_{\beta\alpha}) = \phi_{\beta\alpha} \cdot \tr(\bmSigma_Z) \cdot(1+o_p(1))$, 
$\tr(\bmSigma_X \bmPhi_{\beta\beta}) = \phi_{\beta}^2 \cdot \tr(\bmSigma_X) \cdot(1+o_p(1))$, and 
$\tr(\bmSigma_X\bmSigma_Z \bmSigma_X \bmPhi_{\beta\beta}) = \phi_{\beta}^2 \cdot \tr(\bmSigma_X^2 \bmSigma_Z) \cdot(1+o_p(1))$. Then, we have 
\begin{equation*}
\var(G_{\beta\alpha}^M)
=O_p\Big(\mbox{max}(\frac{1}{n}, \frac{1}{n_z},
\resizebox{0.61\hsize}{!}{$
\frac{
\tr(\bmSigma_Z\bmSigma_X\bmPhi_{\beta\alpha}\bmSigma_Z\bmSigma_X\bmPhi_{\beta\alpha})+
\tr(\bmSigma_Z\bmPhi_{\alpha\alpha}\bmSigma_Z\bmSigma_X\bmPhi_{\beta\beta}\bmSigma_X)
+
\sum_{i=1}^p C_{\beta\alpha_i} (\bmSigma_Z\bmSigma_X)^2_{ii}
}{\tr(\bmPhi_{\alpha\alpha}) \cdot  \tr(\bmPhi_{\beta\beta})}) \Big),
$
}
\end{equation*}
where $C_{\beta\alpha_i}=\tE[\alpha_i^2 \beta_i^2] - 2 (\tE[ \alpha_i \beta_i ])^2 - \tE[ \alpha_i^2 ]\tE[ \beta_i^2 ]$.} 
The variance of $G_{\beta\alpha}^M$ depends on the sample sizes for Population-I and Population-II GWAS, as well as the degree of signal sparsity related to $O(\tr(\bmPhi_{\beta\alpha})^{-1})$, {\bxz $O(\tr(\bmPhi_{\beta\beta})^{-1})$, and 
$O(\tr(\bmPhi_{\alpha\alpha})^{-1})$. The exact form of the asymptotic limit of $\var(G_{\beta\alpha}^M)$ is provided in the supplementary file.} 
The variance of $G_{\beta\alpha}^M$ increases as signals become sparser. 
Because   $n$ is much larger than $n_z$ in most cases,   $\var(G_{\beta\alpha}^M)$   has a scale of $O(1/n_z)$ when the genetic signals are not very sparse, quantifying by {\bxz
$\{\tr(\bmSigma_Z\bmSigma_X\bmPhi_{\beta\alpha}\bmSigma_Z\bmSigma_X\bmPhi_{\beta\alpha})+$
$\tr(\bmSigma_Z\bmPhi_{\alpha\alpha}\bmSigma_Z\bmSigma_X\bmPhi_{\beta\beta}\bmSigma_X)$
$+
\sum_{i=1}^p C_{\beta\alpha_i} (\bmSigma_Z\bmSigma_X)^2_{ii}\}$
$/ \{\tr(\bmPhi_{\alpha\alpha})\cdot \tr(\bmPhi_{\beta\beta})\}>n_z$.}
Thus, our estimator is reliable for polygenic or omnigentic traits \citep{timpson2018genetic} with a large number of causal variants. 


\subsection{Effects of LD heterogeneity on trans-ancestry analysis}\label{sec3.2}


In this subsection, we systematically evaluate the effects of LD heterogeneity on trans-ancestry analysis.
In practice, such LD heterogeneity play a critical role in 
  the transferability of GWAS results across populations. 
Specifically, it is widely observed that the performance of using genetic prediction of complex traits is  substantially reduced  when European GWAS results are used to predict non-European cohorts \citep{weissbrod2021leveraging}. 
Such performance drop may be partially explained by the differences in allele-normalized genetic effects between the two populations, which can be quantified by   $|\varphi_{\beta\alpha}|<1$. However, even when the genetic variants have highly similar effects in the two populations,  that is, $\varphi_{\beta\alpha}\approx 1$,  reduced prediction performance can still be observed.
For example, the trans-ancestry genetic correlation of schizophrenia is reported to be $0.98$ between East Asian and European populations, {but the prediction can be $50\%$ more accurate in within-European analysis than in European-Asian analysis \citep{lam2019comparative}.}  These remaining discrepancies may be caused by the LD heterogeneity. 

{\bxz First, we introduce a LD-related shrinkage factor as follows:  $$S_{\beta\alpha}=\Big[\frac{b_1(\bmSigma_X^2\bmSigma_Z)}{b_1^2(\bmSigma_X\bmSigma_Z)}
+ \frac{\omega}{\h^2_{\beta} \cdot b_1(\bmSigma_X\bmSigma_Z) }\Big]^{-1/2}.$$
Directly applying Theorem~\ref{thm1} shows that
smaller $S_{\beta\alpha}$ indicates more serious shrinkage and smaller $G_{\beta\alpha}$. 
To study the effects of LD heterogeneity on  trans-ancestry analysis, 
we further consider a generalized version of  Theorem~\ref{thm1} by defining $\bmSigma(t)=t\bmSigma_X+(1-t)\bmSigma_Z$, $t \in [0,1]$. 
With $\widehat{\bmbeta}=n^{-1}\X^T\y$ and SNP data $\Z(t)=\Z_0\bmSigma(t)^{1/2}$, the genetic-predicted values generated on the secondary GWAS is $\Z(t)\widehat{\bmbeta}$, which results in the estimator $G_{\beta\alpha}(t)$. 
Then we have a generalized version of $S_{\beta\alpha}$
as follows: 
\begin{flalign*}
&S_{\beta\alpha}(t)=
\Big[\frac{t\cdot \{ b_3(\bmSigma_X)+\omega/\h^2_{\beta} \cdot b_2(\bmSigma_X) \} +(1-t) \cdot \{ b_1(\bmSigma_X^2\bmSigma_Z)+\omega/\h^2_{\beta} \cdot b_1(\bmSigma_X\bmSigma_Z) \} }{ \{t\cdot b_2(\bmSigma_X) +(1-t)\cdot b_1(\bmSigma_X\bmSigma_Z)\}^2}
\Big]^{-1/2}, 
\end{flalign*}
where $b_2(\bmSigma_X)=\tE_{H_{X}}(t^2)$ and $b_3(\bmSigma_X)=\tE_{H_{X}}(t^3)$.
Here $t=0$ and $t=1$ are two special cases. When $t=0$, we have $S_{\beta\alpha}(t)=S_{\beta\alpha}$, which characterizes the shrinkage when predicting complex traits in Population-II by using the results from the Population-I GWAS. On the other hand, when $t=1$, $S_{\beta\alpha}(t)$ represents the LD-related shrinkage factor when preforming prediction between two Population-I GWAS. 

We study the effect of LD heterogeneity on $G_{\beta\alpha}(t)$ by taking the first-order derivative of $S_{\beta\alpha}(t)$ with respect to $t$, which is given by 
\begin{flalign*}
&\dot{S}_{\beta\alpha}(t)=
\frac{a(ct-d)+2bc}{2(at+b)^{3/2}},
\end{flalign*}
where $a=b_3(\bmSigma_X)-b_1(\bmSigma_X^2\bmSigma_Z)+\omega/\h^2_{\beta}\cdot\{b_2(\bmSigma_X)- b_1(\bmSigma_X\bmSigma_Z)\}$, $b=b_1(\bmSigma_X^2\bmSigma_Z)+\omega/\h^2_{\beta}\cdot b_1(\bmSigma_X\bmSigma_Z)$, $c=b_2(\bmSigma_X)- b_1(\bmSigma_X\bmSigma_Z)$, and $d=b_1(\bmSigma_X\bmSigma_Z)$.
For many real GWAS studies with large $\omega$ (that is,  $n\h^2_{\beta}$ is typically much smaller than $p$), we may have
$a\approx\omega/\h^2_{\beta}\cdot c$ and $b\approx\omega/\h^2_{\beta}\cdot d$. It follows that 
\begin{flalign*}
&\dot{S}_{\beta\alpha}(t) \approx 
\frac{c}{2(\omega/\h^2_{\beta})^{1/2}\cdot (ct+d)^{1/2}}.
\end{flalign*}
Thus, if $b_2(\bmSigma_X)> b_1(\bmSigma_X\bmSigma_Z)$, then $\dot{S}_{\beta\alpha}(t)>0$ for $t \in [0,1]$ and ${S}_{\beta\alpha}(t)$ has the largest value at $t=1$. Otherwise, if $b_2(\bmSigma_X)< b_1(\bmSigma_X\bmSigma_Z)$, then ${S}_{\beta\alpha}(t)$ has the largest value at $t=0$.
These results suggest that whether 
GWAS trans-ancestry prediction between two different populations has a lower accuracy than the within-population prediction depends on the eigenvalues of $\bmSigma_X$ and $\bmSigma_Z$. Specifically, this is largely quantified by the difference between the first moment of the LSD of $\bmSigma_X\bmSigma_Z$ and the second moment of the LSD of $\bmSigma_X$.
Moreover, since  $b_1(\bmSigma_X\bmSigma_Z) < \mbox{max} \{b_2(\bmSigma_X),b_2(\bmSigma_Z)\}$, GWAS trans-ancestry prediction between two different populations has a lower accuracy than the best within-population predictions. In the supplementary file, we also discuss the  effect of LD heterogeneity in the classical low-dimensional setting with $\omega=0$ (that is,  $n\h^2_{\beta}$ is much larger than $p$). Briefly, when sample size is much larger than the number of features, we find the the LD mismatch impacts the performance of the cross-population estimates in a more complicated way.
}

\begin{figure}[t]
\includegraphics[page=1,width=1\linewidth]{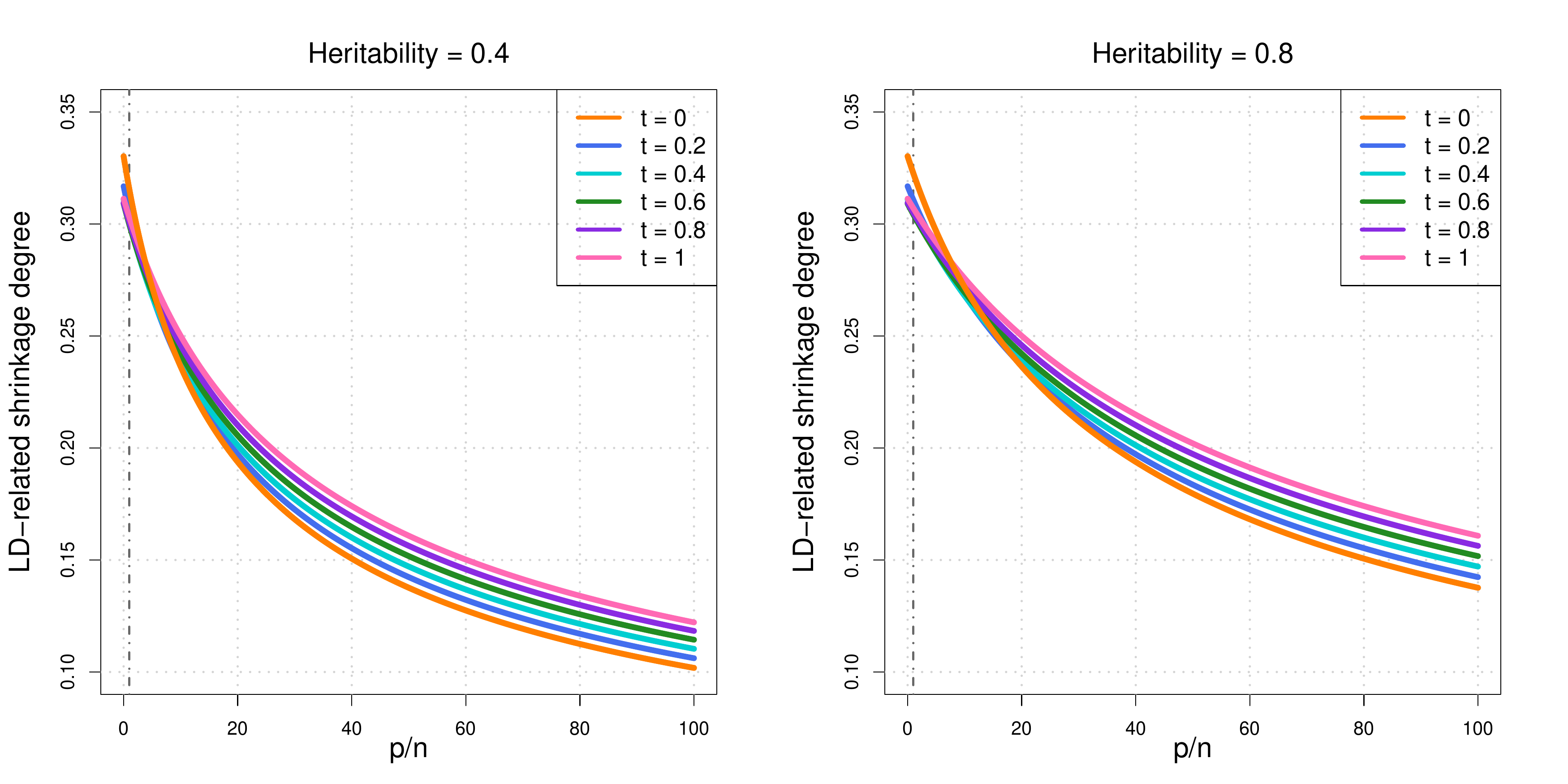}
  \caption{\bxz Illustration of the LD-related shrinkage factor $S_{\beta\alpha}(t)$ using UK Biobank genetic data. 
  The $\bmSigma_X$ is estimated from the European subjects and the $\bmSigma_Z$ is estimated from the Asian subjects. See Section~\ref{sec5.2} for more details of analysis. 
  The $x$-axis displays $\omega=p/n$ and the $y$-axis displays  shrinkage factors. Smaller values indicate more serious  shrinkage. 
We set heritability $\h^2_{\beta}=0.4$ and $0.8$ in the left and right panels, respectively. The vertical dash lines correspond to $\omega=1$.}
\label{fig1}
\end{figure}

We   provide a simulation study to illustrate our findings by setting $\h^2_{\beta}=0.5$ and $\omega\in [10^{-3}, 50]$. 
Moreover, we set  the 
$(i, j)-$entries of $\bmSigma_X$ and $\bmSigma_Z$  to be $\rho_X^{|i-j|}$ and  $\rho_Z^{|i-j|}$, respectively, where $\rho_X$ and $\rho_Z$ are positive auto-correlation coefficients. 
Larger autocorrelation coefficient indicates stronger correlations among predictors. For instance, 
for $\rho_X > \rho_Z$,   training data predictors have overall stronger correlations than those in testing data.
{\bxz Supplementary} Figure~1 illustrates the LD-related shrinkage factor {\bxz$S_{\beta\alpha}(t)$} versus $\omega=p/n$ for $\rho_X=0.5$ and $\rho_Z=0.1$  (Case~I) and for  $\rho_X=0.5$ and 
$\rho_Z=0.9$  (Case~II) {\bxz at different $t$ values}. 
In Case~I,  {\bxz we have $b_2(\bmSigma_X)> b_1(\bmSigma_X\bmSigma_Z)$ and  $S_{\beta\alpha}(t)$ increases with $t$ for relatively large $\omega$ (say $>10$). These results indicate that lower level of correlations among predictors in the testing data may decrease the prediction performance. 
In Case~II, we have $b_2(\bmSigma_X)< b_1(\bmSigma_X\bmSigma_Z)$. Thus, $S_{\beta\alpha}(t)$ decreases as $t$ increases when $\omega$ is large, suggesting that higher correlations in the testing data improves prediction performance. More numerical results can be found in Supplementary Figure~2. 

To evaluate our results in real data sets, we also examine the 
$S_{\beta\alpha}(t)$ using UK Biobank genetic data \citep{bycroft2018uk}. Details of our UK Biobank data analysis are presented in Section~\ref{sec5.2}. Briefly, European subjects in the UK Biobank are used to estimate $\bmSigma_X$ and Asian subjects are used to estimate $\bmSigma_Z$. The estimates for $b_2(\bmSigma_X)$ and $b_1(\bmSigma_X\bmSigma_Z)$ are $4.41$ and $2.86$, respectively. Therefore, we may have $b_2(\bmSigma_X)> b_1(\bmSigma_X\bmSigma_Z)$ if we use European GWAS results to generate genetic-predicted values for Asian subjects.
Figure~\ref{fig1} displays the pattern of $S_{\beta\alpha}(t)$. As expected, $S_{\beta\alpha}(t)$ has the largest value at $t = 1$ for relatively large $\omega$. 
Due to the LD difference between European and Asian subjects, European GWAS results may be less accurate in predicting Asian cohorts than they are in predicting European cohorts. 
In summary, our analysis shows that LD heterogeneity may significantly affect downstream analyses and predictions.}

\section{Reference panels in trans-ancestry analysis}\label{sec4}
When using GWAS marginal summary statistics, population-specific genotype reference panels are frequently used to account for LD patterns in within-population analyses \citep{pasaniuc2017dissecting}. 
Such reference panels are often estimated by using an external independent database, such as the 1000 Genomes reference panel \citep{10002015global},  which matches the population of interest.
In this section, we study the reference panel-based approaches in a unified framework. Specifically, we will show that   the shrinkage exists in the naive genetic correlation estimator even after  adjusting for LD with a reference panel. Moreover, we will discuss how to choose   LD reference panels in trans-ancestry analysis. 
For instance, it is not clear how to apply reference panels when the LD patterns in training and testing GWAS differ.

Let $\W =(\w_{1},\ldots,\w_{p}) \in \bbR^{n_w\times p}$ be a reference panel  database, which is independent of both   $(\X,\y)$ and   $(\Z,\y_z)$.
For trans-ancestry analysis, we examine   three different reference panels $\W$ as follows:   
\begin{itemize}
\item Reference panel-I: $\W=\W_0\bmSigma_X^{1/2}$, where the entries of $\W_0 \in \bbR^{n_w\times p}$ are   i.i.d. random variables with mean zero, variance one and a finite $4$th order moment. 
\item Reference panel-II: $\W=\W_0\bmSigma_Z^{1/2}$. 
\item Reference panel-III: $\W^T=[\bmSigma_X^{1/2}\W_{01}^T$ $\bmSigma_Z^{1/2}\W_{02}^T]$, 
where $\W_{01}^T$ and $\W_{02}^T$ are sub-matrices of 
$\W_0^T$ such that
$\W_0^T=[\W_{01}^T$ $\W_{02}^T]$, $\W_{01} \in \bbR^{n_{w_1}\times p}$, $\W_{02} \in \bbR^{n_{w_2}\times p}$, and  $n_w=n_{w_1}+n_{w_2}$. 
\end{itemize}
Reference panel-I and Reference panel-II are the panels matched to the LD of Population-I GWAS and that  of  Population-II GWAS, respectively.   Reference panel-III is a mixed reference panel corresponding to both populations.

Let $\widehat{\bmSigma}_W=n_w^{-1} \W^T\W$ be the estimated LD matrix from the reference panel $\W$.
The ridge-type reference panel-adjusted GWAS summary statistics is 
$\widehat{\bmbeta}_W=(\widehat{\bmSigma}_W+\lambda \I_p)^{-1}\widehat{\bmbeta}$, where $\lambda \in (0,\infty)$ is a ridge-type tuning parameter. 
Then the predicted trait in Population-II GWAS is 
$\widehat{\y}_{\beta_W}=\Z\widehat{\bmbeta}_W$ and the corresponding  trans-ancestry genetic correlation estimator is denoted by $G_{\beta\alpha}^W=\y_{z}^T\widehat{\y}_{\beta_W}/(\big\Vert\y_{z}\big\Vert\cdot\big\Vert\widehat{\y}_{\beta_W}\big\Vert)$.  To investigate the asymptotic limit of 
$G_{\beta\alpha}^W$, we need to impose an additional condition on the reference panel data as follows.  

\begin{condition}\label{con3}
As $n_w \to \infty$, we assume $p/n_w \to \omega_w \in (0,\infty)$. 
We assume 
$\tr\{\widehat{\bmSigma}_X(\widehat{\bmSigma}_W+\lambda\I_p)^{-1}\bmSigma_Z(\widehat{\bmSigma}_W+\lambda\I_p)^{-1}\widehat{\bmSigma}_X\bmPhi_{\beta\beta}\}=\phi^2_{\beta}\cdot\tr\{\widehat{\bmSigma}_X^2(\widehat{\bmSigma}_W+\lambda\I_p)^{-1}\bmSigma_Z(\widehat{\bmSigma}_W+\lambda\I_p)^{-1}\}  \cdot(1+o_p(1))$ and $\tr\{\bmSigma_X(\widehat{\bmSigma}_W+\lambda\I_p)^{-1}\bmSigma_Z\bmPhi_{\beta\alpha}\}=\phi_{\beta\alpha}\cdot \tr\{\bmSigma_X(\widehat{\bmSigma}_W+\lambda\I_p)^{-1}\bmSigma_Z\}  \cdot(1+o_p(1))$.  
\end{condition}
Similar to   Condition~\ref{con2}~(b), since we have non-i.i.d random effects, the entries of $\bmSigma_X(\widehat{\bmSigma}_W+\lambda\I_p)^{-1}\bmSigma_Z$ and $\widehat{\bmSigma}_X(\widehat{\bmSigma}_W+\lambda\I_p)^{-1}\bmSigma_Z(\widehat{\bmSigma}_W+\lambda\I_p)^{-1}\widehat{\bmSigma}_X$ {\bxz need to be} balanced. 
Then,  the asymptotic limit of $G^W_{\beta\alpha}$ is provided in the following theorem. 

\begin{thm}\label{thm2}
Under polygenic model~(\ref{equ2.1}) and Conditions~\ref{con1},~\ref{con2},~and~\ref{con3},
as $\mbox{min}(n$, $n_z$, $n_w$, $p)\rightarrow\infty$,
for any $\omega,\omega_w, \omega_z, \lambda \in (0,\infty)$, $\h_{\beta}^2, \h_{\alpha}^2\in (0,1]$, and  $\varphi_{\beta\alpha} \in [-1,1]$,   
we have 
\begin{flalign*}
&G_{\beta\alpha}^W=\varphi_{\beta\alpha}\cdot \h_{\alpha} \cdot 
\Big[
\frac{V_1(\lambda)\cdot \h_\beta^2}{
V_2(\lambda) \cdot \omega +V_3(\lambda) \cdot \h^2_\beta}
\Big]^{1/2}
+o_p(1),
\end{flalign*}
where 
$V_1(\lambda)=p^{-1}\tr\{\bmSigma_Z(\widehat{\bmSigma}_W+\lambda \I_p)^{-1}\bmSigma_X\}$,
$V_2(\lambda)=p^{-1}\tr\{(\widehat{\bmSigma}_W+\lambda \I_p)^{-1}\bmSigma_Z(\widehat{\bmSigma}_W+\lambda \I_p)^{-1}\bmSigma_X\}$, and 
$V_3(\lambda)=p^{-1}\tr\{(\widehat{\bmSigma}_W+\lambda \I_p)^{-1}\bmSigma_Z(\widehat{\bmSigma}_W+\lambda \I_p)^{-1}\bmSigma_X^2\}$.  

\end{thm}
Theorem~\ref{thm2} shows that the reference panel-adjusted estimator $G_{\beta\alpha}^W$ is still a shrinkage estimator of $\varphi_{\beta\alpha}$.
In addition to the sample size of the Population-I GWAS and the heritability measures of both populations, the shrinkage is jointly determined by $V_1(\lambda)$, $V_2(\lambda)$, and $V_3(\lambda)$, which are functions of the LD structures in $\X$, $\Z$, and   $\W$.  
Similar to $G_{\beta\alpha}$, we can construct a consistent estimator of $\varphi_{\beta\alpha}$ based on $G_{\beta\alpha}^W$ as follows: 
\begin{flalign*}
& G_{\beta\alpha}^{M_W}=G_{\beta\alpha}^W\cdot \Big[
\frac{V_2(\lambda) \cdot \omega +V_3(\lambda) \cdot \h^2_\beta}{
V_1(\lambda)\cdot \h_\beta^2\cdot \h_\alpha^2}
\Big]^{1/2}
=\varphi_{\beta\alpha}+o_p(1).
\end{flalign*}


\begin{figure}[t]
\includegraphics[page=2,width=1\linewidth]{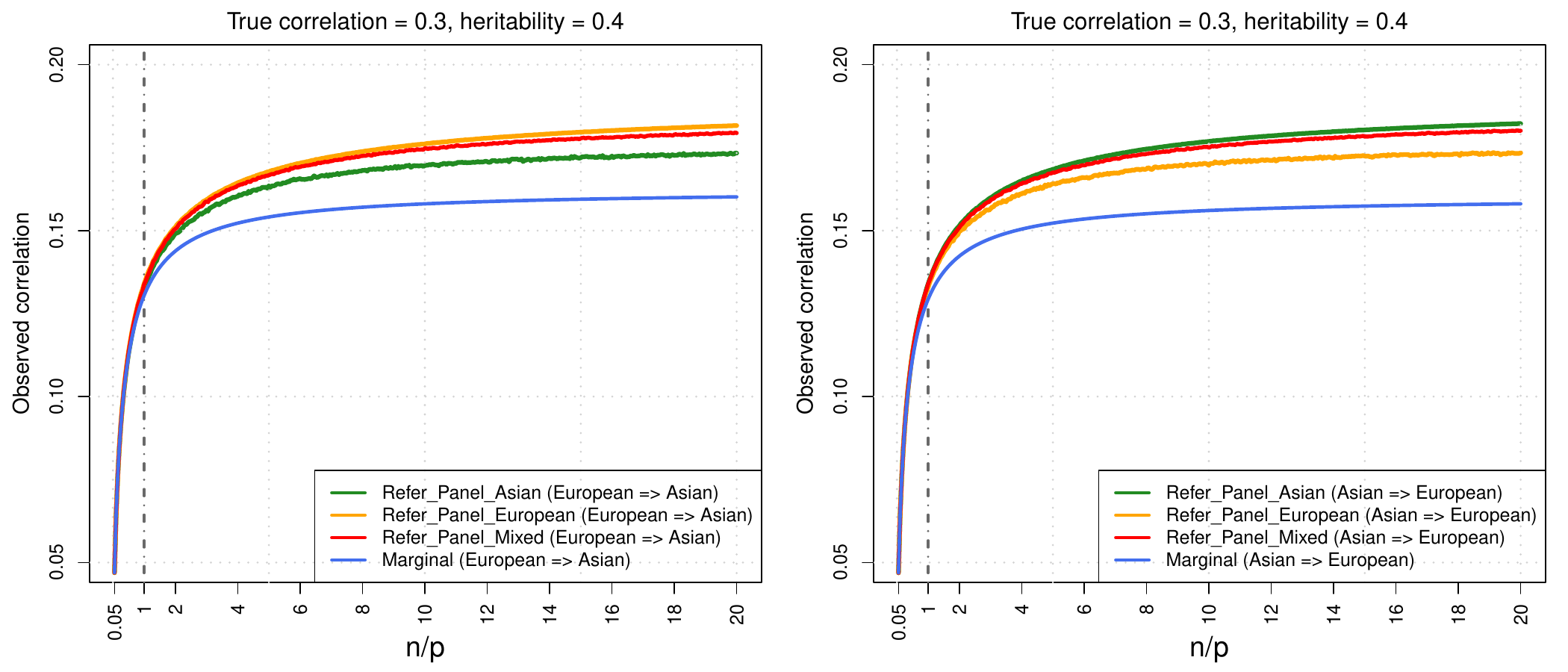}
  \caption{Comparing the naive (uncorrected) trans-ancestry genetic correlation estimators. 
   {\bxz We set $\varphi_{\beta\alpha}=0.3$ and  $\h^2_{\alpha}=\h^2_{\beta}=0.4$. 
   $\bmSigma_X$ and $\bmSigma_Z$ are estimated from the 1000 Genome reference panel. In the left panel, 
   $\bmSigma_X$ and $\bmSigma_Z$ correspond to European and (East) Asian populations, respectively, (European $=>$ Asian).
   In the right panel, 
   $\bmSigma_X$ and $\bmSigma_Z$ correspond to Asian and European populations, respectively, (Asian $=>$ European). In both panels, we include 
   Refer\_Panel\_Asian, $G_{\beta\alpha}^W$ with Asian reference panel;
   Refer\_Panel\_European, $G_{\beta\alpha}^W$ with European reference panel;
   Refer\_Panel\_Mixed, $G_{\beta\alpha}^W$ with a reference panel  having equally-mixed Asian and European samples; and Marginal, $G_{\beta\alpha}$. 
   The vertical line represents $\omega=1$.}}
\label{fig2}
\end{figure}

We use a numerical example to compare the three reference panel approaches {\bxz as well as the marginal estimator.} 
We simulate the data by setting $\varphi_{\beta\alpha}=0.3$, $\h^2_{\alpha}=\h^2_{\beta}=0.4$, $n=n_w$, and $\omega$ ranging from $0.05$ to $20$.
{\bxz
Moreover, we estimate $\bmSigma_X$ and $\bmSigma_Z$ using real genotype data from the 1000 Genome reference panel \citep{10002015global}. Specifically, we randomly select one genomic region (bp 40-50m on chromosome one) and estimate $\bmSigma_X$ and $\bmSigma_Z$ separately from the same $2,000$ genetic variants in two different populations. We consider two cases.
In Case~I,  $\bmSigma_X$ is estimated from European subjects and $\bmSigma_Z$ is estimated from (East) Asian subjects. 
Case~II represents the opposite situation, in which  $\bmSigma_X$ is estimated from the Asian subjects and $\bmSigma_Z$ is estimated from European subjects. 
In each of the two cases, we consider three reference panel options: 1) a reference panel matching the $\bmSigma_X$, the Population-I GWAS population; 2) a reference panel matching the $\bmSigma_Z$, the Population-II GWAS population; and 3) a mixed
reference panel with equally-mixed Asian and European
samples. 
}
 
Figure~\ref{fig2} illustrates trans-ancestry genetic correlation estimators for the two cases. All of these  estimators are smaller than the   true genetic correlation $\varphi_{\beta\alpha}$. {\bxz Moreover, we find the reference panel matching the Population-I GWAS generally has better performance in trans-ancestry analysis. For example,
Case~I represents a scenario in which the effects of genetic variants are estimated from European population and the genetic-predicted values are constructed in Asian population.  
In this case, the use of a reference panel that matches with the European population can greatly improve the estimation accuracy over $G_{\beta\alpha}$. 
The mixed reference panel performs very similar to the European reference panel, whereas a reference panel matching the Asian LD pattern has worse performance.
The opposite situation is shown in Case~II, in which the genetic effects are estimated from Asian population and Asian reference panel outperforms  European reference panel.
In summary, there are two important observations for reference panel approaches in trans-ancestry analysis. 
First, choosing a reference panel whose LD structure matching the Population-I GWAS may be preferred. 
Second, although reference panel-based estimator $G_{\beta\alpha}^W$ can outperform the naive estimator $G_{\beta\alpha}$, the shrinkage may still exist in $G_{\beta\alpha}^W$. Supplementary Figure~3 provides more numerical examples and a discussion of more scenarios can be found in the supplementary file. 
}

\section{Simulation and real data analysis}\label{sec5}
\subsection{Simulated genotype data}\label{sec5.1}
We numerically evaluate our theoretial results in Theorems~\ref{thm1}~and~\ref{thm2} by using  simulated genotype data sets. We set 
$n=p=14,000$, $n_w=5000$, and $n_z=500$. 
The minor allele frequency (MAF) $f$ of each genetic variant  is independently sampled from Uniform $[0.05, 0.45]$. Then 
each entry of $\X_0$, $\W_0$, and $\Z_0$ is independently generated from $\{0,1,2\}$ with probabilities $\{(1-f)^2,2f(1-f),f^2\}$, respectively. 
To mimic the real LD patterns, we construct $\bmSigma_X$ and $\bmSigma_Z$ to be block-diagonal matrices, each with $7$ big blocks. There are 
$2,000$ genetic variants in each block. 
{\bxz Similar to Figure~\ref{fig2}, correlations} 
 among the genetic variants in each block are estimated from one genomics region on chromosome one using the 1000 Genome reference panel \citep{10002015global}, while there is no correlation among genetic variants from different blocks. 
The $\bmSigma_X$ and $\bmSigma_Z$ are estimated from European and East Asian samples, respectively. 
We consider the three reference panels: i) $\W=\W_0\bmSigma_X^{1/2}$ (Ref-X),
ii) $\W=\W_0\bmSigma_Z^{1/2}$ (Ref-Z), and iii) $\W^T=[\bmSigma_X^{1/2}\W^T_{01}$ $\bmSigma_Z^{1/2}\W^T_{02}]$, where   $\W_{01}$ includes the first half samples in $\W_{0}$ and $\W_{02}$ contains the second half (Ref-Mixed). 

We simulate complex traits using model~(\ref{equ2.1}) with $\h^2_{\beta}=\h^2_{\alpha}=$ $0.2$, $0.4$, or $0.6$, reflecting from low- to high- level of heritability. 
To generate sparse  and  dense genetic signals, the proportion of variants with non-zero causal genetic effects ranges from $0.05$, $0.1$, $0.3$, to $0.5$. The causal genetic effects in $\bmbeta$ and $\bmalpha$ are sampled from normal distribution $N(0,1/p)$  with the true genetic correlation $\varphi_{\beta\alpha}$ being $0$, $0.3$ or $0.6$. 
We consider both uncorrected and corrected estimators of $\varphi_{\beta\alpha}$. 
The four naive (uncorrected) estimators of $\varphi_{\beta\alpha}$ include 
i) $G_{\beta\alpha}$ (Marginal);
ii)  $G_{\beta\alpha}^W$ estimated by Ref-X (Ref-X); 
iii) $G_{\beta\alpha}^W$ estimated by Ref-Z  (Ref-Z); and 
iv) $G_{\beta\alpha}^W$ estimated by Ref-Mixed  (Ref-Mixed). 
The four corrected estimators  $\varphi_{\beta\alpha}$ include   $G_{\beta\alpha}^M$ and the three versions of $G_{\beta\alpha}^{M_W}$. 
A total of $200$ replications are conducted for each scenario. 


The simulation results are summarized in {\bxz Supplementary Figures~4-9}. 
For nonzero $\varphi_{\beta\alpha}$, the naive estimators of $\varphi_{\beta\alpha}$ ($G_{\beta\alpha}$ and $G_{\beta\alpha}^{W}$) are all much smaller than $\varphi_{\beta\alpha}$, indicating substantial bias in the estimated genetic correlations {\bxz (Supplementary Figures~5-6)}. 
As expected, the corrected estimators $G_{\beta\alpha}^{M}$ and $G_{\beta\alpha}^{M_W}$ are very close to $\varphi_{\beta\alpha}$ in all settings, regardless of the heritability and signal sparsity {\bxz (Supplementary Figures~7-9)}. 
Among the three reference panel-based estimators,   $G_{\beta\alpha}^{M_W}$ corresponding to Ref-Z consistently has larger variance than that corresponding to Ref-X. 
In summary, these results strongly support our theoretical results, highlighting the importance of correcting for the downstream estimation bias induced by high-dimensional prediction. Since   $G_{\beta\alpha}^{M}$ performs very similarly to   $G_{\beta\alpha}^{M_W}$ and is easier to implement, we   focus on   $G_{\beta\alpha}^{M}$ in later sections when analyzing large-scale real GWAS data. 

\subsection{UK Biobank data analysis}\label{sec5.2}
\subsubsection{Implementation on real genotype data}
In this subsection, we calculate the corrected genetic correlation estimator  $G_{\beta\alpha}^M$ based  on  genotype data obtained  from the UK Biobank (UKB) study  \citep{bycroft2018uk}. 
We download the UKB genotype data and apply the following standard quality control procedures: 
excluding subjects with more than $10\%$ missing genotypes, only including SNPs with MAF $>0.01$, genotyping rate $>90\%$, and passing Hardy-Weinberg test ($p$-value $> 1\times 10^{-7}$).
After quality control, there are  $461,488$ genetic variants on $488,371$ subjects.
Based on the ethnic background information (Data-Field 21000), we focus on White (European) and Asian subjects in our analysis, which are the top two largest ancestry groups in the UKB. The sample sizes are $459,699$ and $9188$ for White and Asian groups, respectively. 
Therefore, we treat the White individuals as the Population-I GWAS with large sample size, and the Asian individuals as the Population-II GWAS with much smaller sample size. 

The major difficulty of calculating $G_{\beta\alpha}^M$ is to estimate   $b_1(\bmSigma_X^2\bmSigma_Z)$ and $b_1(\bmSigma_X\bmSigma_Z)$. 
The high dimensionality of $\bmSigma_X$ and $\bmSigma_Z$ poses major challenges to estimating their functions, such as $b_1(\bmSigma_X\bmSigma_Z)$   \citep{bickel2008regularized}.
The empirical patterns of LD in GWAS data have been shown to have a block diagonal structure: physically close genetic variants can be highly correlated, while genetic variants far from each other are typically independent \citep{pritchard2001linkage}. 
Thus,   $\bmSigma_X$ and $\bmSigma_Z$ can be assumed to be banded covariance matrices \citep{cai2016estimating}. 
Based on this assumption, we perform a simultaneous block-diagonal approximation for the two LD structures from both populations. 
Specifically, we define trans-ancestry independent LD blocks between European and Asian populations. 
We start from the previous results in \cite{berisa2016approximately}, in which  $1701$ and $1445$ independent LD blocks are defined in European and Asian populations, respectively. 
We then manually examine these LD blocks and merge them into $L=253$ trans-ancestry independent LD blocks, which tend to have larger block sizes than population-specific LD blocks.  The principle is that genetic variants in two different trans-ancestry blocks are independent in both populations, and the variants within the same block are correlated in at least one population. 
\begin{figure}[t]
\includegraphics[page=1,width=0.6\linewidth]{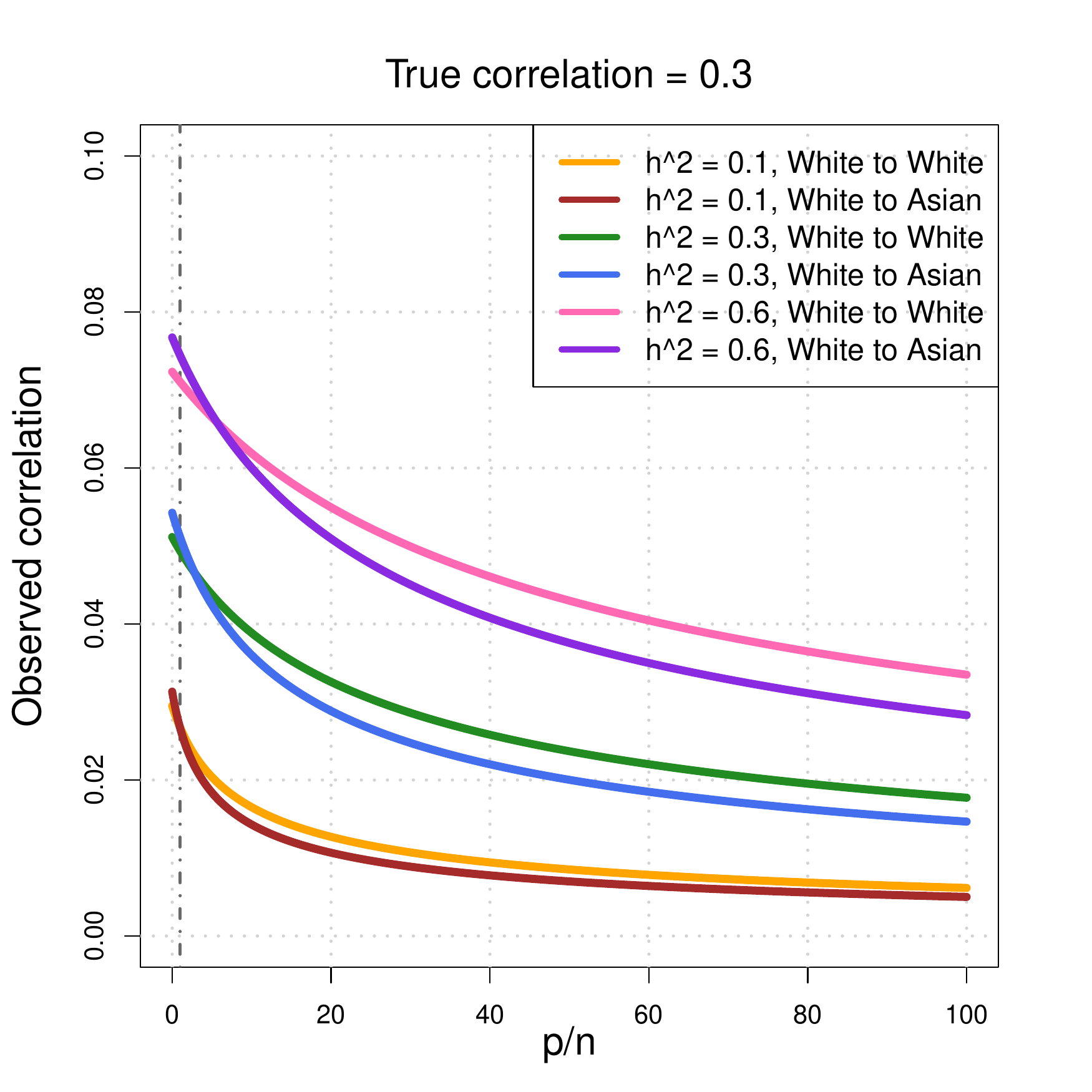}
  \caption{
  Comparing the naive (uncorrected) genetic correlation estimator $G_{\beta\alpha}$ in White-Asian (White to Asian) and within-White (White to White) analyses at different levels of heritability. The underlying true genetic correlation $\varphi_{\beta\alpha}$ is $0.3$. The $b_1(\bmSigma_X^2\bmSigma_Z)$, $b_1(\bmSigma_X\bmSigma_Z)$, $b_3(\bmSigma_X)$, and $b_2(\bmSigma_X)$ are estimated from UKB genotype data based on the LD block approximation. 
  The vertical line represents $\omega=1$.
}
\label{fig3}
\end{figure}

We estimate $\tr(\bmSigma_{X_i}\bmSigma_{Z_i})$ and 
$\tr(\bmSigma_{X_i}^2\bmSigma_{Z_i})$ as follows. 
After column-standardizing all genetic variants, for  $i = 1,\ldots,L$, we have $\widehat{\bmSigma}_{X_i}=n^{-1} \X_i^T\X_i$ and $\widehat{\bmSigma}_{Z_i}=n^{-1} \Z_i^T\Z_i$, where $\X_i$ and $\Z_i$ are the genetic variants within the $i$th block in White and Asian samples, respectively. 
We randomly select $10,000$ unrelated White individuals 
to estimate $\widehat{\bmSigma}_{X_i}$ and use all unrelated Asian subjects to estimate $\widehat{\bmSigma}_{Z_i}$.
If follows from the reasonings in \citep{yao2015sample} that
under Condition~\ref{con1}, we have  
\begin{flalign*}
&\tr(\widehat{\bmSigma}_{X_i}\widehat{\bmSigma}_{Z_i})=
\tr(\bmSigma_{X_i}\bmSigma_{Z_i}) \cdot (1+o_p(1)) \quad \mbox{and} \\
& \tr(\widehat{\bmSigma}_{X_i}^2\widehat{\bmSigma}_{Z_i})=\{\tr(\bmSigma_{X_i}^2\bmSigma_{Z_i})+n^{-1}\tr(\bmSigma_{X_i}\bmSigma_{Z_i})\tr(\bmSigma_{X_i})\} \cdot (1+o_p(1)). 
\end{flalign*}
Then,  we have  $b_1(\bmSigma_X^2\bmSigma_Z)=\sum_{i=1}^{L} \tr(\bmSigma_{X_i}^2\bmSigma_{Z_i})/p$ and $b_1(\bmSigma_X\bmSigma_Z)=\sum_{i=1}^{L}
\tr(\bmSigma_{X_i}\bmSigma_{Z_i})/p$.  
Furthermore, we approximate  
$b_1(\bmSigma_X^{1/2}\bmSigma_Z^{1/2})$ by using  $ \tr(\widehat{\bmSigma}_{X_i}^{1/2}\widehat{\bmSigma}_{Z_i}^{1/2})=
\tr(\bmSigma_{X_i}^{1/2}\bmSigma_{Z_i}^{1/2}) \cdot (1+o_p(1))$
and $b_1(\bmSigma_X^{1/2}\bmSigma_Z^{1/2})=\sum_{i=1}^{L}
\tr(\bmSigma_{X_i}^{1/2}\bmSigma_{Z_i}^{1/2})/p$.

We also estimate $b_3(\bmSigma_X)$ and $b_2(\bmSigma_X)$ in order to quantify the asymptotic shrinkage factor of  $G_{\beta\alpha}$ in within-White analysis.  
Figure~\ref{fig3} {\bxz presents}  $G_{\beta\alpha}$ values in White-Asian and within-White analyses. In all settings, $G_{\beta\alpha}$ is much smaller than the underlying true genetic correlation $\varphi_{\beta\alpha}=0.3$. 
When $\omega$ is large (say $>10$), $G_{\beta\alpha}$ in within-White analysis is larger (therefore, has smaller bias) than that in  White-Asian analysis. 
These results indicate that the LD heterogeneity between UKB White and Asian populations may lead to smaller genetic correlation estimates in trans-ancestry analysis. 
In addition, the difference between the results in within-White and White-Asian analyses decreases as the White GWAS sample size $n$ moves up (that is, $\omega$ becomes smaller towards one). When $\omega \approx 1$, the trans-ancestry genetic correlation estimator can become similar to or even slightly larger than the within-White estimator. {\bxz This observation is related to our discussion on $S_{\beta\alpha}(t)$ for small $\omega$, which can be found in the supplementary file.}  
Overall, these UKB genotype data results provide more insights into the effect of LD heterogeneity on trans-ancestry analysis. 
\subsubsection{Simulation on real genotype data}
We next perform additional simulations to examine the corrected estimator $G_{\beta\alpha}^M$ based on the trans-ancestry LD block approximation. In the White population cohort, there are $366,335$ unrelated White British subjects, of whom $350,000$ or $50,000$ are randomly selected as training GWAS samples. 
Then $1000$ unrelated Asian subjects are randomly selected to construct the genetic-predicted values. 
The proportion of causal genetic variants is set to $0.001$, $0.01$, and $0.1$, respectively. 
The causal variants are randomly selected and the nonzero genetic effects are independently derived from $N(0,1/p)$ using the GCTA \citep{yang2011gcta}.  
We set heritability $\h_{\beta}^2=\h_{\eta}^2=0.3$ and $\varphi_{\beta\eta}=0.25$; or $\h_{\beta}^2=\h_{\eta}^2=0.6$ and $\varphi_{\beta\eta}=0.5$. 
By using the summary statistics from the training GWAS, we generate genetic-predicted traits on Asian individuals. We   estimate $G_{\beta\alpha}$ and $G_{\beta\alpha}^M$ for each simulated data set. 
Each simulation setting is replicated $500$  times.

Table~\ref{tab1} summaries the simulation results. 
The naive  estimator $G_{\beta\alpha}$ is much smaller than $\varphi_{\beta\alpha}$ and their gap depends on the training GWAS sample size. 
For example, when $n=350,000$ and $\varphi_{\beta\alpha}=0.25$, the range of the mean of $G_{\beta\alpha}$ is $[0.040,0.045]$, and the average value of all sparsity levels is $0.042$ (standard error $=0.038$). These results show that the estimated genetic correlation in this setting is about $5$ times smaller than the true genetic correlation. 
Similarly, for $\varphi_{\beta\alpha}=0.5$, the mean of $G_{\beta\alpha}$ is $0.126$ (standard error $=0.049$), which is about $3$ times smaller than the true value. 
The corrected estimator $G_{\beta\alpha}^M$ is much closer to the $\varphi_{\beta\alpha}$ in all settings, with the mean being $0.253$ (standard error $=0.227$) for $\varphi_{\beta\alpha}=0.25$ and $0.515$ (standard error $=0.20$) for $\varphi_{\beta\alpha}=0.5$. 
Similar results are observed for the $n=50,000$ cases.

Our simulation results show that the trans-ancestry LD block approximation approach performs well with $G_{\beta\alpha}^M$ significantly outperforming  $G_{\beta\alpha}$ in real genotype data.
Specifically,  the variances of $G_{\beta\alpha}^M$ and $G_{\beta\alpha}$ increase with a reduction in heritability and sparsity, matching our theoretical results on $\var(G_{\beta\alpha})$. 
In addition, when the signal is very sparse, the genetic correlation might be slightly overestimated by $G_{\beta\alpha}^M$. 
It may be due to our random effect model assumptions being sensitive to very sparse genetic signals \citep{wang2021estimation}. In practice, we can first estimate the sparsity of genetic signals (that is, the polygenicity) \citep{o2019extreme} and our estimator is more robust for traits with higher polygenicity. 

\begin{table}
\centering
\caption{{Simulation results of the naive (uncorrected) estimator $G_{\beta\alpha}$ and corrected estimator $G_{\beta\alpha}^M$ on UKB genotype data. We perform simulation across a wide variety settings of heritability ($\h_{\beta}^2$, $\h_{\alpha}^2$), genetic correlation ($\varphi_{\beta\alpha}$), Population-I GWAS sample size ($n$), and genetic signal sparsity ($0.1$, $0.01$, $0.001$). 
We display the mean of estimates across $500$ simulation replications with corresponding standard errors in brackets. The ``mean'' column shows the average of the three signal sparsity levels.}}
\scalebox{0.9}{
\begin{tabular}{N|N|N|N|N|N|N|N|N|}
\cline{2-9}
     & \multicolumn{4}{c|}{$\h_{\beta}^2=\h_{\alpha}^2=0.3$ and} & \multicolumn{4}{c|}{$\h_{\beta}^2=\h_{\alpha}^2=0.6$ and} \\  
     & \multicolumn{4}{c|}{$\varphi_{\beta\alpha}=0.25$} & \multicolumn{4}{c|}{$\varphi_{\beta\alpha}=0.5$} \\  \hline
\multicolumn{1}{|c|}{Sparsity} &  $0.1$ & $0.01$ & $0.001$ & mean& $0.1$  & $0.01$ & $0.001$ & mean\\ \hline
\multicolumn{1}{|c|}{$G_{\beta\alpha}$,  $n=350k$} & $0.040$ ($0.036$)  & $0.041$ ($0.037$) & $0.045$ ($0.041$)& $0.042$ ($0.038$) & $0.121$ ($0.047$) & $0.125$ ($0.047$)  & $0.134$ ($0.054$) & $0.126$ ($0.049$) \\ \hline
\multicolumn{1}{|c|}{$G_{\beta\alpha}, n=50k$} & $0.030$ ($0.034$)  & $0.031$ ($0.037$) & $0.031$ ($0.039$)& $0.031$ ($0.037$) & $0.098$ ($0.042$) & $0.100$ ($0.044$)  & $0.107$ ($0.046$) & $ 0.102$ ($0.044$)  \\ \hline
\multicolumn{1}{|c|}{$G_{\beta\alpha}^M, n=350k$} & $0.239$ ($0.217$) & $0.245$ ($0.222$) & $0.272$ ($0.242$) & $0.253$ ($0.227$) & $0.491$ ($0.191$)   
& $0.508$ ($0.193$) & $0.544$ ($0.218$)& $0.515$ ($0.200$) \\ \hline
\multicolumn{1}{|c|}{$G_{\beta\alpha}^M, n=50k$} & $0.248$ ($0.281$) & $0.254$ ($0.304$) & $0.256$ ($0.319$) & $0.253$ ($0.301$) & $0.485$ ($0.209$)
& $0.498$ ($0.219$) & $0.529$ ($0.227$)& $0.504$ ($0.218$) \\ \hline
\end{tabular}
}
\label{tab1}
\end{table}

\subsection{Real data applications}
To evaluate the finite sample performance of $G_{\beta\alpha}^M$, 
we consider $30$ complex traits from different trait domains in the UKB study, similar to those used in \cite{kichaev2019leveraging}. 
The training GWAS is performed on these phenotypes of the unrelated White British subset in the whole White population. 
The adjusted covariates include the top $20$ genetic principal components, age, sex, age-squared, age-sex interaction, and age-squared-sex interaction. After sub-setting to subjects with complete data of genetic variants, covariates, and phenotypes, the average sample size per trait is  $n=281,506$. 
We construct the genetic-predicted values based on two independent UKB datasets. The first is a set of White but non-British subjects  ($n=19,224$) and the second is a group of Asian subjects ($n=9188$). 
White non-British and White British groups are known to have similar LD patterns.  
Accordingly, the White non-British analysis can be viewed as a positive control example, where the underlying genetic correlation is expected to be one for every pair of traits. 
The $G_{\beta\alpha}$ is estimated according to model~(\ref{equ2.1}), while adjusting for the same set of covariates as in the training GWAS data. 
Then,   $G_{\beta\alpha}^M$ is estimated by plugging the per-trait training GWAS sample size, the heritability estimated from GCTA \citep{yang2011gcta}, and the LD-related  functions estimated in within-White and White-Asian analyses detailed in Section~\ref{sec5.2}. 

\begin{figure}[t]
\includegraphics[page=2,width=1\linewidth]{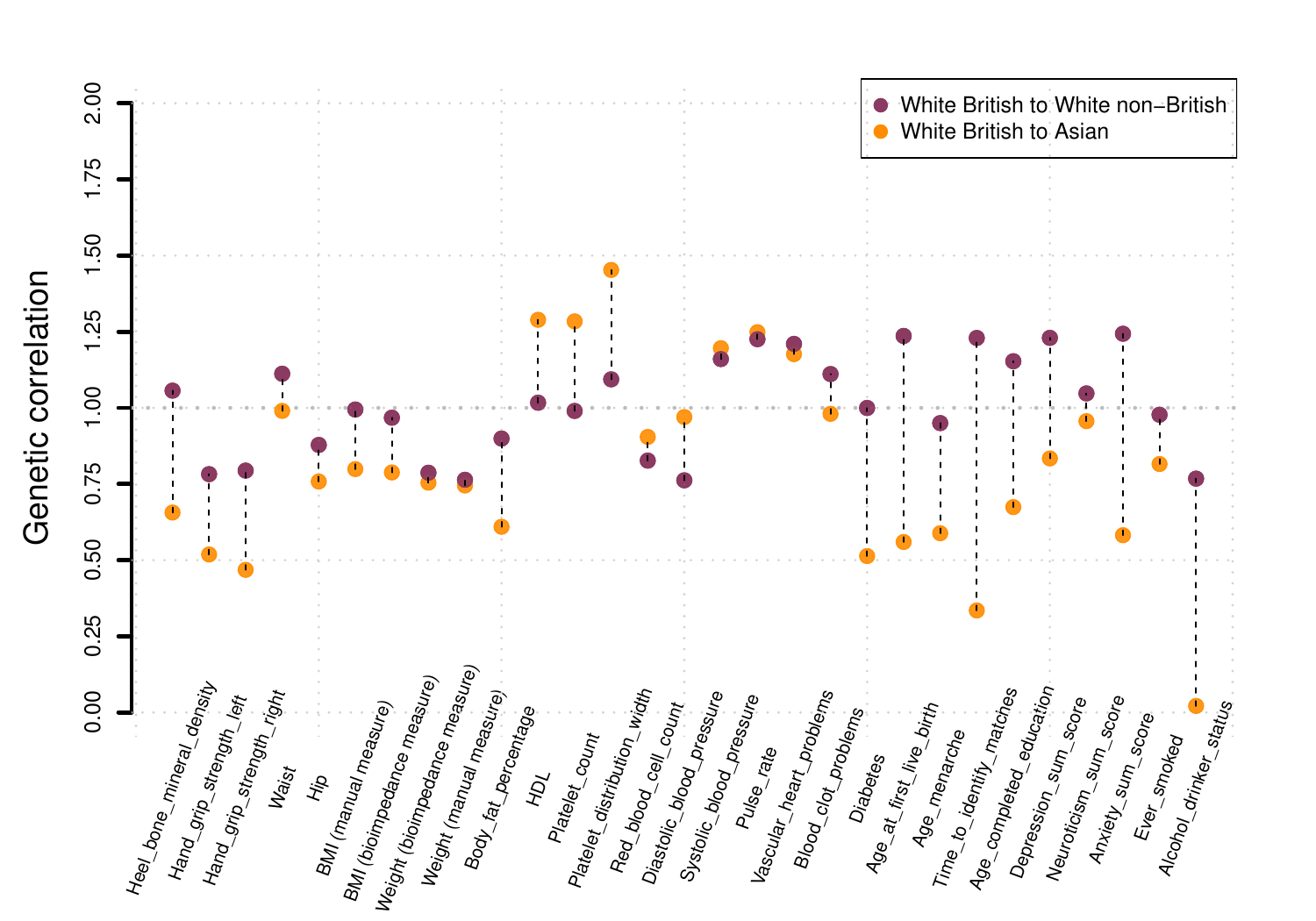}
  \caption{Genetic correlation estimated by the $G_{\beta\alpha}^M$ in White non-British and White Asian analyses across different complex traits. See Supplementary Table~1 for more details. 
}
\label{fig4}
\end{figure}

The data analysis results are summarized in Figure~\ref{fig4}, {\bxz Supplementary Figure~10}, and Supplementary Table~1. 
In the White non-British analysis,   $G_{\beta\alpha}$ ranges from $0.033$ to $0.211$ with mean = $0.133$  across the $30$ complex traits, all of which have significant $T$-test $p$-values after controlling the false discovery rate (FDR) at $5\%$ level ($P <7.57\times 10^{-06}$). The results clearly demonstrate the significant genetic influences on these complex traits. 
The estimated genetic correlations, however, are much smaller than one.
We then correct the genetic correlations and calculate   $G_{\beta\alpha}^M$ for each trait pair.  
The average $G_{\beta\alpha}^M$ of the $30$ complex traits is $1.003$ (range = $[0.762,1.243]$). A genetic correlation close to one is expected in this positive control analysis, indicating the high genetic similarity between White British and White non-British populations. 
Our results confirm our theoretical analysis and provide strong evidence that our proposed estimator can accurately reflect the underlying genetic similarity. 
Furthermore, it indicates that the widely reported naive  estimator $G_{\beta\alpha}$  in the literature, although suggesting significant genetic controls might underestimate the shared genetic co-influences between two traits. 
Next, in the White Asian analysis, the average $G_{\beta\alpha}$ is $0.120$ (range = $[0.002,0.243]$), $26$ of which pass the FDR control at $5\%$ level. 
Similar to the results in the White non-British analysis,   $G_{\beta\alpha}$ may heavily underestimate the similar genetic components between White and Asian populations. 
After correction, the mean $G_{\beta\alpha}^M$ is equal to  $0.809$ with range = $[0.013,1.453]$, which is much closer to one. 
Overall, these results suggest that the genetic architectures of Asian and White populations on these traits are similar but not the same. 
Some traits may have lower genetic similarities than others. As an example, the genetic correlation for alcohol drinker status is $0.768$ for White non-British analysis and $0.021$ for White Asian analysis. Therefore, although alcohol behavior is under genetic control, the associated genetic variants and their effects could be substantially  different between  White and Asian populations. 
In summary, UKB data analysis for a broad range of complex traits support both our theoretical and simulation results. 
Both within-population and trans-ancestry genetic correlation estimates may be improved by using the corrected genetic correlation estimator. 

\section{Discussion}\label{sec6}
The genetic correlation can measure the shared genetic influences between two traits or the same trait between two different populations. 
It provides insight into the transferability of GWAS results from one trait onto another, as well as the generalizability of GWAS findings across populations. 
In this paper, we develop estimators for trans-ancestry genetic correlation based on genetic-predicted values. 
Our estimators only need a large sample size for the GWAS from one population, and the second population may have far smaller GWAS sample size, in line with the current GWAS data resources availability. 
The influence of LD heterogeneity between two populations is quantified and corrected. Additionally, we examine the popular reference panel-based approaches in trans-ancestry settings. 
The proposed estimators are implemented on high-dimensional genetic variant data by defining the trans-ancestry LD blocks. 
We demonstrate the numerical performance of our estimators on simulated and real phenotype data in the UKB study, where our estimators provide reliable estimation in various simulation settings and complex human traits across several trait domains. 
Furthermore, our estimators can also be used to assess the genetic correlation between two complex traits in within-population analysis. In such applications, our results indicate that many reported naive genetic correlation estimates obtained with prediction-based methods are likely to underestimate the underlying genetic similarity. 
 
The proposed estimators do not assume genetic signal sparsity. Thus, they can be applied to a wide range of complex traits with different genetic architectures \citep{timpson2018genetic}.
The variance of our estimator may increase with sparser signals, likely due to the fact that we include more null genetic variants in the estimation when the number of causal variants decreases.  If prior knowledge indicates that the genetic signals of complex traits are relatively sparse, we can incorporate penalty or threshold-based approaches into our estimators, which can reduce the variance and produce more efficient estimates \citep{fan2008sure,jiang2016high}. 
In addition, our analyses are based on random effect models. In contrast to the majority of previous literature (for example, \cite{bulik2015atlas}), we do not have i.i.d assumption, allowing variants to have different genetic effect sizes. In Condition~\ref{con2}~(b), we provide the assumptions for the LD structures when the genetic effects are non-i.i.d. These results may provide insight into the robustness of distributional assumptions frequently used in random effect model-based analysis of GWAS
\citep{yang2011gcta,bulik2015atlas,speed2017reevaluation,bonnet2015heritability}. 

A few interesting problems can be further explored in trans-ancestry analyses.
First, when two populations share a low level of genetic similarity for a trait inherited in both of the populations, it would be of great interest to identify the specific loci contributing to the genetic differences between the two populations. 
For example, it is helpful to identify the genomic regions where the genetic effects are not zero in two populations, but with heterogeneous effect sizes. 
Second, trans-ancestry genetic correlation reveals the genetic similarity of traits across different populations. Such information can be incorporated in transfer learning methods to merge multiple datasets \citep{li2020transfer} and/or set up side information \citep{li2021transfer,ren2020knockoffs}. 
Finally, our analysis of LD heterogeneity provides an example of how the   heterogeneity of covariance matrix may affect estimation and prediction in high-dimensional data. The data distribution shift due to covariance structure differences can be further examined in various data types in future studies \citep{koh2021wilds}. 
\section*{Acknowledgement}
We would like to thank Ziliang Zhu, Yue Yang, and Fei Zou for helpful discussions.
This research has been conducted using the UK Biobank resource (application number $22783$), subject to a data transfer agreement.
We thank the individuals represented in the UK Biobank for their participation and the research teams for their work in collecting, processing and disseminating these datasets for analysis. We would like to thank the University of North Carolina at Chapel Hill and Purdue University and their Research Computing groups for providing computational resources and support that have contributed to these research results. 

\bibliographystyle{rss}
\bibliography{sample.bib}

\end{document}


\maketitle
\date{}
\vspace{-3ex}

\section{\large Appendix: main proofs}\label{sec1}
In this section, we highlight the key steps to prove our main theorems. 
\begin{proposition.s}\label{pop.s1}
Under polygenic model~(1) and Conditions~1 and~2, 
as $\mbox{min}(n$, $n_z$, $m_{\beta\alpha}$, $p)\rightarrow\infty$,
for any $\omega, \omega_z \in (0,\infty)$, $\h_{\beta}^2, \h_{\alpha}^2\in (0,1]$,  $\bmSigma_X$, and $\bmSigma_Z$,
we have 
\begin{flalign*}
&\frac{\big(\Z\bmalpha+\bmeps_{z}\big)^T\big(\Z\bmalpha+\bmeps_{z}\big)}{n_z\cdot\tr(\bmSigma_Z\bmPhi_{\alpha\alpha})/p +n_z\cdot{\sigma^2_{\epsilon_{z}}}} =1 + o_p(1),  \\
&\frac{\big(\X\bmbeta+\bmeps\big)^T\X\Z^T\Z\X^T\big(\X\bmbeta+\bmeps\big)}{nn_z\cdot \{nb_1(\bmSigma_X^2\bmSigma_Z)+pb_1(\bmSigma_X\bmSigma_Z)\}\cdot{\tr(\bmPhi_{\beta\beta})}/p+nn_zpb_1(\bmSigma_X\bmSigma_Z)\cdot{\sigma^2_{\epsilon}}} = 1 + o_p(1),
\end{flalign*}
and
\begin{flalign*}
\frac{\big(\Z\bmalpha+\bmeps_{z}\big)^T\Z\X^T\big(\X\bmbeta+\bmeps\big)}{nn_zb_1(\bmSigma_X\bmSigma_Z)\cdot{\tr(\bmPhi_{\beta\alpha})}/p} =1 + o_p(1)\mathbf{.}
\end{flalign*}
\end{proposition.s}
By continuous mapping theorem, we have
\begin{flalign*}
G_{\beta\alpha}
&=\varphi_{\beta\alpha}\cdot 
\frac{b_1(\bmSigma_X\bmSigma_Z)\cdot \h_{\alpha}^2}{[b_1(\bmSigma_X^2\bmSigma_Z)+b_1(\bmSigma_X\bmSigma_Z)\omega/\h^2_{\beta}]^{1/2}}+o_p(1) \\
&=\varphi_{\beta\alpha}\cdot \h_{\alpha} \cdot
\Big[\frac{b_1(\bmSigma_X^2\bmSigma_Z)}{b_1^2(\bmSigma_X\bmSigma_Z)}
+ \frac{\omega}{\h^2_{\beta} \cdot b_1(\bmSigma_X\bmSigma_Z) }\Big]^{-1/2}+o_p(1).
\end{flalign*}

Next, we have the results for reference panel estimator. 

\begin{proposition.s}\label{pop.s2}
Under polygenic model~(1) and Conditions~1,~2, and ~3, 
as $\mbox{min}(n$, $n_z$, $n_w$, $p)\rightarrow\infty$,
for any $\omega,\omega_w, \omega_z, \lambda \in (0,\infty)$, $\h_{\beta}^2, \h_{\alpha}^2\in (0,1]$,  $\bmSigma_X$, and $\bmSigma_Z$,
we have 
\begin{flalign*}
&\frac{\bmbeta^T\X^T\X(\W^T\W+\lambda n_w\I_p)^{-1}\Z^T\Z(\W^T\W+\lambda n_w\I_p)^{-1}\X^T\X\bmbeta}{n^2n_zn_w^{-2}\cdot (C_2+ \omega\cdot C_1)\cdot \tr(\bmPhi_{\beta\beta})/p^2} =1 + o_p(1),
\end{flalign*}
where
\begin{flalign*}
C_1=\tr[(\widehat{\bmSigma}_W+\lambda\I_p)^{-1}\bmSigma_Z(\widehat{\bmSigma}_W+\lambda\I_p)^{-1}\bmSigma_X] \quad \mbox{and} \quad C_2=\tr[(\widehat{\bmSigma}_W+\lambda\I_p)^{-1}\bmSigma_Z(\widehat{\bmSigma}_W+\lambda\I_p)^{-1}\bmSigma_X^2].
\end{flalign*}
Moreover,
\begin{flalign*}
&\frac{\bmeps^T\X(\W^T\W+\lambda n_w\I_p)^{-1}\Z^T\Z(\W^T\W+\lambda n_w\I_p)^{-1}\X^T\bmeps}{nn_zn_w^{-2}\cdot C_1\cdot{\sigma^2_{\epsilon}}} = 1 + o_p(1),
\end{flalign*}
and
\begin{flalign*}
\frac{\big(\Z\bmalpha+\bmeps_{z}\big)^T\Z(\W^T\W+\lambda n_w\I_p)^{-1}\X^T\big(\X\bmbeta+\bmeps\big)}{nn_zn_w^{-1}\cdot \tr[\bmSigma_X(\widehat{\bmSigma}_W+\lambda\I_p)^{-1}\bmSigma_Z]\cdot{\tr(\bmPhi_{\beta\alpha})}/p^2} =1 + o_p(1)\mathbf{.}
\end{flalign*}
\end{proposition.s}
By continuous mapping theorem, we have
\begin{flalign*}
G_{\beta\alpha}^W
&=\varphi_{\beta\alpha}\cdot \h_{\alpha} \cdot
\frac{\h_{\beta}\cdot\tr[\bmSigma_X(\widehat{\bmSigma}_W+\lambda\I_p)^{-1}\bmSigma_Z]/p }{[C_2/p\cdot \h^2_{\beta}+\omega\cdot C_1/p]^{1/2}}+o_p(1).
\end{flalign*}
\section{Theoretical details}\label{sec2}
\subsection{Results of $G_{\beta\alpha}$}
For any positive integer $k$, let $H(t)$ be the limiting spectral distribution (LSD) of a generic $p\times p$ variance-covariance matrix $\bmSigma$, and the $M(t)$ be the corresponding LSD of a estimator of $\bmSigma$, denoted as $\widehat{\bmSigma}$. 
Define the $k$th moment of $H(t)$ as $b_k(\bmSigma)=\int_{\bbR}t^kdH(t)=p^{-1}\tr(\bmSigma^k)$, and the $k$th moment of $M(t)$ as $b_k(\widehat{\bmSigma})=\int_{\bbR}t^kdM(t)=p^{-1}\tr(\widehat{\bmSigma}^k)$. (Note: change the notation of $b_1$ and integral with respect to $M(t)$)
Then, we have the following Lemma on the two sets of moments.  
\begin{lem}\label{lemma3}
[Lemma~2.16 of \cite{yao2015sample}]
Under Condition~1, as $\mbox{min}(n,p) \to \infty$, for any positive integer $k$, $b_k(\widehat{\bmSigma})$ is a function of $b_l(\bmSigma)$, for $0<l \le k$, and $\omega$. Specifically, the first three moments of $H(t)$ and the first three moments of  $M(t)$ are linked as
$b_1(\widehat{\bmSigma})=b_1(\bmSigma)$, 
$b_2(\widehat{\bmSigma})=b_2(\bmSigma)+\omega b_1(\bmSigma)^2$, and
$b_3(\widehat{\bmSigma})=b_3(\bmSigma)+3\omega b_1(\bmSigma)b_2(\bmSigma)+\omega^2 b_1(\bmSigma)^3$. 
\end{lem}

Lemma~\ref{lemma3} suggests that
$b_k(\bmSigma)$ and $b_k(\widehat{\bmSigma})$ are bounded for any $\omega \in (0,\infty)$ and  positive integer $k$. Next, we prove a lemma for the concentration of quadratic forms, which can be viewed as a generalized version for Lemma~B.26 of \cite{bai2010spectral}. 

\begin{lem} 
\label{bai-LB26-gen}
[A generalized version for Lemma~B.26 of \cite{bai2010spectral}]
Let $\A = (a_{ij})$ be an $p \times p$ nonrandom matrix and $\bmalpha = (\alpha_1, \cdots, \alpha_p)^T$ be a random vector with distribution $F(\mathbf{0}, \bmPhi_{\alpha})$, where $\bmPhi_{\alpha} \in \bbR^{p \times p}$ is positive semi-definite. Assuming that $\tE \left[ |\alpha_j|^{\ell} \right] \leq \nu_{\ell}$. Then, for any $q \geq 1$, 
\begin{flalign*}
\tE\Big[\big\{\bmalpha^T\A\bmalpha - \tr(\A \bmPhi_{\alpha})\big\}^q \Big] 
\le 
C_q \cdot \left[ \left( \nu_4 \cdot \tr(\A \bmPhi_{\alpha} \A^{\top} \bmPhi_{\alpha}) \right)^\frac{q}{2} + \nu_{2q} \cdot \tr \left( \left( \A \bmPhi_{\alpha} \A^T \bmPhi_{\alpha} \right)^{q/2} \right)
\right],
\end{flalign*}
where $C_q$ is a constant depending on $q$ only.
\end{lem}

\paragraph{Proof of Lemma~\ref{bai-LB26-gen}}
Since $\bmPhi_{\alpha}$ is positive semi-definite (thus symmetric), there exists positive semi-definite, symmetric matrix $\B$ such that $\bmPhi_{\alpha} = \B^2$. Let $\bmPhi_{\alpha}^{1/2}$ denote $\B$.
Let $\bmalpha_0$ be a random vector of dimension $p$ with distribution $F(\mathbf{0}, \I_p)$, then $\bmPhi_{\alpha}^{1/2} \bmalpha_0$ has distribution $F(\mathbf{0}, \bmPhi_{\alpha})$. It follows that $\bmPhi_{\alpha}^{1/2} \bmalpha_0 = \bmalpha$. By applying the Lemma~B.26 of \cite{bai2010spectral}, we have

\begin{flalign*}
&\tE\Big[\big\{\bmalpha^T\A\bmalpha - \tr(\A \bmPhi_{\alpha})\big\}^q \Big] \\
=& 
\tE\Big[\big\{\bmalpha_0^T (\bmPhi_{\alpha}^{1/2})^{T} \A \bmPhi_{\alpha}^{1/2} \bmalpha_0 - \tr((\bmPhi_{\alpha}^{1/2})^{T} \A \bmPhi_{\alpha}^{1/2})\big\}^q \Big] \\
\leq& 
C_q \left\{ \left( \nu_4 \cdot \tr \left( (\bmPhi_{\alpha}^{1/2})^{T} \A \bmPhi_{\alpha}^{1/2} \left[ (\bmPhi_{\alpha}^{1/2})^{T} \A \bmPhi_{\alpha}^{1/2} \right]^T \right) \right)^{q/2} \right. \\
&+ \left.
\nu_{2q} \cdot \tr \left( \left( \left[ (\bmPhi_{\alpha}^{1/2})^{T} \A \bmPhi_{\alpha}^{1/2} \right] \left[ (\bmPhi_{\alpha}^{1/2})^{T} \A \bmPhi_{\alpha}^{1/2} \right]^T \right)^{q/2} \right) \right\} \\
=& C_q \left\{ \left( \nu_4 \cdot \tr \left( \A \bmPhi_{\alpha} \A^T \bmPhi_{\alpha} \right) \right)^{q/2} 
+ 
\nu_{2q} \cdot \tr \left( \left( \bmPhi_{\alpha}^{1/2} \A \bmPhi_{\alpha} \A^T \bmPhi_{\alpha}^{1/2} \right)^{q/2} \right)
\right\}, 
\end{flalign*}
where $C_q$ is an absolute constant that only depends on $q$.

Besides, note that $\lambda \left( \bmPhi_{\alpha}^{1/2} \A \bmPhi_{\alpha} \A^T \bmPhi_{\alpha}^{1/2} \right) = \lambda \left( \A \bmPhi_{\alpha} \A^T \bmPhi_{\alpha} \right)$, and it is easy to see that both $\bmPhi_{\alpha}^{1/2} \A \bmPhi_{\alpha} \A^T \bmPhi_{\alpha}^{1/2}$ and $\A \bmPhi_{\alpha} \A^T \bmPhi_{\alpha}$ are positive-definite. Thus, the $k-$th power of each eigenvalue of $\A \bmPhi_{\alpha} \A^T \bmPhi_{\alpha}$ is an eigenvalue of $\left( \A \bmPhi_{\alpha} \A^T \bmPhi_{\alpha} \right)^k$, where $k \in \bbR_+$. A similar argument holds for $\left( \bmPhi_{\alpha}^{1/2} \A \bmPhi_{\alpha} \A^T \bmPhi_{\alpha}^{1/2} \right)^{k}$. That is,
\begin{flalign*}
\lambda \left( \left( \A \bmPhi_{\alpha} \A^T \bmPhi_{\alpha} \right)^k \right)
=
\lambda^k \left( \A \bmPhi_{\alpha} \A^T \bmPhi_{\alpha}\right)
=
\lambda^k \left( \bmPhi_{\alpha}^{1/2} \A \bmPhi_{\alpha} \A^T \bmPhi_{\alpha}^{1/2} \right)
= \lambda \left( \left( \bmPhi_{\alpha}^{1/2} \A \bmPhi_{\alpha} \A^T \bmPhi_{\alpha}^{1/2} \right)^k \right). \\
\end{flalign*}
Thus, we have 
\begin{flalign*}
\tE\Big[\big\{\bmalpha^T\A\bmalpha - \tr(\A \bmPhi_{\alpha})\big\}^q \Big] 
\leq&
C_q \left\{ \left( \nu_4 \tr \left( \A \bmPhi_{\alpha} \A^T \bmPhi_{\alpha} \right) \right)^{q/2} 
+ 
\nu_{2q} \tr \left( \left( \A \bmPhi_{\alpha} \A^T \bmPhi_{\alpha}\right)^{q/2} \right)
\right\}. 
\end{flalign*}

\paragraph{Remark} Observe that the upper bound in Lemma~\ref{bai-LB26-gen} depends on $\tr \left( \A \bmPhi_{\alpha} \A^T \bmPhi_{\alpha} \right)$ and $\tr \left( \left( \A \bmPhi_{\alpha} \A^T \bmPhi_{\alpha} \right)^{q/2} \right)$. Comparing with the upper bound in Lemma~B.26 of \cite{bai2010spectral}, the order of $\tr \left( \A \bmPhi_{\alpha} \A^T \bmPhi_{\alpha} \right)$ will affect the order of the upper bound. Thus, it is desired to control its order to be the same as $\tr(\A \A^{T})$, possibly with further assumption on $\bmPhi_{\alpha}$.
In order to do that, we first state a well-known fact about eigenvalues of positive semi-definite matrices, and it could be easily derived from Problem III.6.5 in \cite{bhatia_1997}.
\begin{lem} \label{eval-prod-ub}
Consider positive semi-definite matrices $\A, \B \in \bbR^{n \times n}$. Let $\lambda_i(\A), \lambda_i(\A)$ be the eigenvalues of $\A$, $\B$ in decreasing order, respectively. Then the following holds for all $1 \leq i, j \leq n$ such that $i + j \leq n + 1$:
\begin{flalign*}
\lambda_{i+j-1}(\A \B) \leq \lambda_i(\A) \lambda_j(\B).
\end{flalign*}
\end{lem}
Now we use Lemma~\ref{eval-prod-ub} to prove the following lemma, which is an variant of Lemma~\ref{bai-LB26-gen} whose bound does not depend on $\bmPhi_{\bmalpha}$. 

\begin{lem} \label{bai-LB26-gen-2} [A generalized version for Lemma~B.26 of \cite{bai2010spectral} that does not depend on $\bmPhi_{\bmalpha}$]
Let $\A = (a_{ij})$ be an $p \times p$ nonrandom matrix and $\bmalpha = (\alpha_1, \cdots, \alpha_p)'$ be a random vector with distribution $F(\mathbf{0}, \bmPhi_{\alpha})$, where $\bmPhi_{\alpha} \in \bbR^{p \times p}$ is positive definite. Let the eigenvalues of $\bmPhi_{\alpha}$ be uniformly bounded, that is, there exists some absolute constant $C_{\alpha \alpha} > 0$ such that $\lambda_{\max}(\bmPhi_{\alpha}) \leq C_{\alpha \alpha}$. Assume that $\tE \left[ |\alpha_j|^{\ell} \right] \leq \nu_{\ell}$. Then, for any $q \geq 1$, 
\begin{flalign*}
\tE\Big[\big\{\bmalpha^T\A\bmalpha - \tr(\A \bmPhi_{\alpha})\big\}^q \Big] 
\le 
C_q \cdot \left[ \left( \nu_4 \cdot C_{\alpha \alpha}^2 \cdot \tr(\A \A^{\top} ) \right)^\frac{q}{2} + \nu_{2q} \cdot C_{\alpha \alpha}^q \cdot \tr \left( \left( \A  \A^T \right)^{q/2} \right)
\right],
\end{flalign*}
where $C_q$ is a constant depending on $q$ only.
\end{lem}

\paragraph{Proof of Lemma~\ref{bai-LB26-gen-2}} For $\tr \left( \A \bmPhi_{\alpha} \A^T \bmPhi_{\alpha} \right)$, use Lemma~\ref{eval-prod-ub}, and note that $\lambda_{\max}(\bmPhi_{\alpha}) = \lambda_1(\bmPhi_{\alpha}) \leq C_{\alpha \alpha}$, we have
\begin{flalign*}
\tr \left( \A \bmPhi_{\alpha} \A^T \bmPhi_{\alpha} \right) 
&= \sum_{i=1}^p \lambda_i \left( \A \bmPhi_{\alpha} \A^T \bmPhi_{\alpha} \right) 
\leq \lambda_{\max}(\bmPhi_{\alpha}) \sum_{i=1}^p \lambda_i \left( \A \bmPhi_{\alpha} \A^T \right) \\
&= \lambda_{\max}(\bmPhi_{\alpha}) \sum_{i=1}^p \lambda_i \left( \bmPhi_{\alpha} \A^T \A  \right) 
\leq \lambda_{\max}^2(\bmPhi_{\alpha}) \sum_{i=1}^p \lambda_i \left( \A^T \A  \right) 
\leq C_{\alpha \alpha}^2 \tr \left( \A \A^T \right).
\end{flalign*}
For $\tr \left( \left( \A \bmPhi_{\alpha} \A^T \bmPhi_{\alpha} \right)^{q/2} \right)$, use Lemma~\ref{eval-prod-ub} and the fact that $\lambda_i^k(\A) = \lambda_i(\A^k)$ for any positive semi-definite $\A \in \R^{n \times n}$, any $1 \leq i \leq n$ and any $k \geq 0$, get
\begin{flalign*}
\tr \left( \left( \A \bmPhi_{\alpha} \A^T \bmPhi_{\alpha} \right)^{q/2} \right)
&= \sum_{i=1}^p \lambda_i \left( \left( \A \bmPhi_{\alpha} \A^T \bmPhi_{\alpha} \right)^{q/2} \right) 
= \sum_{i=1}^p \lambda_i^{q/2} \left( \A \bmPhi_{\alpha} \A^T \bmPhi_{\alpha} \right)  \\
&\leq \lambda_{\max}^{q/2}(\bmPhi_{\alpha}) \sum_{i=1}^p \lambda_i^{q/2} \left( \A \bmPhi_{\alpha} \A^T \right) 
= \lambda_{\max}^{q/2}(\bmPhi_{\alpha}) \sum_{i=1}^p \lambda_i^{q/2} \left( \bmPhi_{\alpha} \A^T \A  \right) \\
&\leq \lambda_{\max}^q(\bmPhi_{\alpha}) \sum_{i=1}^p \lambda_i^{q/2} \left( \A^T \A  \right) 
= \lambda_{\max}^q(\bmPhi_{\alpha}) \sum_{i=1}^p \lambda_i^{q/2} \left( \A \A^T  \right) \\
&= \lambda_{\max}^q(\bmPhi_{\alpha}) \sum_{i=1}^p \lambda_i \left( \left( \A \A^T \right)^{q/2} \right)
\leq C_{\alpha \alpha}^q \tr \left( \left( \A \A^T \right)^{q/2} \right).
\end{flalign*}
The following lemma can be viewed as corollary to Lemma~\ref{bai-LB26-gen} (or Lemma~\ref{bai-LB26-gen-2}). 
\begin{lem}[Corollary to Lemma~\ref{bai-LB26-gen}] \label{quadratic-form-plim}
{\bxz Consider a random vector $\bmalpha \in \bbR^p$ with $\bmalpha \sim F(\mathbf{0}, \bmSigma_{\alpha})$, $\bmSigma_{\alpha}$ is positive semi-definite, and a nonrandom {\xcy positive semi-definite} matrix $\A \in \bbR^{p \times p}$.} {\xcy Further assume $\tr(\A \bmSigma_{\alpha}) \neq 0$.} Then we have
\begin{flalign*}
\frac{\bmalpha^{T} \A \bmalpha}{\tr(\A \bmSigma_{\alpha})} \stackrel{p}{\to} 1 \text{, as } p \to \infty.
\end{flalign*}
Moreover, 
consider random vectors $\bmalpha \in \bbR^{p}$ and $\bmbeta \in \bbR^{p}$, where 
$$
\begin{pmatrix} 
\bmalpha \\
\bmbeta
\end{pmatrix}
\stackrel{}{\sim} F 
\left (
\begin{pmatrix} 
{\bf 0}\\
{\bf 0} 
\end{pmatrix},
\begin{pmatrix} 
\bmSigma_{\alpha \alpha} & \bmSigma_{\alpha \beta} \\
\bmSigma_{\beta \alpha}^T & \bmSigma_{\beta \beta},
\end{pmatrix}
\right ),$$ 
and a nonrandom {\xcy positive semi-definite} matrix $\A \in \bbR^{2p}$. Then we have 
\begin{flalign*}
\frac{\bmalpha^T \A \bmbeta}{ \tr(\A \bmSigma_{\beta \alpha}) } \stackrel{p}{\to} 1,\ \text{as } p \to \infty.
\end{flalign*}
\end{lem}
\paragraph{Proof of Lemma~\ref{quadratic-form-plim}}
{\xcy We first prove the result about $\bmalpha^{T} \A \bmalpha$.
Apply Lemma~\ref{bai-LB26-gen} and let $q=2$, then we have
\begin{flalign*}
\tE\Big[\big\{\bmalpha^T\A\bmalpha - \tr(\A \bmPhi_{\alpha})\big\}^2 \Big] 
\le 
C_2 \cdot \left[ \left( \nu_4 \cdot \tr(\A \bmPhi_{\alpha} \A^{\top} \bmPhi_{\alpha}) \right) + \nu_{4} \cdot \tr \left( \left( \A \bmPhi_{\alpha} \A^T \bmPhi_{\alpha} \right) \right)
\right].
\end{flalign*}
Since $\tE[\bmalpha^T\A\bmalpha] = \tr(\A \bmPhi_{\alpha})$, we have
\begin{flalign*}
\var(\bmalpha^T\A\bmalpha)
\le 
2 \cdot C_2 \cdot \nu_4 \cdot \tr(\A \bmPhi_{\alpha} \A^{\top} \bmPhi_{\alpha}).
\end{flalign*}
Thus 
\begin{flalign*}
\var \left( \frac{\bmalpha^T\A\bmalpha}{\tr(\A \bmPhi_{\alpha})} \right)
\le 
2 \cdot C_2 \cdot \nu_4 \cdot \frac{\tr(\A \bmPhi_{\alpha} \A^{\top}\bmPhi_{\alpha})}{[\tr(\A \bmPhi_{\alpha})]^2}. 
\end{flalign*}
Apply Markov's inequality. For any constant $c > 0$, 
\begin{flalign*}
P\left(
\abs*{ \frac{\bmalpha^T\A\bmalpha - \tr(\A \bmPhi_{\alpha})}{\tr(\A \bmPhi_{\alpha})} } \geq c
\right)
\leq 
\frac{\var \left( \frac{\bmalpha^T\A\bmalpha}{\tr(\A \bmPhi_{\alpha})} \right)}{c^2} \to 0 \text{ as }p \to \infty.
\end{flalign*}
This is precisely the definition of ${\bmalpha^{T} \A \bmalpha} / {\tr(\A \bmSigma_{\alpha})} \stackrel{p}{\to} 1$.
}

{\xcy Now} prove the results of $\bmalpha^T \A \bmbeta$. 
Define a matrix $\B \in \bbR^{(2p) \times (2p)}$ and a vector $\z \in \bbR^{(2p)}$ such that 
\begin{flalign*}
\B = \begin{bmatrix}
\B_{11} & \A \\
\A^{\top} & \B_{22}
\end{bmatrix}, 
\bmgamma = \begin{bmatrix} 
\bmalpha \\
\bmbeta 
\end{bmatrix} \sim F(\mathbf{0}, \bmSigma_{\gamma}),
\end{flalign*}
with $\B_{11} \in \bbR^{p \times p}, \B_{22} \in \bbR^{p \times p}$ being any positive semi-definite matrices {\xcy such that $\B$ is positive semi-definite}. Then by Lemma~\ref{bai-LB26-gen}, $\tr(\B \bmSigma_{\gamma}) / p$ and $\bmgamma^{\top} \B \bmgamma/p$ converge in probability to the same limit, similarly for $\tr(\B_{11} \bmSigma_{\alpha}) / p$ and $\bmalpha^{\top} \B_{11} \bmalpha/p$, and for $\tr(\B_{22} \bmSigma_{\beta}) / p$ and $\bmbeta^{\top} \B_{22} \bmbeta/p$. {\bxz Let $\plim_{p \to \infty}$ denote converge in probability, we have}
\begin{flalign*}
\plim_{p \to \infty} \frac{\bmalpha^{\top} \A \bmbeta}{p}
&= \plim_{p \to \infty} \frac{1}{2p} (\z^{\top} \B \z - \bmalpha^{\top} \B_{11} \bmalpha - \bmbeta^{\top} \B_{22} \bmbeta) \\
&= \plim_{p \to \infty} \frac{1}{2p} ( \tr(\B \bmSigma_{\gamma}) - \tr(\B_{11} \bmSigma_{\alpha}) - \tr(\B_{22} \bmSigma_{\beta}) ) \\
&= \plim_{p \to \infty} \frac{\tr(\A \bmSigma_{\beta \alpha})}{p}. 
\end{flalign*}
It follows that 
\begin{flalign*}
\frac{\bmalpha^{\top} \A \bmbeta}{\tr(\A \bmSigma_{\beta \alpha})} \stackrel{p}{\to} 1.
\end{flalign*}
Based on the above lemmas,  we have the key lemma used in our analysis as follows. 
\begin{lem}
\label{lemma4}
Let $\widehat{\bmSigma}_{\X}=n^{-1}\X^T\X$,
$\widehat{\bmSigma}_{\Z}=n_z^{-1}\Z^T\Z$, $\B_{k_1,k_2}=\widehat{\bmSigma}_{\X}^{k_1}\widehat{\bmSigma}_{\Z}^{k_2}$,
and define $\A^0=\I$ for any matrix $\A$. 
Moreover, let $\bmalpha$ be a $p$-dimensional random vector of independent elements with mean zero, variance $\bmPhi_{\alpha}=\mbox{Diag}(\phi^2_1,\cdots,\phi^2_p)$, and finite fourth order moments, we have 
\begin{flalign*}
\bmalpha^T\B_{k_1,k_2}\bmalpha= \tr(\B_{k_1,k_2}\bmPhi_{\alpha} )\cdot\{1+o_p(1)\}.
\end{flalign*}
\end{lem}
Lemma~\ref{lemma4} indicates that the quadratic forms of $\B_{k_1,k_2}$ concentrate around their means. Finally, the proposition below summarizes the results on the mean of quadratic forms.
The consistency of quadratic forms in Proposition~S\ref{pop.s1} follows from Lemma~\ref{lemma4}. 
\begin{proposition.s}\label{pop.i1}
Under the same conditions as in Propositions~S\ref{pop.s1}, we have 
\begin{flalign*}
&\tE \Big\{\big(\Z\bmalpha+\bmeps_{z}\big)^T\big(\Z\bmalpha+\bmeps_{z}\big) \Big\} = n_z\cdot\tr(\bmSigma_Z\bmPhi_{\alpha\alpha})/p +n_z\cdot{\sigma^2_{\epsilon_{z}}}, &\\
&\tE\Big\{\big(\X\bmbeta+\bmeps\big)^T\X\Z^T\Z\X^T\big(\X\bmbeta+\bmeps\big)\Big\}\\
&=nn_z\cdot \{nb_1(\bmSigma_X^2\bmSigma_Z)+pb_1(\bmSigma_X\bmSigma_Z)\}\cdot{\tr(\bmPhi_{\beta\beta})}/p+nn_zpb_1(\bmSigma_X\bmSigma_Z)\cdot{\sigma^2_{\epsilon}}, &\\
&\tE\Big\{\big(\Z\bmalpha+\bmeps_{z}\big)^T\Z\X^T\big(\X\bmbeta+\bmeps\big)\Big\}=nn_zb_1(\bmSigma_X\bmSigma_Z)\cdot{\tr(\bmPhi_{\beta\alpha})}/p.&
\end{flalign*}
\end{proposition.s}

\subsection{{Variance of $G_{\beta\alpha}$}}
In this section, we study the variance of $G_{\beta\alpha}$. 
Consider 
\begin{flalign*}
G_{\beta\alpha}=\frac{\y_{\alpha}^T\widehat{\y}_\beta}{\big\Vert\y_{\alpha}\big\Vert\cdot\big\Vert\widehat{\y}_\beta\big\Vert} \quad \mbox{and} \quad
G_{\beta\alpha}^2=\frac{(\y_{\alpha}^T\widehat{\y}_\beta)^2}{\big\Vert\y_{\alpha}\big\Vert^2\cdot\big\Vert\widehat{\y}_\beta\big\Vert^2},
\end{flalign*}
where
\begin{flalign*}
\big\Vert\y_{\alpha}\big\Vert^2&= \y_{\alpha}^T\y_{\alpha}=\big(\Z\bmalpha+\bmeps_{z}\big)^T\big(\Z\bmalpha+\bmeps_{z}\big)\\
&=
\big\{n_z\cdot\tr(\bmSigma_Z\bmPhi_{\alpha\alpha})/p +n_z\cdot{\sigma^2_{\epsilon_{z}}}\big\} \cdot\{1+o_p(1)\}
=n_z\cdot\tr(\bmPhi_{\alpha\alpha})/(p\cdot\h^2_{\alpha})\cdot\{1+o_p(1)\},
\end{flalign*}
and 
\begin{flalign*}
\big\Vert\widehat{\y}_\beta\big\Vert^2&= \widehat{\y}_\beta^T\widehat{\y}_\beta
=\big(\X\bmbeta+\bmeps\big)^T\X\Z^T\Z\X^T\big(\X\bmbeta+\bmeps\big)\\
&=\Big[
nn_z\cdot \{nb_1(\bmSigma_X^2\bmSigma_Z)+pb_1(\bmSigma_X\bmSigma_Z)\}\cdot{\tr(\bmPhi_{\beta\beta})}/p+nn_zpb_1(\bmSigma_X\bmSigma_Z)\cdot{\sigma^2_{\epsilon}}
\Big]
\cdot\{1+o_p(1)\} \\
&=\Big[
n^2n_z\cdot{\tr(\bmPhi_{\beta\beta})}\cdot \{b_1(\bmSigma_X^2\bmSigma_Z)+ \omega \cdot b_1(\bmSigma_X\bmSigma_Z)/\h^2_{\beta}\}/p 
\Big]
\cdot\{1+o_p(1)\}.
\end{flalign*}
In addition, 
\begin{flalign*}
&{\xcy\tE}[(\y_{\alpha}^T\widehat{\y}_\beta)^2]\\
&=\big\{\big(\Z\bmalpha+\bmeps_{z}\big)^T\Z\X^T\big(\X\bmbeta+\bmeps\big)\big\}^2\\
&={\xcy\tE} \left[ \big\{\bmalpha^T\Z^T\Z\X^T\X\bmbeta+
\bmalpha^T\Z^T\Z\X^T\bmeps+
\bmeps_{z}^T\Z\X^T\X\bmbeta+ 
\bmeps_{z}^T\Z\X^T\bmeps  \big\}^2 \right] \\
&= {\xcy\tE} \left[ \bmalpha^T\Z^T\Z\X^T\X\bmbeta\bmalpha^T\Z^T\Z\X^T\X\bmbeta
+\bmalpha^T\Z^T\Z\X^T\X\bmbeta\bmalpha^T\Z^T\Z\X^T\bmeps
+ \bmalpha^T\Z^T\Z\X^T\X\bmbeta\bmeps_{z}^T\Z\X^T\X\bmbeta+ \right.\\
&\bmalpha^T\Z^T\Z\X^T\X\bmbeta\bmeps_{z}^T\Z\X^T\bmeps+
\bmalpha^T\Z^T\Z\X^T\bmeps\bmalpha^T\Z^T\Z\X^T\X\bmbeta+
\bmalpha^T\Z^T\Z\X^T\bmeps\bmalpha^T\Z^T\Z\X^T\bmeps+\\
&\bmalpha^T\Z^T\Z\X^T\bmeps\bmeps_{z}^T\Z\X^T\X\bmbeta+
\bmalpha^T\Z^T\Z\X^T\bmeps\bmeps_{z}^T\Z\X^T\bmeps+
\bmeps_{z}^T\Z\X^T\X\bmbeta\bmalpha^T\Z^T\Z\X^T\X\bmbeta+\\
&\bmeps_{z}^T\Z\X^T\X\bmbeta\bmalpha^T\Z^T\Z\X^T\bmeps+
\bmeps_{z}^T\Z\X^T\X\bmbeta\bmeps_{z}^T\Z\X^T\X\bmbeta+
\bmeps_{z}^T\Z\X^T\X\bmbeta\bmeps_{z}^T\Z\X^T\bmeps +\\
&\left. \bmeps_{z}^T\Z\X^T\bmeps\bmalpha^T\Z^T\Z\X^T\X\bmbeta+
\bmeps_{z}^T\Z\X^T\bmeps\bmalpha^T\Z^T\Z\X^T\bmeps+
\bmeps_{z}^T\Z\X^T\bmeps\bmeps_{z}^T\Z\X^T\X\bmbeta+
\bmeps_{z}^T\Z\X^T\bmeps\bmeps_{z}^T\Z\X^T\bmeps \right]\\
&={\xcy\tE} \big(\bmalpha^T\Z^T\Z\X^T\X\bmbeta\bmalpha^T\Z^T\Z\X^T\X\bmbeta+
\bmalpha^T\Z^T\Z\X^T\bmeps\bmalpha^T\Z^T\Z\X^T\bmeps+\bmeps_{z}^T\Z\X^T\X\bmbeta\bmeps_{z}^T\Z\X^T\X\bmbeta+\\
&\bmeps_{z}^T\Z\X^T\bmeps\bmeps_{z}^T\Z\X^T\bmeps
\big)\cdot\{1+o_p(1)\}=\big(\bm{A_1}+\bm{A_2}+\bm{A_3}+\bm{A_4}
\big)\cdot\{1+o_p(1)\}.
\end{flalign*}

From now on, we ignore the $1/n^2$ in $(\y_{\alpha}^T\widehat{\y}_\beta)^2$ for simplicity. After careful calculation, we have the following results on $\tE[(\y_{\alpha}^T\widehat{\y}_\beta)^2]$ and $\tE[\| \y_{\alpha} \|_2^2 \cdot \| \widehat{\y}_{\beta} \|_2^2]$, which will be used in the calculation of the rate of $Var(G_{\beta \alpha})$.

First, the dominant terms of $\bm{A_1} = \tE [\tr( \bmalpha^T \Z^T \Z \X^T \X \bmbeta \bmalpha^T \Z^T \Z \X^T \X \bmbeta)]$ are 
\begin{flalign*}
&\tE [\tr( \bmalpha^T \Z^T \Z \X^T \X \bmbeta \bmalpha^T \Z^T \Z \X^T \X \bmbeta)] \\
=& \frac{1}{n^2} \cdot \Big\{  n_z (n_z + 1) \cdot n \cdot \tr(\bmSigma_X \bmSigma_{\alpha}) \cdot \tr(\bmSigma_Z \bmSigma_X \bmSigma_Z \bmSigma_{\beta})  \\
&+  n_z (n_z + 1) \cdot n (n+1) \cdot \{ \tr(\bmSigma_Z \bmSigma_X \bmSigma_{\beta \alpha} \bmSigma_Z \bmSigma_X \bmSigma_{\beta \alpha} ) + \tr(\bmSigma_Z \bmSigma_{\alpha} \bmSigma_Z \bmSigma_X \bmSigma_{\beta} \bmSigma_X ) \\
& + \{ \tr(\bmSigma_Z \bmSigma_X \bmSigma_{\beta \alpha}) \}^2 
+  \sum_{i=1}^p C_{\alpha \beta_i} \cdot (\bmSigma_Z \bmSigma_X)_{ii}^2 \} \Big \} \cdot \{ 1 + o_p(1) \}, 
\end{flalign*}
where {\bxz $C_{\alpha \beta_i} = \tE[\alpha_i^2 \beta_i^2] - 2 (\tE[ \alpha_i \beta_i ])^2 - \tE[ \alpha_i^2 ] \cdot \tE[ \beta_i^2 ]$}. 
By Condition 2(a), $\bmSigma_{\alpha} = p^{-1} \cdot \bmPhi_{\alpha \alpha}, \bmSigma_{\beta} = p^{-1} \cdot \bmPhi_{\beta \beta}$, and $\bmSigma_{\beta \alpha} = p^{-1} \cdot \bmPhi_{\beta \alpha}$. 
Note that {\bxz $C_{\alpha \beta_i} = 0$} if $\bmalpha$ and  $\bmbeta$ are normally distributed, and under Condition 2(a), we have $C_{\alpha \beta_i} \propto p^{-2}$. 

Second, the dominant terms of $\bm{A_2} = \tE[ \tr( \bmalpha^T \Z^T \Z \X^T \bmeps \bmalpha^T \Z^T \Z \X^T \bmeps) ]$ are 
\begin{flalign*}
&\tE[ \tr( \bmalpha^T \Z^T \Z \X^T \bmeps \bmalpha^T \Z^T \Z \X^T \bmeps) ] \\
&= \frac{1}{n^2} \cdot \left\{ n_z(n_z + 1) \cdot n \cdot \sigma_{\epsilon}^2 \cdot \tr(\bmSigma_X \bmSigma_Z \bmSigma_{\alpha} \bmSigma_Z) + n_z \cdot n \cdot \sigma_{\epsilon}^2 \cdot \tr(\bmSigma_Z \bmSigma_X) \cdot \tr(\bmSigma_Z \bmSigma_{\alpha}) \right\} \cdot \{ 1 + o_p(1) \} \\
&= \frac{1}{n^2} \cdot \left\{ n_z \cdot n \cdot \sigma_{\epsilon}^2 \cdot \{ (n_z + 1) \cdot \tr(\bmSigma_X \bmSigma_Z \bmSigma_{\alpha} \bmSigma_Z) + \tr(\bmSigma_Z \bmSigma_X) \cdot \tr(\bmSigma_Z \bmSigma_{\alpha}) \} \right\} \cdot \{ 1 + o_p(1) \} \\
&= \frac{1}{n^2} \cdot \left\{ n_z \cdot n \cdot \frac{1 - h_{\beta}^2}{h_{\beta}^2} \cdot \tr(\bmSigma_X \bmSigma_{\beta}) \cdot \{ n_z \cdot \tr(\bmSigma_X \bmSigma_Z \bmSigma_{\alpha} \bmSigma_Z) + \tr(\bmSigma_Z \bmSigma_X) \cdot \tr(\bmSigma_Z \bmSigma_{\alpha}) \} \right\} \cdot \{ 1 + o_p(1) \}, 
\end{flalign*}
where the equality follows from $h_{\beta}^2 = \left( \tr(\bmSigma_X \bmSigma_{\beta}) / \{ \tr(\bmSigma_X \bmSigma_{\beta}) + \sigma_{\epsilon}^2 \} \right) \cdot \{ 1 + o_p(1) \}$. 

Next, the dominant terms of $\bm{A_3} = \tE[ \tr( \bmeps_{z}^T \Z \X^T \X \bmbeta \bmeps_{z}^T \Z \X^T \X \bmbeta )]$ are 
\begin{flalign*}
&\tE[ \tr( \bmeps_{z}^T \Z \X^T \X \bmbeta \bmeps_{z}^T \Z \X^T \X \bmbeta )] \\
=& \frac{1}{n^2} \cdot \left\{ n_z \cdot n(n+1) \cdot \sigma_{\epsilon_z}^2 \cdot \tr(\bmSigma_Z \bmSigma_{\alpha} \bmSigma_Z \bmSigma_X) + n_z \cdot n \cdot \sigma_{\epsilon_z}^2 \cdot \tr(\bmSigma_Z \bmSigma_{\alpha}) \cdot \tr(\bmSigma_Z \bmSigma_X) \right\} \cdot \{ 1 + o_p(1) \} \\
=& \frac{1}{n^2} \cdot \left\{ n_z \cdot n \cdot \sigma_{\epsilon_z}^2 \cdot \{ (n+1) \cdot \tr(\bmSigma_Z \bmSigma_{\alpha} \bmSigma_Z \bmSigma_X) + \tr(\bmSigma_Z \bmSigma_{\alpha}) \cdot \tr(\bmSigma_Z \bmSigma_X) \} \right\} \cdot \{ 1 + o_p(1) \} \\
=& \frac{1}{n^2} \cdot \left\{ n_z \cdot n \cdot \frac{1 - h_{\alpha}^2}{h_{\alpha}^2} \cdot \tr(\bmSigma_Z \bmSigma_{\alpha}) \cdot \{ n \cdot \tr(\bmSigma_Z \bmSigma_{\alpha} \bmSigma_Z \bmSigma_X) + \tr(\bmSigma_Z \bmSigma_{\alpha}) \cdot \tr(\bmSigma_Z \bmSigma_X) \} \right\} \cdot \{ 1 + o_p(1) \},
\end{flalign*}
where the last equality follows from $h_{\alpha}^2 = \left( \tr(\bmSigma_Z \bmSigma_{\alpha}) / \{ \tr(\bmSigma_Z \bmSigma_{\alpha}) + \sigma_{\epsilon_z}^2 \} \right) \cdot \{ 1 + o_p(1) \}$. 

Finally, we have 
\begin{flalign*}
\bm{A_4} &= \tE[ \tr( \bmeps_{z}^T \Z \X^T \bmeps \bmeps_{z}^T \Z \X^T \bmeps ) ] \\
&= \frac{1}{n^2} \cdot \left\{ n_z \cdot n \cdot \sigma_{\epsilon}^2 \cdot \sigma_{\epsilon_z}^2 \cdot \tr(\bmSigma_Z \bmSigma_X) \right\} \cdot \{ 1 + o_p(1) \} \\
&= \frac{1}{n^2} \cdot \left\{ n_z \cdot n \cdot \frac{1 - h^2}{h^2} \cdot \tr(\bmSigma_X \bmSigma_{\beta}) \cdot \frac{1 - h_z^2}{h_z^2} \cdot \tr(\bmSigma_Z \bmSigma_{\alpha}) \cdot \tr(\bmSigma_Z \bmSigma_X) \right\} \cdot \{ 1 + o_p(1) \}.
\end{flalign*}


It is also easy to see that
\begin{flalign*}
( \tE[ \y_{\alpha}^T\widehat{\y}_\beta ] )^2 = \left[ \frac{1}{n} \tr(\bmSigma_Z \bmSigma_X \bmSigma_{\beta \alpha}) \cdot n \cdot n_z \right]^2, 
\end{flalign*}
matching with one of the terms in the dominant terms of $\bm{A_1}$.
With Condition 1(a), we get the dominant terms of $\tE[(\y_{\alpha}^T\widehat{\y}_\beta)^2] - ( \tE[ \y_{\alpha}^T\widehat{\y}_\beta ] )^2$ are 
\begin{flalign*}
\tE[(\y_{\alpha}^T \widehat{\y}_\beta)^2] - ( \tE[ \y_{\alpha}^T\widehat{\y}_\beta ] )^2 
=& (1 + o_p(1)) \cdot \frac{1}{n^2} \cdot \left\{ n_z^2 \cdot n^2 \cdot \tr(\bmSigma_Z \bmSigma_X \bmSigma_{\beta \alpha} \bmSigma_Z \bmSigma_X \bmSigma_{\beta \alpha}) \right. \\
&+ n_z^2 \cdot n^2 \cdot \tr(\bmSigma_Z \bmSigma_{\alpha} \bmSigma_Z \bmSigma_X \bmSigma_{\beta} \bmSigma_X) \\
&+ (n_z^2 \cdot n + n_z \cdot n^2) \cdot [\tr(\bmSigma_Z \bmSigma_X \bmSigma_{\beta \alpha})]^2 \\
&+ n_z^2 \cdot n^2 \cdot \sum_{i=1}^p C_{\alpha \beta_i} \cdot (\bmSigma_Z \bmSigma_X)_{ii}^2 \\
&+ n_z^2 \cdot n \cdot \tr(\bmSigma_X \bmSigma_{\beta}) \cdot \tr(\bmSigma_X \bmSigma_Z \bmSigma_{\alpha} \bmSigma_Z) / h_{\beta}^2 \\
&+ n_z \cdot n^2 \cdot \tr(\bmSigma_Z \bmSigma_{\alpha}) \cdot \tr(\bmSigma_Z \bmSigma_X \bmSigma_{\beta} \bmSigma_X) / h_{\alpha}^2 \\
&+ \left. n_z \cdot n \cdot \tr(\bmSigma_X \bmSigma_{\beta}) \cdot \tr(\bmSigma_Z \bmSigma_{\alpha}) \cdot \tr(\bmSigma_X \bmSigma_Z) / (h_{\beta}^2 \cdot h_{\alpha}^2) \right\}.
\end{flalign*}

For $\tE[ \| \y_{\alpha} \|_2^2 \cdot \| \widehat{\y}_\beta \|_2^2 ]$, after careful calculation, we find that the dominant terms of $\tE[ \| \y_{\alpha} \|_2^2 \cdot \| \widehat{\y}_\beta \|_2^2 ]$ are identical to the dominant terms of $\tE[ \| \y_{\alpha} \|_2^2] \cdot \tE[ \| \widehat{\y}_\beta \|_2^2 ]$, and we have 
\begin{flalign*}
&\tE[ \| \y_{\alpha} \|_2^2 \cdot \| \widehat{\y}_\beta \|_2^2 ] \\
=& \tE[ \| \y_{\alpha} \|_2^2] \cdot \tE[ \| \widehat{\y}_\beta \|_2^2 ] \cdot (1 + o_p(1)) \\
=& (1 + o_p(1)) \cdot \frac{1}{n^2} \cdot \left\{
\tE[\bmeps_z^{\top} \bmeps_z] \cdot \tE[\bmeps^{\top} \X \Z^{\top} \Z \X^{\top} \bmeps] 
+ \E[\bmeps_z^{\top} \bmeps_z] \cdot \tE[\bmbeta^{\top} \X^{\top} \X \Z^{\top} \Z \X^{\top} \X \bmbeta] \right. \\
&+ \left. \tE[\bmalpha^{\top} \Z^{\top} \Z \bmalpha] \cdot \tE[\bmeps^{\top} \X \Z^{\top} \Z \X^{\top} \bmeps]
+ \tE[\bmalpha^{\top} \Z^{\top} \Z \bmalpha] \cdot \tE[\bmbeta^{\top} \X^{\top} \X \Z^{\top} \Z \X^{\top} \X \bmbeta]
\right\} \\
=& (1 + o_p(1)) \cdot \frac{1}{n^2} \cdot \left\{
(n_z \cdot \sigma_{\epsilon_z}^2) \cdot (n_z \cdot n \cdot \sigma_{\epsilon}^2 \cdot \tr(\bmSigma_X \bmSigma_Z)) \right. \\
&+ (n_z \cdot \sigma_{\epsilon_z}^2) \cdot \left[ 
n_z \cdot n^2 \cdot \tr(\bmSigma_Z \bmSigma_X \bmSigma_{\beta} \bmSigma_X) + n_z \cdot n \cdot \tr(\bmSigma_Z \bmSigma_X) \cdot \tr(\bmSigma_{\beta} \bmSigma_X) \right] \\
&+ (n_z \cdot \tr(\bmSigma_{\alpha} \bmSigma_Z)) \cdot (n_z \cdot n \cdot \sigma_{\epsilon}^2 \cdot \tr(\bmSigma_X \bmSigma_Z)) \\
&+ \left. 
(n_z \cdot \tr(\bmSigma_{\alpha} \bmSigma_Z)) \cdot \left[ 
n_z \cdot n^2 \cdot \tr(\bmSigma_Z \bmSigma_X \bmSigma_{\beta} \bmSigma_X) + n_z \cdot n \cdot \tr(\bmSigma_Z \bmSigma_X) \cdot \tr(\bmSigma_{\beta} \bmSigma_X) \right]
\right\} \\
=& (1 + o_p(1)) \cdot \frac{1}{n^2} \cdot \left\{ 
n_z^2 \cdot n \cdot \tr(\bmSigma_Z \bmSigma_{\alpha}) \cdot \tr(\bmSigma_X \bmSigma_{\beta}) \cdot \tr(\bmSigma_X \bmSigma_Z) / (h_{\alpha}^2 \cdot h_{\beta}^2) \right. \\
&+ \left.
n_z^2 \cdot n^2 \cdot \tr(\bmSigma_Z \bmSigma_{\alpha}) \cdot \tr(\bmSigma_Z \bmSigma_X \bmSigma_{\beta} \bmSigma_X) / h_{\alpha}^2
\right\}.
\end{flalign*}

\newpage
\paragraph{Derivation of rate of $Var(G_{\beta \alpha})$ with general $\bmSigma_X$ and  $\bmSigma_Z$.} Using the above results, we will derive the rate of $Var(G_{\beta \alpha})$. 
By definition,
\begin{flalign*}
Var(G_{\beta \alpha}) 
= \tE \left[ \frac{(\y_{\alpha}^{\top} \widehat{\y}_{\beta})^2}{ \| \y_{\alpha} \|_2^2 \cdot \| \widehat{\y}_{\beta} \|_2^2 } \right] 
- \left( \tE \left[ \frac{\y_{\alpha}^{\top} \widehat{\y}_{\beta}}{\| \y_{\alpha} \|_2 \cdot \| \widehat{\y}_{\beta} \|_2} \right] \right)^2.
\end{flalign*}
{\bxz 
Again, let $\plim_{p \to \infty}$ denote converge in probability.}
We focus on the limit in probability of the right-hand side of the definition above and claim
\begin{flalign*}
\plim_{p, n, n_z \to \infty} Var(G_{\beta \alpha}) = \plim_{p, n, n_z \to \infty} \left\{ \frac{(\y_{\alpha}^{\top} \widehat{\y}_{\beta})^2}{ \| \y_{\alpha} \|_2^2 \cdot \| \widehat{\y}_{\beta} \|_2^2 } \right\} 
- \left\{ \plim_{p, n, n_z \to \infty} \left( \frac{\y_{\alpha}^{\top} \widehat{\y}_{\beta}}{\| \y_{\alpha} \|_2 \cdot \| \widehat{\y}_{\beta} \|_2} \right) \right\}^2,
\end{flalign*}
where $p/n \to \omega$ and $p/n_z \to \omega_z$ as $\min(n,n_z) \to \infty$ for $\omega$ and $\omega_z \in (0,\infty)$ by Condition 1(c). 
Recall 
\begin{flalign*}
&\tE[ \| \y_{\alpha} \|_2^2 \cdot \| \widehat{\y}_\beta \|_2^2 ]  \\
=& \frac{1}{n^2} \cdot \left\{
\tE[\bmeps_z^{\top} \bmeps_z \cdot \bmeps^{\top} \X \Z^{\top} \Z \X^{\top} \bmeps] 
+ \tE[\bmeps_z^{\top} \bmeps_z \cdot \bmbeta^{\top} \X^{\top} \X \Z^{\top} \Z \X^{\top} \X \bmbeta] \right. \\
&+ \left. \tE[\bmalpha^{\top} \Z^{\top} \Z \bmalpha \cdot \bmeps^{\top} \X \Z^{\top} \Z \X^{\top} \bmeps]
+ \tE[\bmalpha^{\top} \Z^{\top} \Z \bmalpha \cdot \bmbeta^{\top} \X^{\top} \X \Z^{\top} \Z \X^{\top} \X \bmbeta]
\right\} \\
=& \frac{1}{n^2} \cdot \left\{
\tE[\bmeps_z^{\top} \bmeps_z] \cdot \tE[\bmeps^{\top} \X \Z^{\top} \Z \X^{\top} \bmeps] 
+ \tE[\bmeps_z^{\top} \bmeps_z] \cdot \tE[\bmbeta^{\top} \X^{\top} \X \Z^{\top} \Z \X^{\top} \X \bmbeta] \right. \\
&+ \left. \tE[\bmalpha^{\top} \Z^{\top} \Z \bmalpha \cdot \bmeps^{\top} \X \Z^{\top} \Z \X^{\top} \bmeps]
+ \tE[\bmalpha^{\top} \Z^{\top} \Z \bmalpha \cdot \bmbeta^{\top} \X^{\top} \X \Z^{\top} \Z \X^{\top} \X \bmbeta]
\right\}.
\end{flalign*}
Applying Lemma~\ref{bai-LB26-gen} with $\A$ being the identity matrix of suitable dimension,  we have 
\begin{flalign*}
\bmeps_z^{\top} \bmeps_z & \stackrel{p}{\to} \tr(\bmSigma_{\epsilon_z}) = n_z \cdot \sigma_{\epsilon_z}^2 = n_z \cdot \tr(\bmSigma_Z \bmSigma_{\alpha}) \cdot (1 - h_{\alpha}^2) / h_{\alpha}^2, \\
\bmeps^{\top} \X \Z^{\top} \Z \X^{\top} \bmeps & \stackrel{p}{\to} \tr(Cov(\Z \X^{\top} \bmeps)) = \tE[ \bmeps^{\top} \X \Z^{\top} \Z \X^{\top} \bmeps ], \\
\bmalpha^{\top} \Z^{\top} \Z \bmalpha & \stackrel{p}{\to} \tr(Cov(\Z \bmalpha)) = \tE[ \bmalpha^{\top} \Z^{\top} \Z \bmalpha ],\quad \mbox{and} \\
\bmbeta^{\top} \X^{\top} \X \Z^{\top} \Z \X^{\top} \X \bmbeta & \stackrel{p}{\to} \tr(Cov(\Z \X^{\top} \X \bmbeta)) = \tE[ \bmbeta^{\top} \X^{\top} \X \Z^{\top} \Z \X^{\top} \X \bmbeta ],
\end{flalign*}
where 
\begin{flalign*}
\tr(Cov(\Z \X^{\top} \bmeps)) 
= \sum_{i=1}^{n_z} \tE [ ((\Z \X^{\top} \bmeps)_i )^2 ] 
= \tE \left[ \sum_{i=1}^{n_z}  ((\Z \X^{\top} \bmeps)_i )^2 \right]
= \tE \left[ (\Z \X^{\top} \bmeps)^{\top} (\Z \X^{\top} \bmeps) \right],
\end{flalign*}
and similar reasoning holds for $\bmalpha^{\top} \Z^{\top} \Z \bmalpha$ and $\bmbeta^{\top} \X^{\top} \X \Z^{\top} \Z \X^{\top} \X \bmbeta$. 

By the continuous mapping theorem,  we have 
\begin{flalign*}
(\bmeps_z^{\top} \bmeps_z) \cdot (\bmeps^{\top} \X \Z^{\top} \Z \X^{\top} \bmeps) & \stackrel{p}{\to}
\tr(\bmSigma_{\epsilon_z}) \cdot \tE[ \bmeps^{\top} \X \Z^{\top} \Z \X^{\top} \bmeps ], \\
(\bmeps_z^{\top} \bmeps_z) \cdot (\bmbeta^{\top} \X^{\top} \X \Z^{\top} \Z \X^{\top} \X \bmbeta) & \stackrel{p}{\to} 
\tr(\bmSigma_{\epsilon_z}) \cdot \tE[ \bmbeta^{\top} \X^{\top} \X \Z^{\top} \Z \X^{\top} \X \bmbeta ], \\
(\bmalpha^{\top} \Z^{\top} \Z \bmalpha) \cdot (\bmeps^{\top} \X \Z^{\top} \Z \X^{\top} \bmeps) & \stackrel{p}{\to} 
\tE[ \bmalpha^{\top} \Z^{\top} \Z \bmalpha ] \cdot \tE[ \bmeps^{\top} \X \Z^{\top} \Z \X^{\top} \bmeps ], \quad \mbox{and} \\
(\bmalpha^{\top} \Z^{\top} \Z \bmalpha) \cdot (\bmbeta^{\top} \X^{\top} \X \Z^{\top} \Z \X^{\top} \X \bmbeta) & \stackrel{p}{\to} 
\tE[ \bmalpha^{\top} \Z^{\top} \Z \bmalpha ] \cdot \tE[ \bmbeta^{\top} \X^{\top} \X \Z^{\top} \Z \X^{\top} \X \bmbeta ].
\end{flalign*}
It follows that 
\begin{flalign*}
\| \y_{\alpha} \|_2^2 \cdot \| \widehat{\y}_\beta \|_2^2 
 \stackrel{p}{\to} &
\tr(\bmSigma_{\epsilon_z}) \cdot \tE[ \bmeps^{\top} \X \Z^{\top} \Z \X^{\top} \bmeps ] 
+ \tr(\bmSigma_{\epsilon_z}) \cdot \tE[ \bmbeta^{\top} \X^{\top} \X \Z^{\top} \Z \X^{\top} \X \bmbeta ] \\
&+ \tE[ \bmalpha^{\top} \Z^{\top} \Z \bmalpha ] \cdot \tE[ \bmeps^{\top} \X \Z^{\top} \Z \X^{\top} \bmeps ] 
+ \tE[ \bmalpha^{\top} \Z^{\top} \Z \bmalpha ] \cdot \tE[ \bmbeta^{\top} \X^{\top} \X \Z^{\top} \Z \X^{\top} \X \bmbeta ] \\
=& \tE[ \| \y_{\alpha} \|_2^2] \cdot \tE[ \| \widehat{\y}_\beta \|_2^2 ]. 
\end{flalign*}
Again, by the continuous mapping theorem,  we have
\begin{flalign*}
\| \y_{\alpha} \|_2 \cdot \| \widehat{\y}_\beta \|_2 
 \stackrel{p}{\to} &
\left( \tr(\bmSigma_{\epsilon_z}) \cdot \tE[ \bmeps^{\top} \X \Z^{\top} \Z \X^{\top} \bmeps ] 
+ \tr(\bmSigma_{\epsilon_z}) \cdot \tE[ \bmbeta^{\top} \X^{\top} \X \Z^{\top} \Z \X^{\top} \X \bmbeta ] \right. \\
&+ \tE[ \bmalpha^{\top} \Z^{\top} \Z \bmalpha ] \cdot \tE[ \bmeps^{\top} \X \Z^{\top} \Z \X^{\top} \bmeps ] 
+ \left. \tE[ \bmalpha^{\top} \Z^{\top} \Z \bmalpha ] \cdot \tE[ \bmbeta^{\top} \X^{\top} \X \Z^{\top} \Z \X^{\top} \X \bmbeta ] \right)^{1/2} 
\end{flalign*}
That is, $\plim \{ \| \y_{\alpha} \|_2^2 \cdot \| \widehat{\y}_\beta \|_2^2 \} = ( \plim \{ \| \y_{\alpha} \|_2 \cdot \| \widehat{\y}_\beta \|_2 \})^2$. 
Also recall
\begin{flalign*}
\tE[(\y_{\alpha}^T\widehat{\y}_\beta)^2] 
=& \tE [\tr( \bmalpha^T \Z^T \Z \X^T \X \bmbeta \bmalpha^T \Z^T \Z \X^T \X \bmbeta)] 
+ \tE[ \tr( \bmalpha^T \Z^T \Z \X^T \bmeps \bmalpha^T \Z^T \Z \X^T \bmeps) ] \\
&+ \tE[ \tr( \bmeps_{z}^T \Z \X^T \X \bmbeta \bmeps_{z}^T \Z \X^T \X \bmbeta )]
+ \tE[ \tr( \bmeps_{z}^T \Z \X^T \bmeps \bmeps_{z}^T \Z \X^T \bmeps ) ].
\end{flalign*}
In addition, we have  
\begin{flalign*}
\tE[(\Z^{\top} \Z \bmalpha)_i \cdot (\X^T \X \bmbeta)_j] 
=& [Cov(\Z^{\top} \Z \bmalpha, \X^T \X \bmbeta)]_{ij} \\
=& n_z \cdot n \cdot \sum_{a=1}^p \sum_{c=1}^p (\bmSigma_Z)_{ia} \cdot (\bmSigma_X)_{jc} \cdot \E[\bmalpha_{a} \cdot \bmbeta_{c}] \\
=& n_z \cdot n \cdot (\bmSigma_Z \bmSigma_{\alpha \beta} \bmSigma_X)_{ij}
\end{flalign*}
for $i, j \in [p]$, where the last equality follows from diagonal $\bmSigma_{\alpha \beta}$ in Condition 2(a). By Lemma~\ref{quadratic-form-plim} with $\A = \mathbf{I}_p$, $\bmalpha = \Z^{\top} \Z \bmalpha$, $\bmbeta = \X^T \X \bmbeta$, we have
\begin{flalign*}
\bmalpha^T \Z^T \Z \X^T \X \bmbeta 
\stackrel{p}{\to} \tr(\bmSigma_Z \bmSigma_{\alpha \beta} \bmSigma_X),
\end{flalign*}
and thus 
\begin{flalign*}
\left\{ \plim \left( \y_{\alpha}^T\widehat{\y}_\beta \right) \right\}^2 
= [ n_z \cdot n \cdot \tr(\bmSigma_Z \bmSigma_{\alpha \beta} \bmSigma_X)]^2.
\end{flalign*}
Therefore, we have
\begin{flalign*}
\plim \left\{ (\y_{\alpha}^T\widehat{\y}_\beta)^2 \right\} 
- \left\{ \plim \left( \y_{\alpha}^T\widehat{\y}_\beta \right) \right\}^2 
=& \E[(\y_{\alpha}^T\widehat{\y}_\beta)^2] - (\E[ \y_{\alpha}^T\widehat{\y}_\beta ])^2, 
\end{flalign*}
By the continuous mapping theorem, we have 
\begin{flalign*}
&\plim_{p, n, n_z \to \infty} \left\{ \frac{(\y_{\alpha}^{\top} \widehat{\y}_{\beta})^2}{ \| \y_{\alpha} \|_2^2 \cdot \| \widehat{\y}_{\beta} \|_2^2 } \right\} 
- \left\{ \plim_{p, n, n_z \to \infty} \left( \frac{\y_{\alpha}^{\top} \widehat{\y}_{\beta}}{\| \y_{\alpha} \|_2 \cdot \| \widehat{\y}_{\beta} \|_2} \right) \right\}^2 \\
=& \frac{ \plim \{ (\y_{\alpha}^{\top} \widehat{\y}_{\beta})^2 \} }{ \plim \{ \| \y_{\alpha} \|_2^2 \cdot \| \widehat{\y}_{\beta} \|_2^2 \} } 
- \left\{ \frac{ \plim \{ \y_{\alpha}^{\top} \widehat{\y}_{\beta} \} }{ \plim \{ \| \y_{\alpha} \|_2 \cdot \| \widehat{\y}_{\beta} \|_2 \} } \right\}^2 \\
=& \frac{ \plim \{ (\y_{\alpha}^{\top} \widehat{\y}_{\beta})^2 \} }{ \plim \{ \| \y_{\alpha} \|_2^2 \cdot \| \widehat{\y}_{\beta} \|_2^2 \} } 
- \frac{ ( \plim \{ \y_{\alpha}^{\top} \widehat{\y}_{\beta} \} )^2 }{ \plim \{ \| \y_{\alpha} \|_2^2 \cdot \| \widehat{\y}_{\beta}^2 \|_2^2 \} } \\
=& \left( \E[(\y_{\alpha}^T\widehat{\y}_\beta)^2] - (\E[ \y_{\alpha}^T\widehat{\y}_\beta ])^2 \right) / \left( \E[ \| \y_{\alpha} \|_2^2] \cdot \E[ \| \widehat{\y}_\beta \|_2^2 ] \right). 
\end{flalign*}
Thus, we finally have
\begin{align} \label{eq:var_Gba_general}
\begin{split}
&\plim Var(G_{\beta \alpha}) \\
=& \left\{ n_z^2 \cdot n^2 \cdot \tr(\bmSigma_Z \bmSigma_X \bmSigma_{\beta \alpha} \bmSigma_Z \bmSigma_X \bmSigma_{\beta \alpha}) \right. \\
&+ n_z^2 \cdot n^2 \cdot \tr(\bmSigma_Z \bmSigma_{\alpha} \bmSigma_Z \bmSigma_X \bmSigma_{\beta} \bmSigma_X) \\
&+ (n_z^2 \cdot n + n_z \cdot n^2) \cdot [\tr(\bmSigma_Z \bmSigma_X \bmSigma_{\beta \alpha})]^2 \\
&+ n_z^2 \cdot n^2 \cdot \sum_{i=1}^p C_{\alpha \beta_i} \cdot (\bmSigma_Z \bmSigma_X)_{ii}^2 \\
&+ n_z^2 \cdot n \cdot \tr(\bmSigma_X \bmSigma_{\beta}) \cdot \tr(\bmSigma_X \bmSigma_Z \bmSigma_{\alpha} \bmSigma_Z) / h_{\beta}^2 \\
&+ n_z \cdot n^2 \cdot \tr(\bmSigma_Z \bmSigma_{\alpha}) \cdot \tr(\bmSigma_Z \bmSigma_X \bmSigma_{\beta} \bmSigma_X) / h_{\alpha}^2 \\
&+ \left. n_z \cdot n \cdot \tr(\bmSigma_X \bmSigma_{\beta}) \cdot \tr(\bmSigma_Z \bmSigma_{\alpha}) \cdot \tr(\bmSigma_X \bmSigma_Z) / (h_{\beta}^2 \cdot h_{\alpha}^2) \right\} \\
& \cdot \left\{ 
n_z^2 \cdot n \cdot \tr(\bmSigma_Z \bmSigma_{\alpha}) \cdot \tr(\bmSigma_X \bmSigma_{\beta}) \cdot \tr(\bmSigma_X \bmSigma_Z) / (h_{\alpha}^2 \cdot h_{\beta}^2) \right. \\
&+ \left.
n_z^2 \cdot n^2 \cdot \tr(\bmSigma_Z \bmSigma_{\alpha}) \cdot \tr(\bmSigma_Z \bmSigma_X \bmSigma_{\beta} \bmSigma_X) / h_{\alpha}^2
\right\}^{-1} 
\cdot (1 + o_p(1)) \\
=& \left\{ \frac{
n \cdot [ \tr(\bmSigma_Z \bmSigma_X \bmSigma_{\beta \alpha} \bmSigma_Z \bmSigma_X \bmSigma_{\beta \alpha}) + \tr(\bmSigma_Z \bmSigma_{\alpha} \bmSigma_Z \bmSigma_X \bmSigma_{\beta} \bmSigma_X) + \sum_{i=1}^p C_{\alpha \beta_i} \cdot (\bmSigma_Z \bmSigma_X)_{ii}^2 ]
}{ \tr(\bmSigma_Z \bmSigma_{\alpha}) \cdot \tr(\bmSigma_X \bmSigma_{\beta}) \cdot \tr(\bmSigma_X \bmSigma_Z) / (h_{\alpha}^2 \cdot h_{\beta}^2) 
+ 
n \cdot \tr(\bmSigma_Z \bmSigma_{\alpha}) \cdot \tr(\bmSigma_Z \bmSigma_X \bmSigma_{\beta} \bmSigma_X) / h_{\alpha}^2 } \right. \\
&+ \frac{
[\tr(\bmSigma_Z \bmSigma_X \bmSigma_{\beta \alpha})]^2 + \tr(\bmSigma_X \bmSigma_{\beta}) \cdot \tr(\bmSigma_X \bmSigma_Z \bmSigma_{\alpha} \bmSigma_Z) / h_{\beta}^2
}{ \tr(\bmSigma_Z \bmSigma_{\alpha}) \cdot \tr(\bmSigma_X \bmSigma_{\beta}) \cdot \tr(\bmSigma_X \bmSigma_Z) / (h_{\alpha}^2 \cdot h_{\beta}^2) 
+ 
n \cdot \tr(\bmSigma_Z \bmSigma_{\alpha}) \cdot \tr(\bmSigma_Z \bmSigma_X \bmSigma_{\beta} \bmSigma_X) / h_{\alpha}^2 } \\
&+ \frac{n}{n_z} \cdot \frac{ [\tr(\bmSigma_Z \bmSigma_X \bmSigma_{\beta \alpha})]^2 }{
\tr(\bmSigma_Z \bmSigma_{\alpha}) \cdot \tr(\bmSigma_X \bmSigma_{\beta}) \cdot \tr(\bmSigma_X \bmSigma_Z) / (h_{\alpha}^2 \cdot h_{\beta}^2) 
+ 
n \cdot \tr(\bmSigma_Z \bmSigma_{\alpha}) \cdot \tr(\bmSigma_Z \bmSigma_X \bmSigma_{\beta} \bmSigma_X) / h_{\alpha}^2
} \\
&+ \left. \frac{1}{n_z} \right\} \cdot(1 + o_p(1)). 
\end{split}
\end{align}
These results quantify how the $Var(G_{\beta \alpha})$ is determined by the parameters in the model. 

\paragraph{Rate of $Var(G_{\beta \alpha})$ in terms of $m_{\beta \alpha}, \delta_{\beta \alpha}$, and $\kappa_{\beta \alpha}$ with $\bmSigma_X = \bmSigma_Z = \mathbf{I}_p$.}
To provide more insights into how the sparsity influences the variance of $G_{\beta \alpha}$, we study a special case with $\bmSigma_X = \bmSigma_Z = \mathbf{I}_p$. 
Recall that {\bxz $C_{\alpha \beta_i}= \E[\alpha_i^2 \beta_i^2] - 2 (\E[ \alpha_i \beta_i ])^2 - \E[ \alpha_i^2 ] \cdot \E[ \beta_i^2 ] \propto p^{-2}$ by Condition 2(a). {\xcy Assume that for all $i \in [p]$ such that $[\bmSigma_{\beta 
\alpha}]_{ii} \neq 0$, $C_{\alpha \beta_i} = p^{-2} \cdot \nu_{\alpha \beta}$ for some absolute constant $\nu_{\alpha \beta}$. Otherwise $C_{\alpha \beta_i} = 0$.} } Furthermore, assume the $m_{\alpha}$ nonzero entries of $\bmPhi_{\alpha}$ are all equal to constant $\sigma_{\alpha}^2$, the $m_{\beta}$ nonzero entries of $\bmPhi_{\beta}$ are all equal to constant $\sigma_{\beta}^2$, and the $m_{\beta \alpha}$ nonzero entries of $\bmPhi_{\beta \alpha}$ are all equal to constant $\sigma_{\beta \alpha}$. Then the $Var(G_{\beta \alpha})$ above becomes
\begin{flalign*}
&\plim Var(G_{\beta \alpha}) \\
=&  \left\{ \frac{
n \cdot [ \tr(\bmSigma_{\beta \alpha}^2) + \tr( \bmSigma_{\alpha} \bmSigma_{\beta} ) + {\xcy \nu_{\alpha \beta} \cdot m_{\beta \alpha} /p^2 }]
}{ \tr(\bmSigma_{\alpha}) \cdot \tr( \bmSigma_{\beta}) \cdot p / (h_{\alpha}^2 \cdot h_{\beta}^2) 
+ 
n \cdot \tr(\bmSigma_{\alpha}) \cdot \tr( \bmSigma_{\beta} ) / h_{\alpha}^2 } \right. \\
&+ \frac{
[\tr( \bmSigma_{\beta \alpha})]^2 + \tr( \bmSigma_{\beta}) \cdot \tr( \bmSigma_{\alpha} ) / h_{\beta}^2
}{ \tr( \bmSigma_{\alpha}) \cdot \tr( \bmSigma_{\beta}) \cdot p / (h_{\alpha}^2 \cdot h_{\beta}^2) 
+ 
n \cdot \tr( \bmSigma_{\alpha}) \cdot \tr( \bmSigma_{\beta} ) / h_{\alpha}^2 } \\
&+ \left. \frac{n}{n_z} \cdot \frac{ [\tr( \bmSigma_{\beta \alpha})]^2 }{
\tr( \bmSigma_{\alpha}) \cdot \tr( \bmSigma_{\beta}) \cdot p / (h_{\alpha}^2 \cdot h_{\beta}^2) 
+ 
n \cdot \tr( \bmSigma_{\alpha}) \cdot \tr( \bmSigma_{\beta} ) / h_{\alpha}^2
} 
+ \frac{1}{n_z} \right\} \cdot (1 + o_p(1))  \\
=& \left\{ \frac{
n \cdot [ \tr(\bmPhi_{\beta \alpha}^2) /p^2 + \tr( \bmPhi_{\alpha} \bmPhi_{\beta} ) / p^2 + {\xcy \nu_{\alpha \beta} \cdot m_{\beta \alpha} /p^2 } ]
}{ \tr(\bmPhi_{\alpha}) \cdot \tr( \bmPhi_{\beta}) / (p \cdot h_{\alpha}^2 \cdot h_{\beta}^2) 
+ 
n \cdot \tr(\bmPhi_{\alpha}) \cdot \tr( \bmPhi_{\beta} ) / (p^2 \cdot h_{\alpha}^2) } \right. \\
&+ \frac{
[\tr( \bmPhi_{\beta \alpha})]^2 / p^2 + \tr( \bmPhi_{\beta}) \cdot \tr( \bmPhi_{\alpha} ) / ( p^2 \cdot h_{\beta}^2)
}{ \tr( \bmPhi_{\alpha}) \cdot \tr( \bmPhi_{\beta}) / (p \cdot h_{\alpha}^2 \cdot h_{\beta}^2) 
+ 
n \cdot \tr( \bmPhi_{\alpha}) \cdot \tr( \bmPhi_{\beta} ) / (p^2 \cdot h_{\alpha}^2) } \\
&+ \left. \frac{n}{n_z} \cdot \frac{ [\tr( \bmPhi_{\beta \alpha})]^2 / p^2 }{
\tr( \bmPhi_{\alpha}) \cdot \tr( \bmPhi_{\beta}) / ( p \cdot h_{\alpha}^2 \cdot h_{\beta}^2 ) 
+ 
n \cdot \tr( \bmPhi_{\alpha}) \cdot \tr( \bmPhi_{\beta} ) / (p \cdot h_{\alpha}^2)
} 
+ \frac{1}{n_z} \right\} \cdot(1 + o_p(1)) \\
=& \left\{ \frac{
n \cdot [ m_{\beta \alpha} \cdot \sigma_{\beta \alpha}^2 /p^2 
+ m_{\beta \alpha} \cdot \sigma_{\alpha}^2 \cdot \sigma_{\beta}^2 / p^2 
+ {\xcy \nu_{\alpha \beta} \cdot m_{\beta \alpha} /p^2 } ]
}{ p^{-2} \cdot m_{\alpha} \cdot m_{\beta} \cdot \sigma_{\alpha}^2 \cdot \sigma_{\beta}^2 \cdot [ p / (h_{\alpha}^2 \cdot h_{\beta}^2) 
+ 
n / h_{\alpha}^2 ] } \right. \\
&+ \frac{
[ m_{\beta \alpha} \cdot \sigma_{\beta \alpha} ]^2 / p^2 + m_{\alpha} \cdot m_{\beta} \cdot \sigma_{\alpha}^2 \cdot \sigma_{\beta}^2 / ( p^2 \cdot h_{\beta}^2)
}{ p^{-2} \cdot m_{\alpha} \cdot m_{\beta} \cdot \sigma_{\alpha}^2 \cdot \sigma_{\beta}^2 \cdot [ p / (h_{\alpha}^2 \cdot h_{\beta}^2) 
+ 
n / h_{\alpha}^2 ] }  \\
&+ \left. \frac{n}{n_z} \cdot 
\frac{ [m_{\beta \alpha} \sigma_{\beta \alpha}]^2 / p^2
}{ p^{-2} \cdot m_{\alpha} \cdot m_{\beta} \cdot \sigma_{\alpha}^2 \cdot \sigma_{\beta}^2 \cdot [ p / (h_{\alpha}^2 \cdot h_{\beta}^2) 
+ 
n / h_{\alpha}^2 ] }
+ \frac{1}{n_z} 
\right\} \cdot(1 + o_p(1)) \\
=& \left\{ \left[ \frac{n}{p / (h_{\alpha}^2 \cdot h_{\beta}^2) + n / h_{\alpha}^2} \cdot \frac{m_{\beta \alpha}}{m_{\alpha} \cdot m_{\beta}} \cdot \frac{\sigma_{\beta \alpha}^2}{\sigma_{\alpha}^2 \cdot \sigma_{\beta}^2} 
+ \frac{n}{p / (h_{\alpha}^2 \cdot h_{\beta}^2) 
+ 
n / h_{\alpha}^2} \cdot \frac{m_{\beta \alpha}}{m_{\alpha} \cdot m_{\beta}} \right. \right. \\
&+ \left. \frac{n}{p / (h_{\alpha}^2 \cdot h_{\beta}^2) 
+ 
n / h_{\alpha}^2} \cdot \frac{{\xcy m_{\beta \alpha}}}{m_{\alpha} \cdot m_{\beta}} \cdot \frac{\nu_{\alpha \beta}}{\sigma_{\alpha}^2 \cdot \sigma_{\beta}^2} \right] \\
&+ \left[ \frac{m_{\beta \alpha}^2}{m_{\alpha} \cdot m_{\beta}} \cdot \frac{\sigma_{\beta \alpha}^2}{\sigma_{\alpha}^2 \cdot \sigma_{\beta}^2} \cdot \frac{1}{p / (h_{\alpha}^2 \cdot h_{\beta}^2) + n / h_{\alpha}^2} 
+ \frac{1}{h_{\beta}^2} \cdot \frac{1}{p / (h_{\alpha}^2 \cdot h_{\beta}^2) + n / h_{\alpha}^2} \right] \\
&+ \left. \frac{n}{n_z} \cdot \frac{m_{\beta \alpha}^2}{m_{\alpha} \cdot m_{\beta}} \cdot \frac{\sigma_{\beta \alpha}^2}{\sigma_{\alpha}^2 \cdot \sigma_{\beta}^2} \cdot \frac{1}{p / (h_{\alpha}^2 \cdot h_{\beta}^2) + n / h_{\alpha}^2} + \frac{1}{n_z}
\right\} \cdot (1 + o_p(1)).
\end{flalign*}
By Condition 1(c) and 2(a), we further have 
\begin{flalign*}
&\plim Var(G_{\beta \alpha}) \\
=& \left\{ \left( \left[ \frac{1}{\omega} \cdot \frac{1}{\delta_{\beta \alpha}} \cdot \kappa_{\beta \alpha}^2 \cdot \frac{\sigma_{\beta \alpha}^2}{\sigma_{\alpha}^2 \cdot \sigma_{\beta}^2} 
+ \frac{1}{\omega} \cdot \frac{1}{\delta_{\beta \alpha}} \cdot \kappa_{\beta \alpha}^2 
\right]
+ \left[ \kappa_{\beta \alpha}^2 \cdot \frac{\sigma_{\beta \alpha}^2}{\sigma_{\alpha}^2 \cdot \sigma_{\beta}^2} + \frac{1}{h_{\beta}^2} \right]
\right. \right. \\
&+ \left. \left.  
\frac{\omega_z}{\omega} \cdot \kappa_{\beta \alpha}^2 \cdot \frac{\sigma_{\beta \alpha}^2}{\sigma_{\alpha}^2 \cdot \sigma_{\beta}^2} \right) 
\cdot \frac{h_{\alpha}^2}{p / h_{\beta}^2 + n }
+ {\xcy \frac{\nu_{\alpha \beta}}{\sigma_{\alpha}^2 \cdot \sigma_{\beta}^2} \cdot \kappa_{\beta \alpha}^2 \cdot \frac{n \cdot h_{\alpha}^2}{p / h_{\beta}^2 + n} \cdot \frac{1}{m_{\beta \alpha}} }
+ \frac{1}{n_z} \right\}  \cdot (1 + o_p(1)).
\end{flalign*}

In addition, since $p/n \to \omega$, we have
\begin{flalign*}
&\plim Var(G_{\beta \alpha}) \\
=& \left\{ \left( \left[ \frac{1}{\omega} \cdot \frac{1}{\delta_{\beta \alpha}} \cdot \kappa_{\beta \alpha}^2 \cdot \frac{\sigma_{\beta \alpha}^2}{\sigma_{\alpha}^2 \cdot \sigma_{\beta}^2} 
+ \frac{1}{\omega} \cdot \frac{1}{\delta_{\beta \alpha}} \cdot \kappa_{\beta \alpha}^2 
\right]
+ \left[ \kappa_{\beta \alpha}^2 \cdot \frac{\sigma_{\beta \alpha}^2}{\sigma_{\alpha}^2 \cdot \sigma_{\beta}^2} + \frac{1}{h_{\beta}^2} \right] 
\right. \right. \\
&+ \left. \left. 
\frac{\omega_z}{\omega} \cdot \kappa_{\beta \alpha}^2 \cdot \frac{\sigma_{\beta \alpha}^2}{\sigma_{\alpha}^2 \cdot \sigma_{\beta}^2} \right) 
\cdot {\xcy \frac{h_{\alpha}^2}{\omega / h_{\beta}^2 + 1} \cdot \frac{1}{n} }
+ {\xcy \frac{\nu_{\alpha \beta}}{\sigma_{\alpha}^2 \cdot \sigma_{\beta}^2} \cdot \kappa_{\beta \alpha}^2 \cdot \frac{ h_{\alpha}^2}{\omega / h_{\beta}^2 + 1} \cdot \frac{1}{m_{\beta \alpha}} }
+ \frac{1}{n_z} \right\}  \cdot (1 + o_p(1)).
\end{flalign*}
{\xcy These results suggest that $Var(G_{\beta \alpha})$ depends on the sample size of both populations ($n, n_z$) and the sparsity of $\bmPhi_{\beta \alpha}$). Since Population-I is often much larger than Population-II, $Var(G_{\beta \alpha})$ will mostly have order $O_p(\max( \{ h_{\alpha}^2/ (\omega / h_{\beta}^2 + 1) \} \cdot (\kappa_{\beta \alpha}^2 / m_{\beta \alpha}), 1/n_z))$. Thus, the variance of $G_{\beta \alpha}$ increases as $m_{\beta \alpha}$ decreases, which is the number of SNPs relevant to both traits in both populations. That is, the variance of $G_{\beta \alpha}$ gets larger when the two traits have fewer overlapping signals. 
In addition, $Var(G_{\beta \alpha})$ decreases as $h_{\alpha}^2, h_{\beta}^2$ decrease. This is not what we typically expect. However, as illustrated below, this issue disappears after we introduce the shrinkage factor $S_{\beta \alpha}$ in the consistent estimator $G_{\beta \alpha}^M$. 
}

\paragraph{Rate of $Var(G_{\beta \alpha}^M)$ in terms of $m_{\beta \alpha}, \delta_{\beta \alpha}$, and $\kappa_{\beta \alpha}$ with general $\bmSigma_X$ and $\bmSigma_Z$}
In Section 3.1, we proposed a consistent estimator $G_{\beta \alpha}^M$. Based on the results for $Var(G_{\beta \alpha})$, we will provide the rate of $Var(G_{\beta \alpha}^M)$ under Condition 1(a), 1(c), 2(a) and 2(b) with general $\bmSigma_X$ and $\bmSigma_Z$. Recall 
\begin{flalign*}
Var(G_{\beta \alpha}^M) = Var(G_{\beta \alpha}) \cdot \Big[\frac{b_1(\bmSigma_X^2\bmSigma_Z)}{\h^2_{\alpha} \cdot b_1^2(\bmSigma_X\bmSigma_Z)}
+ \frac{\omega}{\h^2_{\beta}\h^2_{\alpha} \cdot b_1(\bmSigma_X\bmSigma_Z) }\Big].
\end{flalign*}
Under Condition 1(a), 2(b), and further assume $\tr(\bmSigma_Z \bmSigma_{\alpha}) = \tr(\bmSigma_Z) \cdot \tr(\bmSigma_{\alpha}) / p$, $\tr(\bmSigma_X \bmSigma_{\beta}) = \tr(\bmSigma_X) \cdot \tr(\bmSigma_{\beta}) / p$, and $\tr(\bmSigma_Z \bmSigma_X \bmSigma_{\beta} \bmSigma_X) = \tr(\bmSigma_X^2 \bmSigma_Z) \cdot \tr(\bmSigma_{\beta}) / p$, the factor 
\begin{flalign*}
\left\{ \tr(\bmSigma_Z \bmSigma_{\alpha}) \cdot \tr(\bmSigma_X \bmSigma_{\beta}) \cdot \tr(\bmSigma_X \bmSigma_Z) / (h_{\alpha}^2 \cdot h_{\beta}^2) 
+ 
n \cdot \tr(\bmSigma_Z \bmSigma_{\alpha}) \cdot \tr(\bmSigma_Z \bmSigma_X \bmSigma_{\beta} \bmSigma_X) / h_{\alpha}^2 \right\}^{-1}
\end{flalign*}
in the first three terms in $Var(G_{\beta \alpha})$ in equation~\eqref{eq:var_Gba_general} becomes  
\begin{flalign*}
&\frac{ \frac{b_1(\bmSigma_X^2\bmSigma_Z)}{\h^2_{\alpha} \cdot b_1^2(\bmSigma_X\bmSigma_Z)}
+ \frac{\omega}{\h^2_{\beta}\h^2_{\alpha} \cdot b_1(\bmSigma_X\bmSigma_Z) } 
}{
\tr(\bmSigma_Z \bmSigma_{\alpha}) \cdot \tr(\bmSigma_X \bmSigma_{\beta}) \cdot \tr(\bmSigma_X \bmSigma_Z) / (h_{\alpha}^2 \cdot h_{\beta}^2) 
+ 
n \cdot \tr(\bmSigma_Z \bmSigma_{\alpha}) \cdot \tr(\bmSigma_Z \bmSigma_X \bmSigma_{\beta} \bmSigma_X) / h_{\alpha}^2
} \\
=& \frac{ \frac{1}{h_{\alpha}^2} \left[
\frac{ \tr(\bmSigma_X^2\bmSigma_Z) / p}{ (\tr(\bmSigma_X\bmSigma_Z))^2 / p^2}
+ \frac{\omega}{\h^2_{\beta} \cdot \tr(\bmSigma_X\bmSigma_Z) / p } \right] 
}{
\tr(\bmSigma_{\alpha}) \cdot \tr(\bmSigma_{\beta}) \cdot \tr(\bmSigma_X \bmSigma_Z) / (h_{\alpha}^2 \cdot h_{\beta}^2) 
+ 
n \cdot \tr(\bmSigma_{\alpha}) \cdot \tr( \bmSigma_X^2 \bmSigma_Z) \cdot \tr(\bmSigma_{\beta}) / (p \cdot  h_{\alpha}^2 )
} \\
=& \frac{1}{\tr(\bmSigma_{\alpha}) \cdot \tr(\bmSigma_{\beta})} \cdot
\frac{
\frac{p \cdot \omega }{(\tr(\bmSigma_X\bmSigma_Z))^2}
\left[ \frac{\tr(\bmSigma_X^2\bmSigma_Z)}{\omega} + 
\frac{\tr(\bmSigma_X\bmSigma_Z) }{\h^2_{\beta} } \right]
}{
\tr(\bmSigma_X \bmSigma_Z) / h_{\beta}^2 + \tr( \bmSigma_X^2 \bmSigma_Z) / \omega 
} \\
&= \frac{p \cdot \omega}{\tr(\bmSigma_{\alpha}) \cdot \tr(\bmSigma_{\beta}) \cdot (\tr(\bmSigma_X\bmSigma_Z))^2}.
\end{flalign*}
Thus, without applying further simplification to the numerator using Condition 2(a), we have
\begin{flalign*}
&\plim Var(G_{\beta \alpha}^M) \\
=& \left\{ \frac{
n \cdot p \cdot \omega \cdot [ \tr(\bmSigma_Z \bmSigma_X \bmSigma_{\beta \alpha} \bmSigma_Z \bmSigma_X \bmSigma_{\beta \alpha}) + \tr(\bmSigma_Z \bmSigma_{\alpha} \bmSigma_Z \bmSigma_X \bmSigma_{\beta} \bmSigma_X) + \sum_{i=1}^p C_{\alpha \beta_i} \cdot (\bmSigma_Z \bmSigma_X)_{ii}^2 ]
}{ \tr(\bmSigma_{\alpha}) \cdot \tr(\bmSigma_{\beta}) \cdot (\tr(\bmSigma_X\bmSigma_Z))^2 } \right. \\
&+ p \cdot \omega \cdot \frac{
[\tr(\bmSigma_Z \bmSigma_X \bmSigma_{\beta \alpha})]^2 + \tr(\bmSigma_X \bmSigma_{\beta}) \cdot \tr(\bmSigma_X \bmSigma_Z \bmSigma_{\alpha} \bmSigma_Z) / h_{\beta}^2
}{ \tr(\bmSigma_{\alpha}) \cdot \tr(\bmSigma_{\beta}) \cdot (\tr(\bmSigma_X\bmSigma_Z))^2 } \\
&+ \frac{n \cdot p \cdot \omega}{n_z} \cdot \frac{ [\tr(\bmSigma_Z \bmSigma_X \bmSigma_{\beta \alpha})]^2 }{
\tr(\bmSigma_{\alpha}) \cdot \tr(\bmSigma_{\beta}) \cdot (\tr(\bmSigma_X\bmSigma_Z))^2
} \\
&+ \left. \frac{1}{n_z} \cdot \left[\frac{b_1(\bmSigma_X^2\bmSigma_Z)}{\h^2_{\alpha} \cdot b_1^2(\bmSigma_X\bmSigma_Z)}
+ \frac{\omega}{\h^2_{\beta}\h^2_{\alpha} \cdot b_1(\bmSigma_X\bmSigma_Z) } \right]
\right\} \cdot(1 + o_p(1)).
\end{flalign*}

\newpage

\subsection{Results of $G_{\beta\alpha}^W$}
The following lemma is on the concentration of quadratic forms when reference panels are used. 
\begin{lem}
\label{lemma5}
Let $\widehat{\bmSigma}_{\X}=n^{-1}\X^T\X$,
$\widehat{\bmSigma}_{\Z}=n_z^{-1}\Z^T\Z$,
$\widehat{\bmSigma}_{\W}=n_w^{-1}\W^T\W$,
$\C_{k_1,k_2}=\widehat{\bmSigma}_{\X}^{k_1}(\widehat{\bmSigma}_{\W}+\lambda \I_p)^{-1}\widehat{\bmSigma}_{\Z}^{k_2}$,
$\D_{k_1,k_2}=\widehat{\bmSigma}_{\X}^{k_1}(\widehat{\bmSigma}_{\W}+\lambda \I_p)^{-1}\widehat{\bmSigma}_{\Z}^{k_2}(\widehat{\bmSigma}_{\W}+\lambda \I_p)^{-1}$,
and define $\A^0=\I$ for any matrix $\A$. 
Moreover, let $\bmalpha$ be a $p$-dimensional random vector of independent elements with mean zero, variance $\bmPhi_{\alpha}=\mbox{Diag}(\phi^2_1,\cdots,\phi^2_p)$, and finite fourth order moments, we have 
\begin{flalign*}
\bmalpha^T\C_{k_1,k_2}\bmalpha= \tr(\C_{k_1,k_2}\bmPhi_{\alpha} )\cdot\{1+o_p(1)\}
\quad \mbox{and} 
\quad
\bmalpha^T\D_{k_1,k_2}\bmalpha= \tr(\D_{k_1,k_2}\bmPhi_{\alpha} )\cdot\{1+o_p(1)\}.
\end{flalign*}
\end{lem}
The proof of Lemma~\ref{lemma5}  is similar to that of Lemma~\ref{lemma4}. 
Lemma~\ref{lemma5} indicates that the quadratic forms of $\C_{k_1,k_2}$ and $\D_{k_1,k_2}$ concentrate around their means for all positive integers $k_1$ and $k_2$. For our reference panel, we need the results of  $\C_{1,1}$, $\D_{1,1}$, and $\D_{2,1}$.
The proposition below summarizes the results on the mean of quadratic forms when reference panels are used.
Then the consistency of quadratic forms in Proposition~S\ref{pop.s2} follows from Lemma~\ref{lemma5}. 
\begin{proposition.s}\label{pop.i2}
Under the same conditions as in Propositions~S\ref{pop.s2}, we have 
\begin{flalign*}
&\tE \Big\{\bmbeta^T\X^T\X(\W^T\W+\lambda n_w\I_p)^{-1}\Z^T\Z(\W^T\W+\lambda n_w\I_p)^{-1}\X^T\X\bmbeta \Big\} \\
&= n^2n_zn_w^{-2}\cdot (\tr[(\widehat{\bmSigma}_W+\lambda\I_p)^{-1}\bmSigma_Z(\widehat{\bmSigma}_W+\lambda\I_p)^{-1}\bmSigma_X^2]+ \\
&\omega\cdot \tr[(\widehat{\bmSigma}_W+\lambda\I_p)^{-1}\bmSigma_Z(\widehat{\bmSigma}_W+\lambda\I_p)^{-1}\bmSigma_X])\cdot \tr(\bmPhi_{\beta\beta})/p^2,
\end{flalign*}
\begin{flalign*}
&\tE \Big\{\bmeps^T\X(\W^T\W+\lambda n_w\I_p)^{-1}\Z^T\Z(\W^T\W+\lambda n_w\I_p)^{-1}\X^T\bmeps \Big\} \\
&= nn_zn_w^{-2}\cdot \tr[(\widehat{\bmSigma}_W+\lambda\I_p)^{-1}\bmSigma_Z(\widehat{\bmSigma}_W+\lambda\I_p)^{-1}\bmSigma_X]\cdot{\sigma^2_{\epsilon}},
\end{flalign*}
and
\begin{flalign*}
&\tE \Big\{\big(\Z\bmalpha+\bmeps_{z}\big)^T\Z(\W^T\W+\lambda n_w\I_p)^{-1}\X^T\big(\X\bmbeta+\bmeps\big) \Big\} \\
&= nn_zn_w^{-1}\cdot \tr[\bmSigma_X(\widehat{\bmSigma}_W+\lambda\I_p)^{-1}\bmSigma_Z]\cdot{\tr(\bmPhi_{\beta\alpha})}/p^2.
\end{flalign*}
\end{proposition.s}
\section{More discussions on the effect of LD heterogeneity}\label{sec3}
Under the setup in Section~3.2 of the main text, we consider the situation where the sample size $n$ is large and $\omega$ is small. We may have $a \approx b_3(\bmSigma_X)-b_1(\bmSigma_X^2\bmSigma_Z)$, $b \approx b_1(\bmSigma_X^2\bmSigma_Z)$. Let $e = b_3(\bmSigma_X)-b_1(\bmSigma_X^2\bmSigma_Z)$, $f = b_1(\bmSigma_X^2\bmSigma_Z)$, then
\begin{flalign*}
\dot{S}_{\beta\alpha}(t) \approx& 
\frac{1}{2} \cdot \frac{e (ct -d) + 2 cf}{(e t + f)^{3/2}} \\
=& \frac{
(b_3(\bmSigma_X)-b_1(\bmSigma_X^2\bmSigma_Z)) \cdot 
(b_2(\bmSigma_X)-b_1(\bmSigma_X \bmSigma_Z))
}{2 (e t + f)^{3/2}} \cdot t \\
&+ \frac{
2 \cdot b_1(\bmSigma_X^2\bmSigma_Z) \cdot b_2(\bmSigma_X) - 
b_1(\bmSigma_X\bmSigma_Z) \cdot (b_1(\bmSigma_X^2\bmSigma_Z) + b_3(\bmSigma_X))
}{2 (e t + f)^{3/2}}
\end{flalign*}
If $b_1(\bmSigma_X\bmSigma_Z) < b_2(\bmSigma_X)$ and $b_1(\bmSigma_X^2 \bmSigma_Z) < b_3(\bmSigma_X)$, then the first term of $\dot{S}_{\beta\alpha}(t)$ is nonnegative for $t \in [0,1]$. Furthermore, let $b_3(\bmSigma_X) = c_1 \cdot b_2(\bmSigma_X)$ and $b_1(\bmSigma_X^2 \bmSigma_Z) = c_2 \cdot b_1(\bmSigma_X \bmSigma_Z)$, then the numerator of the second term of $\dot{S}_{\beta\alpha}(t)$ can be written as
\begin{flalign*}
& 2 \cdot c_2 \cdot b_1(\bmSigma_X \bmSigma_Z) \cdot b_2(\bmSigma_X) - b_1(\bmSigma_X \bmSigma_Z) \cdot c_2 \cdot b_1(\bmSigma_X \bmSigma_Z) - b_1(\bmSigma_X \bmSigma_Z) \cdot c_1 \cdot b_2(\bmSigma_X) \\
&= b_1(\bmSigma_X \bmSigma_Z) \cdot \left\{
(2 \cdot c_2 - c_1) \cdot b_2(\bmSigma_X) - c_2 \cdot b_1(\bmSigma_X \bmSigma_Z)
\right\}
\end{flalign*}
Thus, when $\omega$ is small, the sign of $\dot{S}_{\beta\alpha}(t)$ not only depends on whether $b_1(\bmSigma_X\bmSigma_Z) < b_2(\bmSigma_X)$ and whether $b_1(\bmSigma_X^2 \bmSigma_Z) < b_3(\bmSigma_X)$, but also depends on the constants $c_1, c_2$ and their interaction with $b_2(\bmSigma_X)$ and $b_1(\bmSigma_X \bmSigma_Z)$. 

If $b_2(\bmSigma_X)$ is much larger than $b_1(\bmSigma_X\bmSigma_Z)$, then we have $(2 \cdot c_2 - c_1) \cdot b_2(\bmSigma_X) - c_2 \cdot b_1(\bmSigma_X \bmSigma_Z) > 0$. Furthermore, if $b_1(\bmSigma_X^2 \bmSigma_Z) < b_3(\bmSigma_X)$, then $\dot{S}_{\beta\alpha}(t) > 0$ for all $t \in [0,1]$, which means ${S}_{\beta\alpha}(t)$ is the largest at $t=1$. Recall $S_{\beta \alpha} = G_{\beta \alpha} / G_{\beta \alpha}^M$. This implies that the within-population estimate based on $\X$ is better than the cross-population estimate based on $\X$ and $\Z$. 

On the other hand, if $b_2(\bmSigma_X)$ is much less than $b_1(\bmSigma_X\bmSigma_Z)$ (which, by the Cauchy-Schwarz inequality, implies that $b_2(\bmSigma_Z)$ is much larger than $b_1(\bmSigma_X\bmSigma_Z)$) and $b_1(\bmSigma_X^2 \bmSigma_Z) > b_3(\bmSigma_X)$, we will have $\dot{S}_{\beta\alpha}(t) < 0$ for all $t \in [0,1]$, and ${S}_{\beta\alpha}(t)$ is the largest at $t=0$. This means the within-population estimate based on $\Z$ is better than the cross-population estimate. 

In summary, when the sample size is much larger than the number of features and when $b_1(\bmSigma_X\bmSigma_Z)$ is much less than $\mbox{max} \{b_2(\bmSigma_X),b_2(\bmSigma_Z)\}$, the best within-population estimate is always better than the cross-population estimate. Otherwise, when $\bmSigma_X$ and $\bmSigma_Z$ are similar enough, the sign of $\dot{S}_{\beta\alpha}(t)$ will be hard to determine and so is the value of $t$ that gives the largest ${S}_{\beta\alpha}(t)$. This shows that, when the sample size is large, the similarity between $\bmSigma_X$ and $\bmSigma_Z$ impacts the performance of the cross-population estimate in a more delicate way.
\section{More discussions on reference panel}\label{sec3}
We use a additional numerical example to provide more insights into the reference panel approaches. 
We simulate the data by setting $\varphi_{\beta\alpha}=0.3$, $\h^2_{\alpha}=\h^2_{\beta}=0.4$ or $0.8$, $n=n_w$, and $\omega$ ranging from $0.05$ to $20$. Moreover, we set  a block-diagonal structure with  $50$ blocks along the diagonal and each with a block size of $20$ 
for both $\bmSigma_X$ and $\bmSigma_Z$. 
Every block has an auto-correlation structure with the correlation coefficients being $\rho_X$ and $\rho_Z$ in $\bmSigma_X$ and $\bmSigma_Z$, respectively. 

Supplementary Figure~\ref{figs20} illustrates trans-ancestry genetic correlation estimators for $(\rho_X, \rho_Z) =(0.8, 0.2)$ (Case~I) as well as  $(\rho_X, \rho_Z)=(0.2, 0.8)$  (Case~II). All of these  estimators are smaller than the   true genetic correlation $\varphi_{\beta\alpha}$. {Similar to Figure~2, we find the reference panel matching the training GWAS has better performance in trans-ancestry analysis.}
Case~I represents a scenario in which the Population-I GWAS genetic variants are in stronger LD than those in the Population-II GWAS. 
In this case, the use of a reference panel that matches with the Population-I GWAS can greatly improve the estimation accuracy over $G_{\beta\alpha}$. 
The mixed reference panel performs very similar to the Population-I reference panel, whereas a reference panel matching the Population-II GWAS performs poorly. The mixed reference panel may even slightly outperform the Population-I reference panel when the heritability is high (say, $0.8$) and sample size is large (say, $\omega=1$). 
Case~II represents the opposite situation, in which the LD in the Population-II GWAS  is stronger than that in the Population-I GWAS. 
In this case, $G_{\beta\alpha}$ performs similarly to the estimator using the Population-I reference panel. Surprisingly, Population-II and mixed reference panels may perform even worse than $G_{\beta\alpha}$, possibly indicating the negative effects of over-correcting LD.  In summary, our example suggests that over-correcting LD can introduce more bias than not correcting LD.  Moreover, the mixed reference panel may work well for highly heritable traits with summary statistics estimated from  large-scale GWAS. 
\section{Supplementary figures and tables}\label{sec4}
\clearpage
\begin{suppfigure}
\includegraphics[page=1,width=1\linewidth]{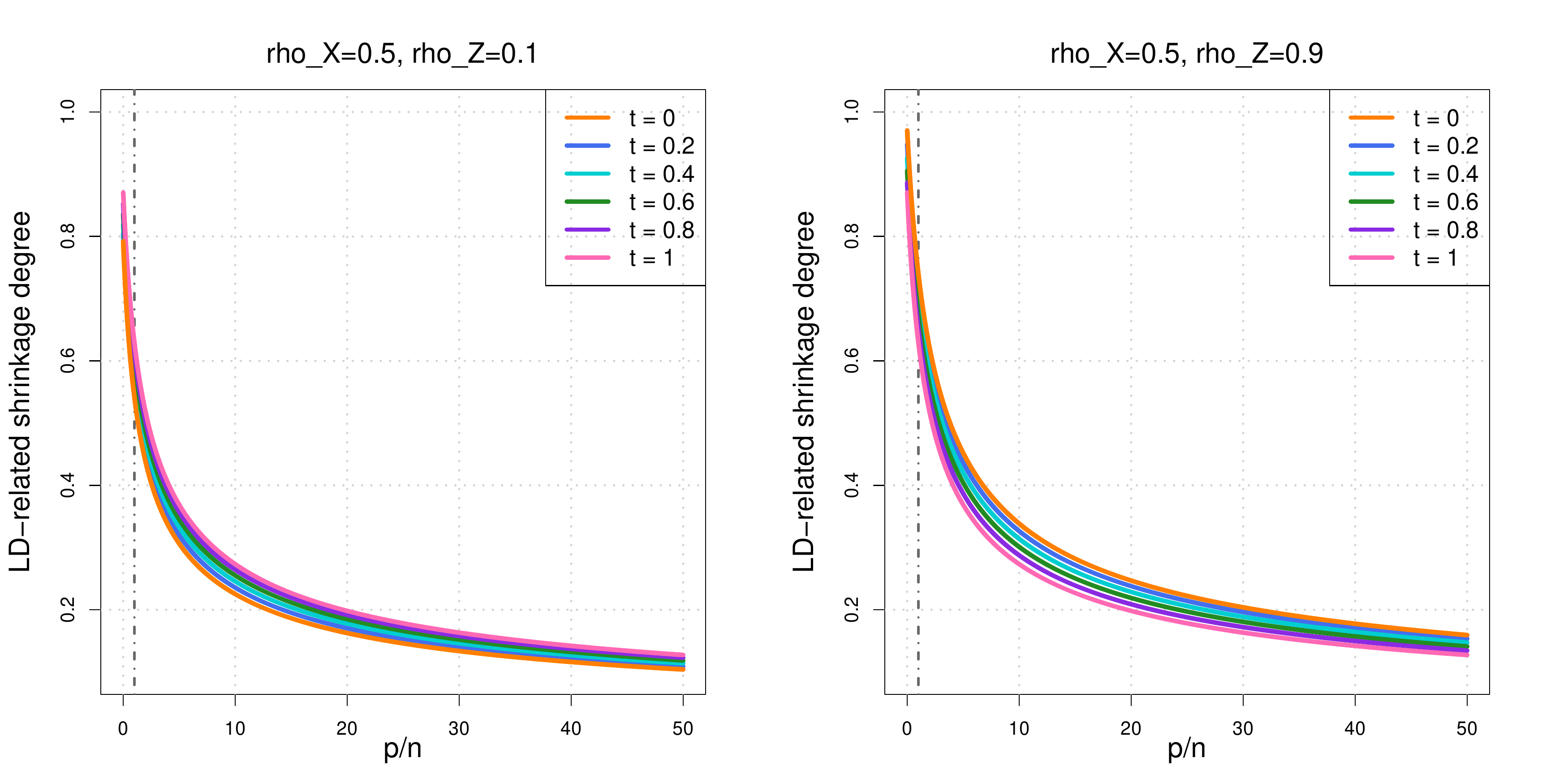}
  \caption{Illustration of the LD-related shrinkage factor $S_{\beta\alpha}(t)$ for $\rho_X=0.5$ and $\rho_Z=0.1$  (Case~I, left panel) and for  $\rho_X=0.5$ and 
$\rho_Z=0.9$  (Case~II, right panel). 
The $x$-axis displays $\omega=p/n$ and the $y$-axis displays  shrinkage degrees. Smaller values indicate more serious  shrinkage. The vertical line represents $\omega=1$.
}
\label{figs1}
\end{suppfigure}
\begin{suppfigure}
\includegraphics[page=1,width=1\linewidth]{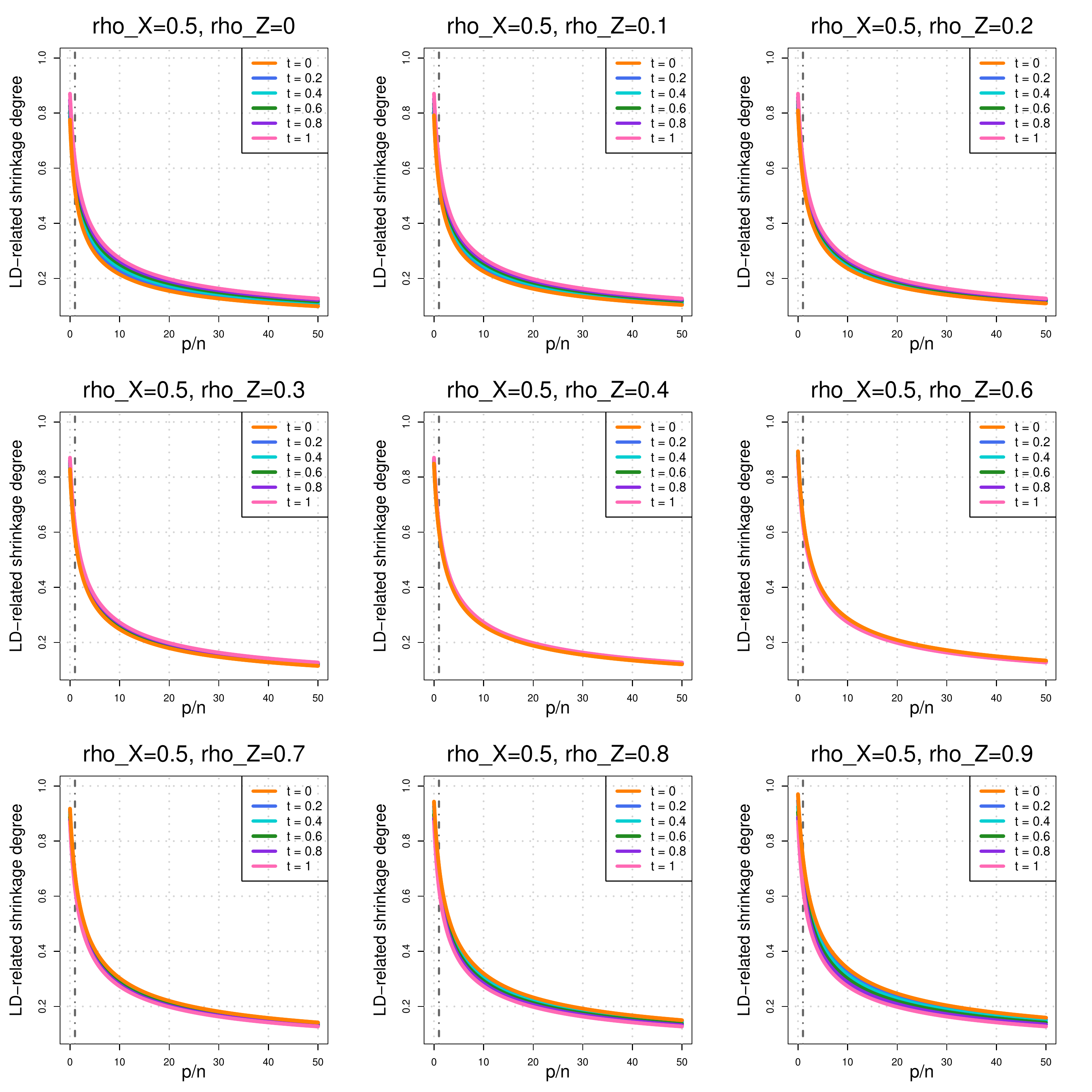}
  \caption{Illustration of the LD-related shrinkage factor $S_{\beta\alpha}(t)$. 
The $x$-axis displays $\omega=p/n$ and the $y$-axis displays  shrinkage degrees. Smaller values indicate more serious  shrinkage. We simulate the data with $\rho_X=0.5$ and vary $\rho_Z$ from $0$ to $0.9$.  The vertical line represents $\omega=1$.
}
\label{figs2}
\end{suppfigure}

\begin{suppfigure}[t]
\includegraphics[page=2,width=1\linewidth]{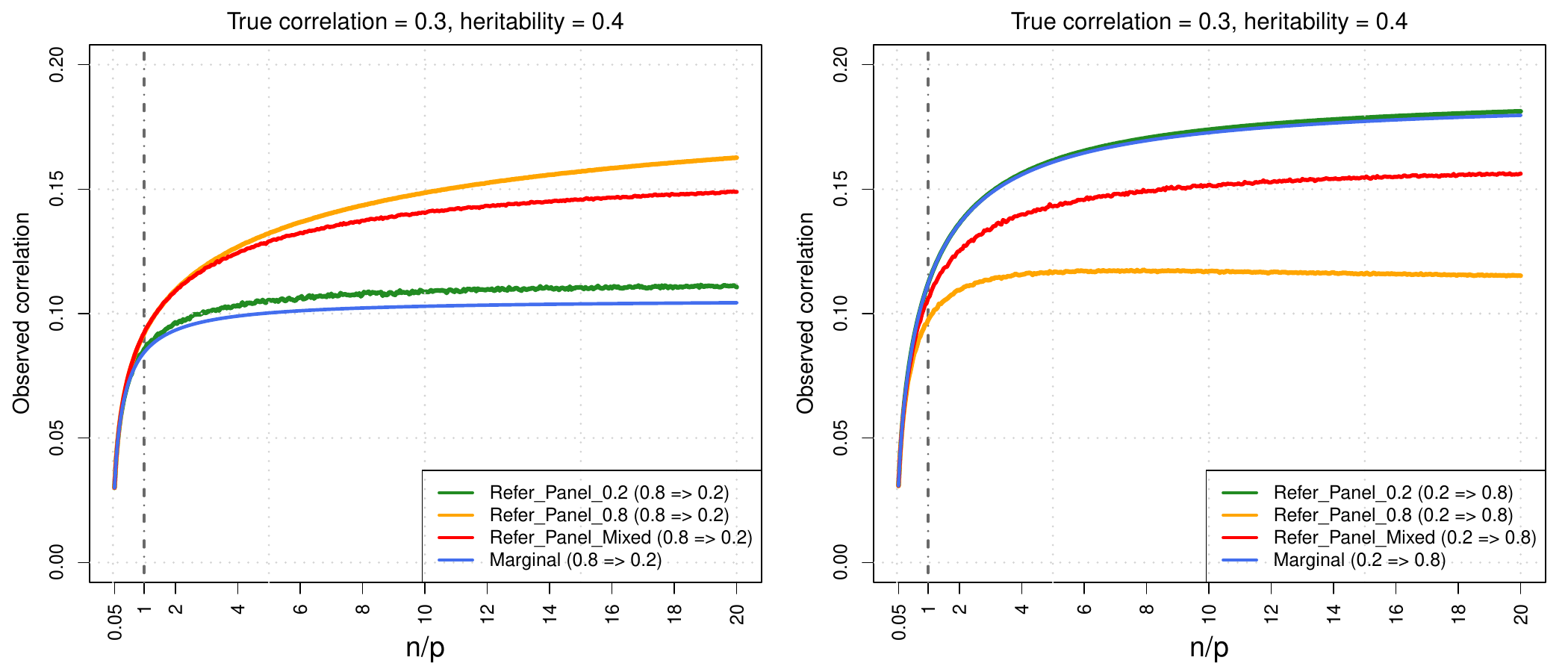}
\includegraphics[page=4,width=1\linewidth]{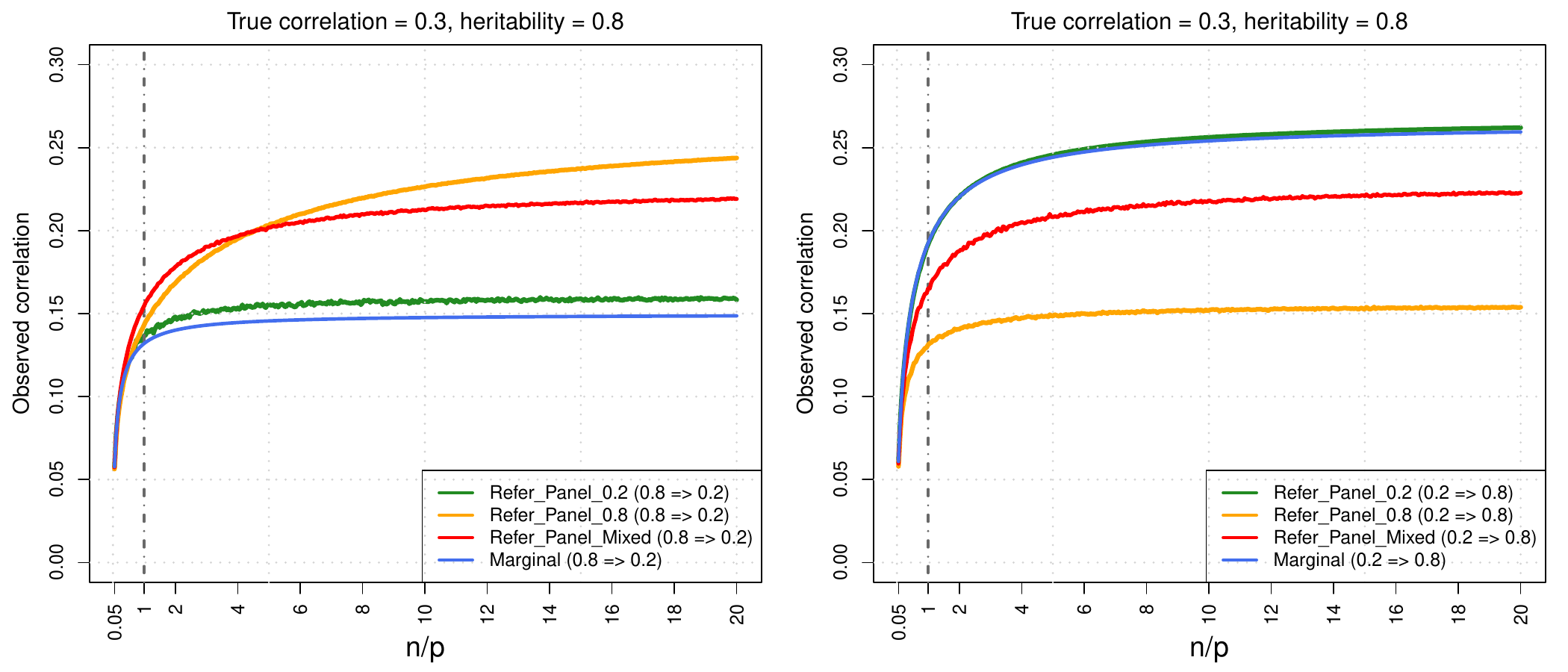}
  \caption{Comparing the naive (uncorrected) trans-ancestry genetic correlation estimators. 
   We set $\varphi_{\beta\alpha}=0.3$, $\h^2_{\alpha}=\h^2_{\beta}=0.4$ (Upper panels) or $0.8$ (Lower panels). $\bmSigma_X$ and $\bmSigma_Z$ are block diagonal matrices, with correlation coefficients $\rho_X=0.8$ and $\rho_Z=0.2$ in left two panels ($0.8 => 0.2$), and $\rho_X=0.2$ and $\rho_Z=0.8$ in right two panels ($0.2 => 0.8$).
   Refer\_Panel\_0.2, $G_{\beta\alpha}^W$ with $\rho_W=0.2$;
   Refer\_Panel\_0.8, $G_{\beta\alpha}^W$ with $\rho_W=0.8$;
   Refer\_Panel\_Mixed, $G_{\beta\alpha}^W$ with equally-mixed $\rho_W=0.2$ and $0.8$ samples; and Marginal, $G_{\beta\alpha}$.
   The vertical line represents $\omega=1$.
}
\label{figs20}
\end{suppfigure}

\begin{suppfigure}
\includegraphics[page=1,width=1\linewidth]{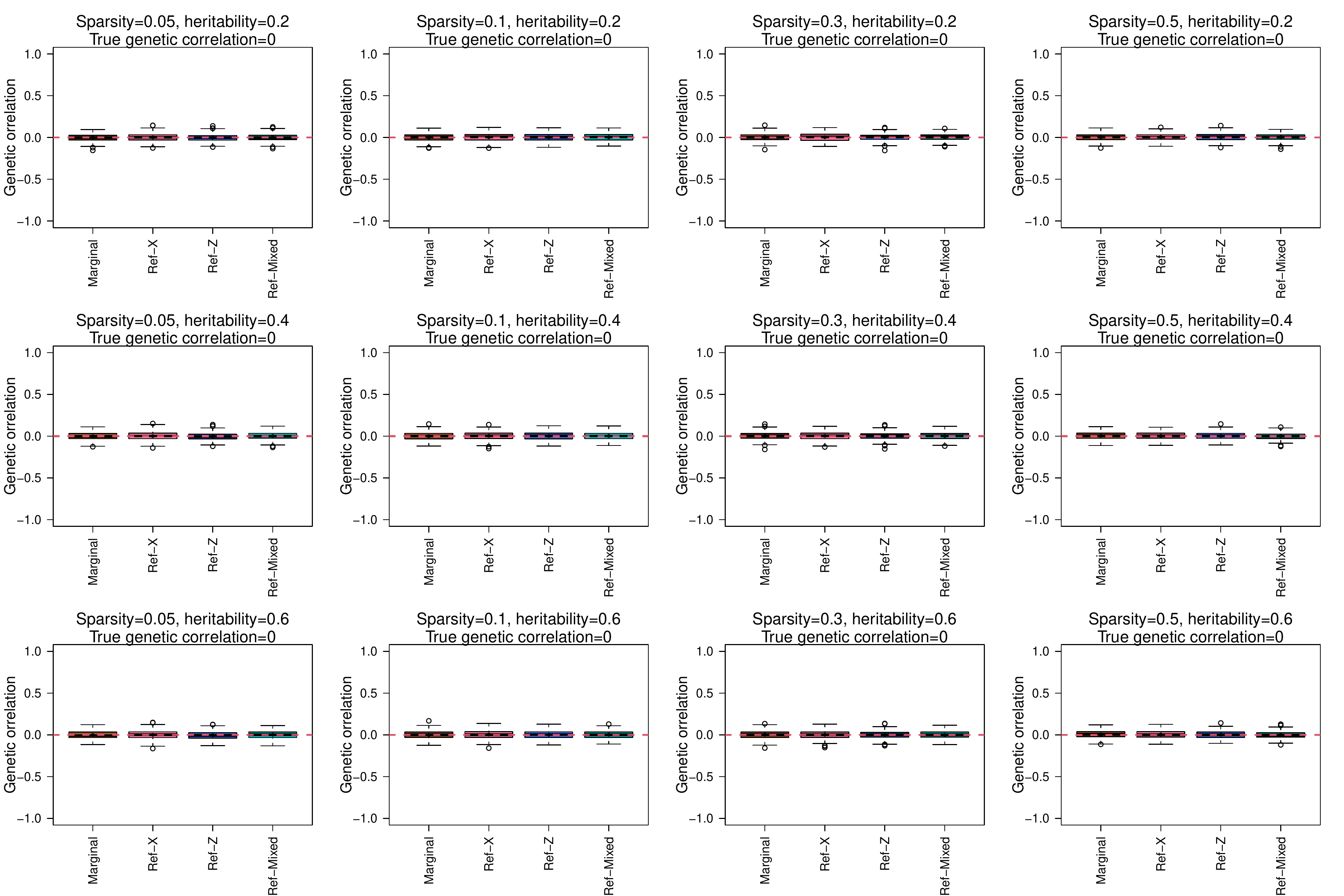}
  \caption{Uncorrected genetic correlations across different signal sparsity and heritability settings. The $x$-axis displays the four genetic correlation estimators based on marginal estimator (Marginal,  $G_{\alpha\beta}$) or additionally using reference panels ($G_{\alpha\beta}^W$, three versions including Ref-X, Ref-Z, and Ref-Mixed). 
We simulate the data with $n=p=14,000$, $n_w=5000$, and $n_z=500$.  The horizontal line represents the true genetic correlation $\varphi_{\beta\alpha}=0$.
}
\label{figs3}
\end{suppfigure}
\begin{suppfigure}
\includegraphics[page=2,width=1\linewidth]{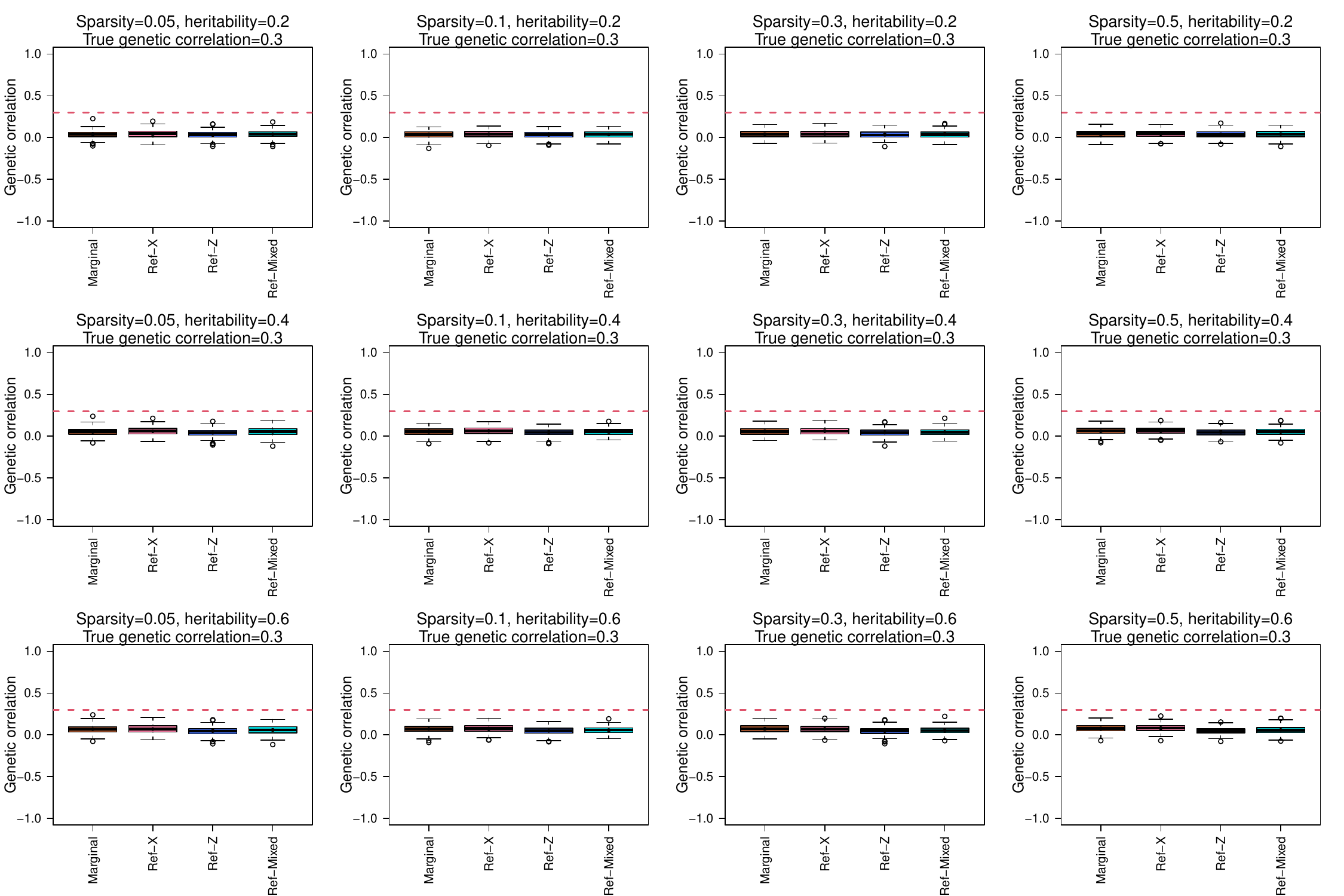}
  \caption{Uncorrected genetic correlations across different signal sparsity and heritability settings. The $x$-axis displays the four genetic correlation estimators based on marginal estimator (Marginal,  $G_{\alpha\beta}$) or additionally using reference panels ($G_{\alpha\beta}^W$, three versions including Ref-X, Ref-Z, and Ref-Mixed). 
We simulate the data with $n=p=14,000$, $n_w=5000$, and $n_z=500$.  The horizontal line represents the true genetic correlation $\varphi_{\beta\alpha}=0.3$.
}
\label{figs4}
\end{suppfigure}
\begin{suppfigure}
\includegraphics[page=3,width=1\linewidth]{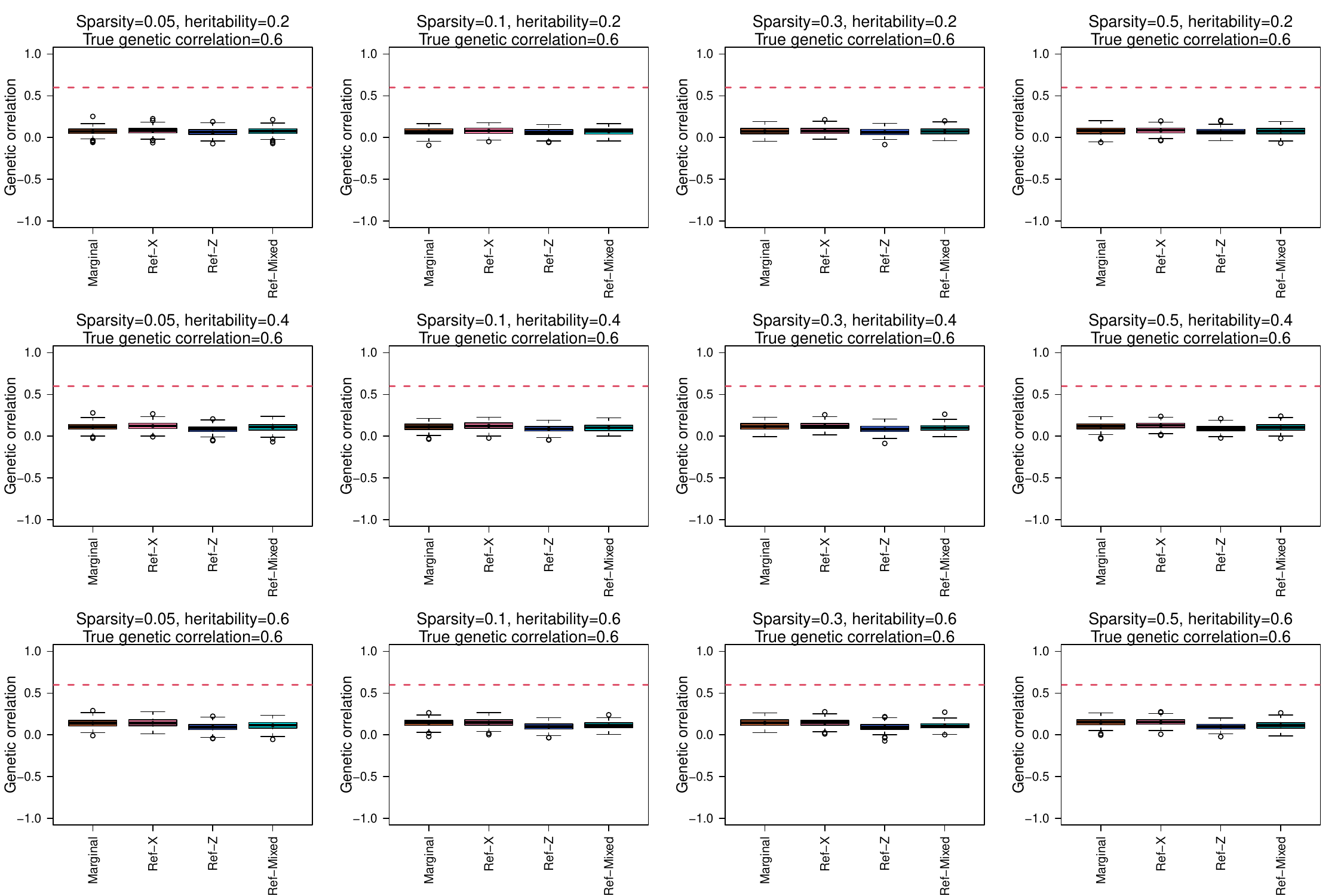}
  \caption{Uncorrected genetic correlations across different signal sparsity and heritability settings. The $x$-axis displays the four genetic correlation estimators based on marginal estimator (Marginal,  $G_{\alpha\beta}$) or additionally using reference panels ($G_{\alpha\beta}^W$, three versions including Ref-X, Ref-Z, and Ref-Mixed). 
We simulate the data with $n=p=14,000$, $n_w=5000$, and $n_z=500$.  The horizontal line represents the true genetic correlation $\varphi_{\beta\alpha}=0.6$.
}
\label{figs5}
\end{suppfigure}
\begin{suppfigure}
\includegraphics[page=1,width=1\linewidth]{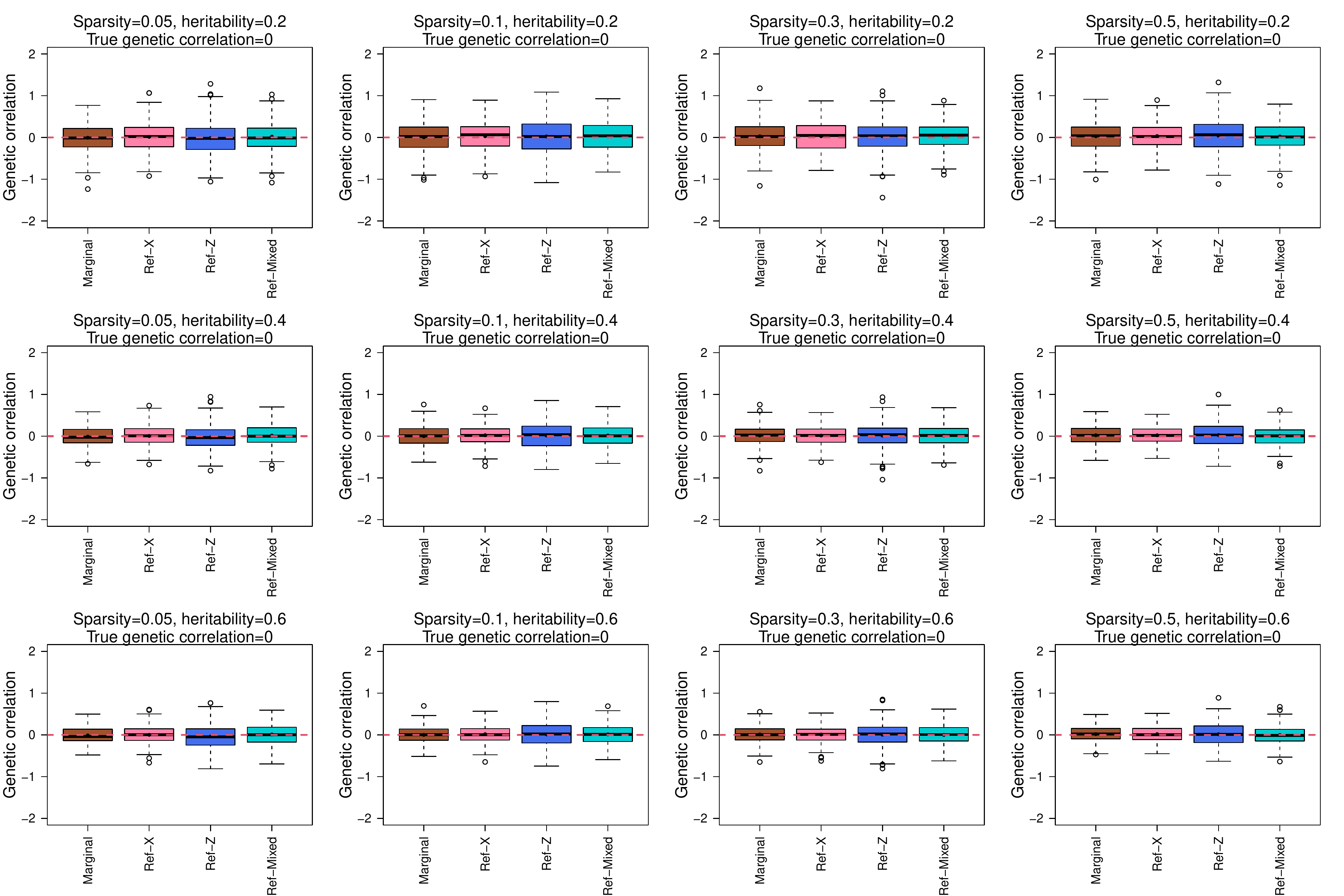}
  \caption{Corrected genetic correlations across different signal sparsity and heritability settings. The $x$-axis displays the four genetic correlation estimators based on marginal estimator (Marginal,  $G_{\alpha\beta}^M$) or additionally using reference panels ($G_{\alpha\beta}^{M_W}$, three versions including Ref-X, Ref-Z, and Ref-Mixed). 
We simulate the data with $n=p=14,000$, $n_w=5000$, and $n_z=500$. 
More details can be found in Section~5.1 of the main text. The horizontal line represents the true genetic correlation $\varphi_{\beta\alpha}=0$.
}
\label{figs6}
\end{suppfigure}
\begin{suppfigure}
\includegraphics[page=2,width=1\linewidth]{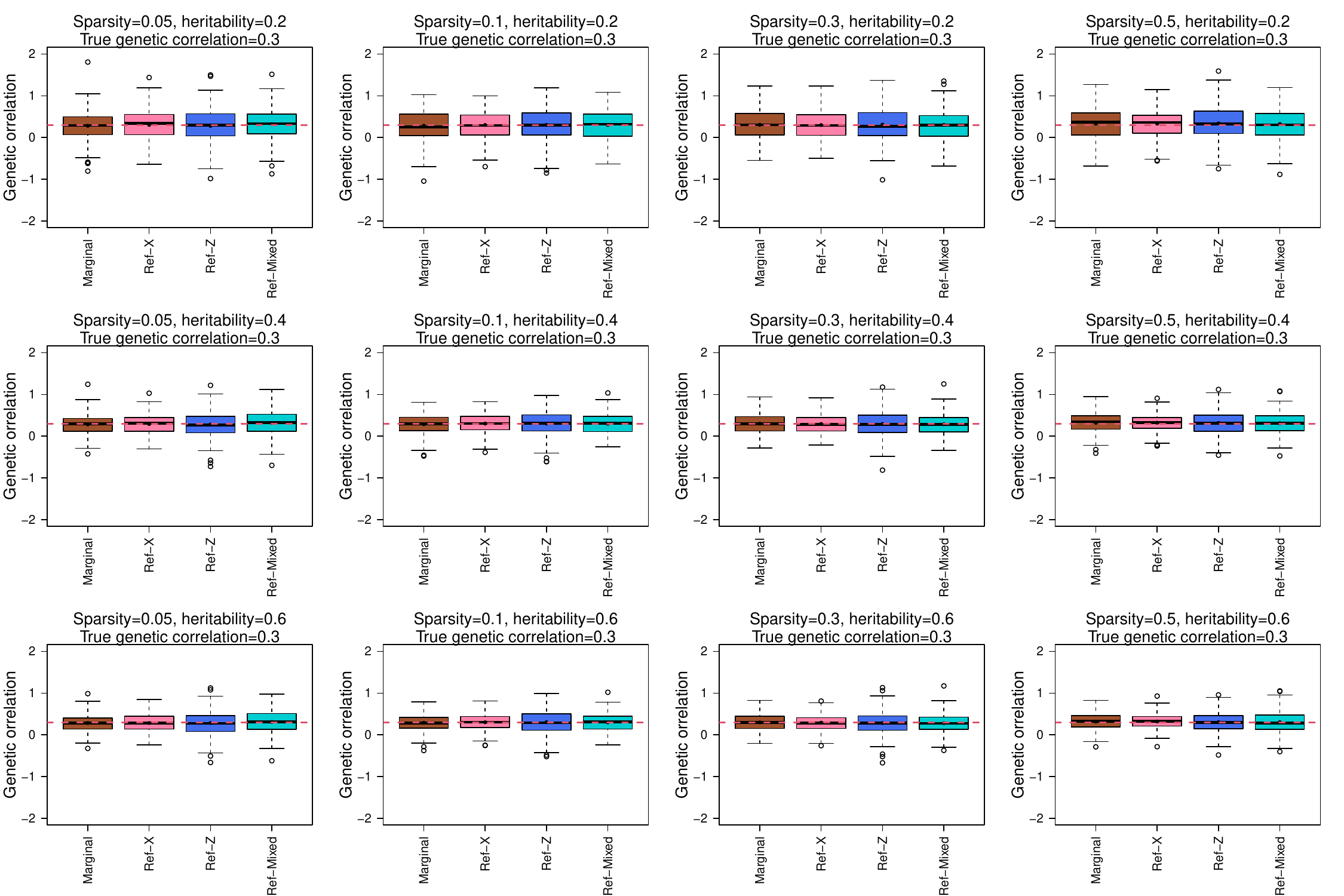}
  \caption{Corrected genetic correlations across different signal sparsity and heritability settings. The $x$-axis displays the four genetic correlation estimators based on marginal estimator (Marginal,  $G_{\alpha\beta}^M$) or additionally using reference panels ($G_{\alpha\beta}^{M_W}$, three versions including Ref-X, Ref-Z, and Ref-Mixed). 
We simulate the data with $n=p=14,000$, $n_w=5000$, and $n_z=500$. 
More details can be found in Section~5.1 of the main text. The horizontal line represents the true genetic correlation $\varphi_{\beta\alpha}=0.3$.
}
\label{figs7}
\end{suppfigure}
\begin{suppfigure}
\includegraphics[page=3,width=1\linewidth]{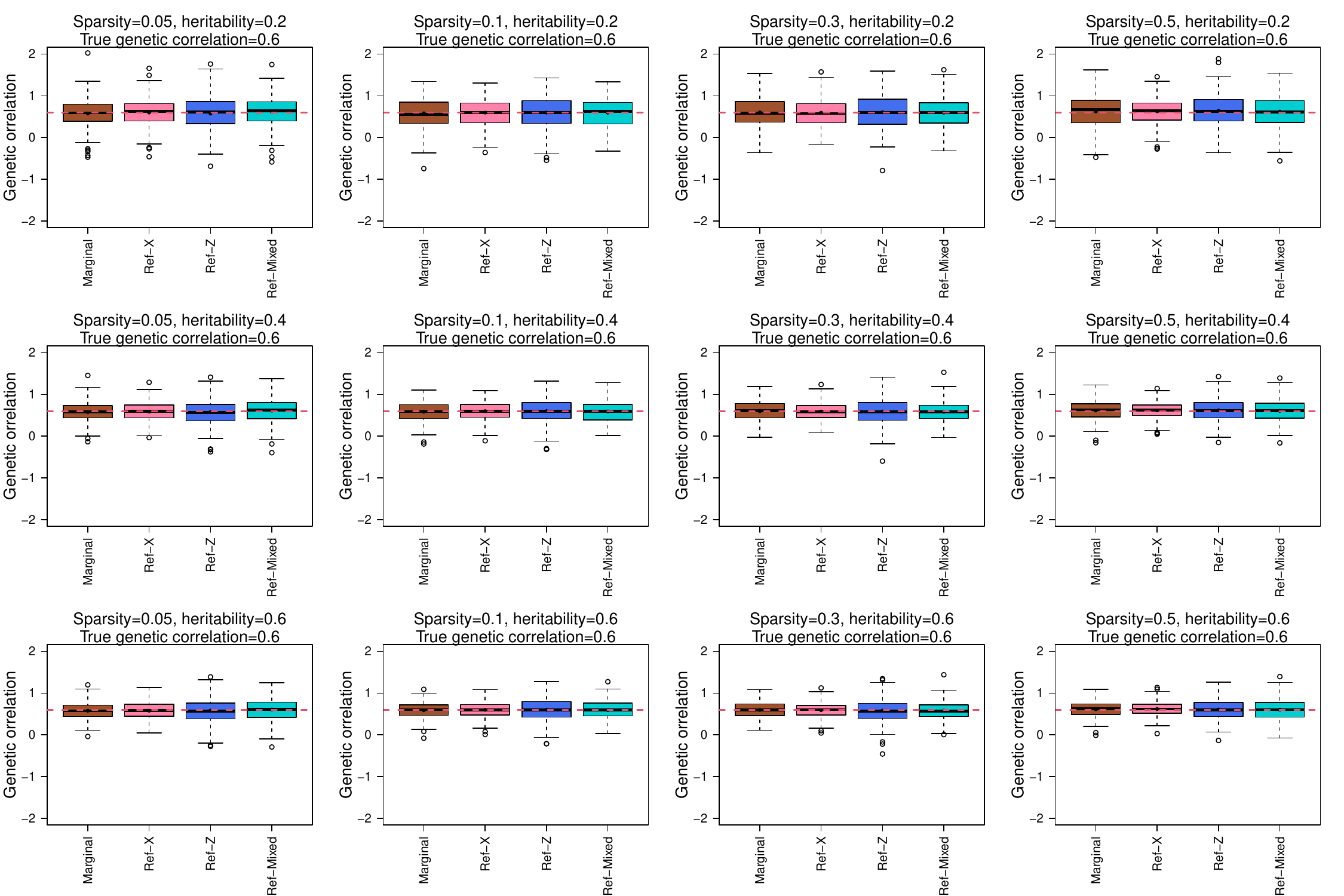}
  \caption{Corrected genetic correlations across different signal sparsity and heritability settings. The $x$-axis displays the four genetic correlation estimators based on marginal estimator (Marginal,  $G_{\alpha\beta}^M$) or additionally using reference panels ($G_{\alpha\beta}^{M_W}$, three versions including Ref-X, Ref-Z, and Ref-Mixed). 
We simulate the data with $n=p=14,000$, $n_w=5000$, and $n_z=500$. 
More details can be found in Section~5.1 of the main text. The horizontal line represents the true genetic correlation $\varphi_{\beta\alpha}=0.6$.
}
\label{figs8}
\end{suppfigure}
\begin{suppfigure}
\includegraphics[page=1,width=1\linewidth]{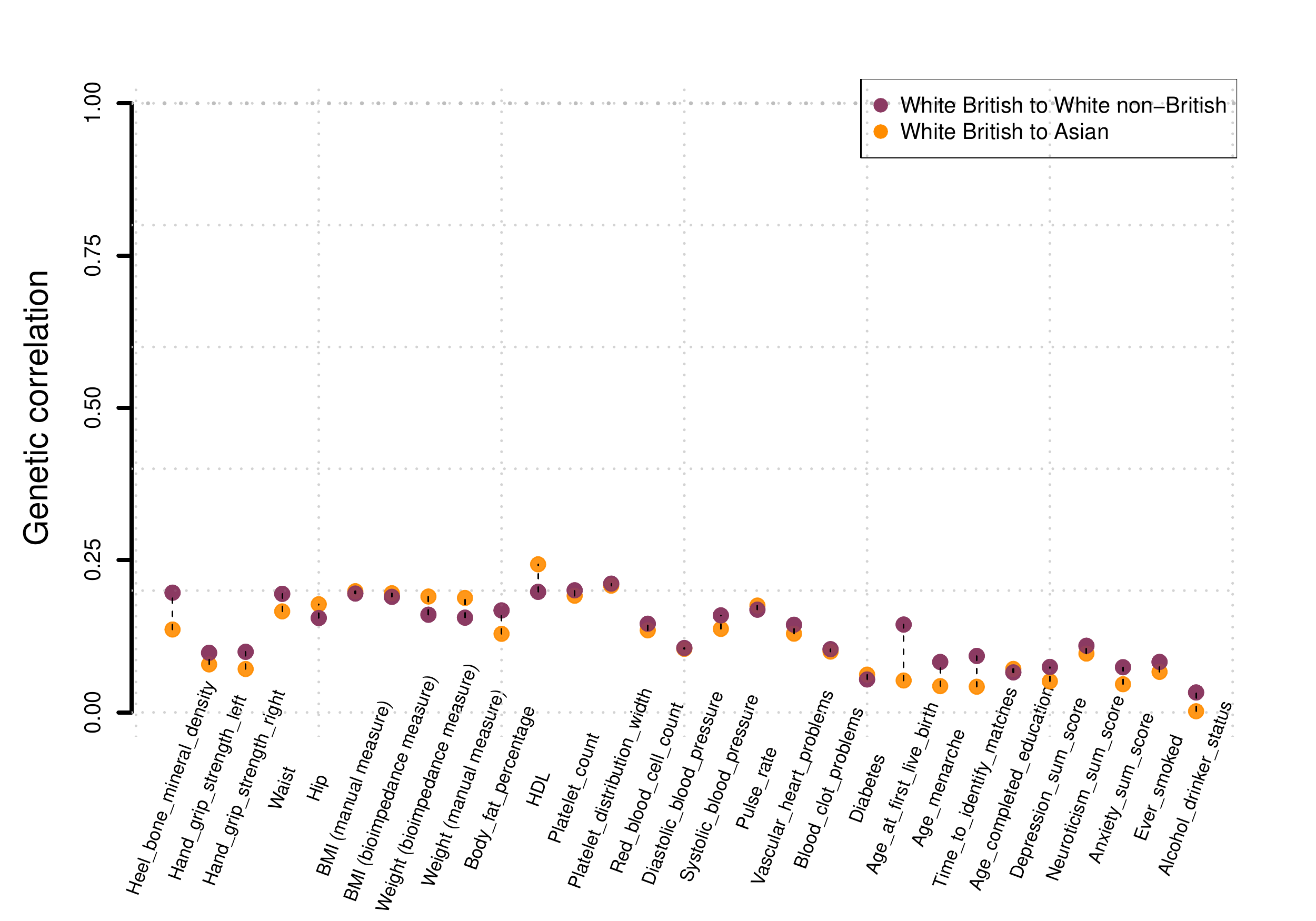}
  \caption{Genetic correlation estimated by $G_{\beta\alpha}$ in White non-British and White Asian analyses across different complex traits. See Supplementary Table~\ref{tabs1} for more details. 
}
\label{figs9}
\end{suppfigure}

\clearpage

\begin{supptable}
\centering
\scalebox{1}{
\begin{tabular}{N|N|N|N|N|N|N|N|N|}
\cline{2-9}
     & \multicolumn{4}{c|}{$\h_{\beta}^2=\h_{\alpha}^2=0.3$ and} & \multicolumn{4}{c|}{$\h_{\beta}^2=\h_{\alpha}^2=0.6$ and} \\  
     & \multicolumn{4}{c|}{$\varphi_{\beta\alpha}=0.25$} & \multicolumn{4}{c|}{$\varphi_{\beta\alpha}=0.5$} \\  \hline
\multicolumn{1}{|c|}{Sparsity} &  $0.1$ & $0.01$ & $0.001$ & mean& $0.1$  & $0.01$ & $0.001$ & mean\\ \hline
\multicolumn{1}{|c|}{$G_{\beta\alpha}$,  $n=350k$} & $0.040$ ($0.036$)  & $0.041$ ($0.037$) & $0.045$ ($0.041$)& $0.042$ ($0.038$) & $0.121$ ($0.047$) & $0.125$ ($0.047$)  & $0.134$ ($0.054$) & $0.126$ ($0.049$) \\ \hline
\multicolumn{1}{|c|}{$G_{\beta\alpha}, n=50k$} & $0.030$ ($0.034$)  & $0.031$ ($0.037$) & $0.031$ ($0.039$)& $0.031$ ($0.037$) & $0.098$ ($0.042$) & $0.100$ ($0.044$)  & $0.107$ ($0.046$) & $ 0.102$ ($0.044$)  \\ \hline
\multicolumn{1}{|c|}{$G_{\beta\alpha}^M, n=350k$} & $0.239$ ($0.217$) & $0.245$ ($0.222$) & $0.272$ ($0.242$) & $0.253$ ($0.227$) & $0.491$ ($0.191$)   
& $0.508$ ($0.193$) & $0.544$ ($0.218$)& $0.515$ ($0.200$) \\ \hline
\multicolumn{1}{|c|}{$G_{\beta\alpha}^M, n=50k$} & $0.248$ ($0.281$) & $0.254$ ($0.304$) & $0.256$ ($0.319$) & $0.253$ ($0.301$) & $0.485$ ($0.209$)
& $0.498$ ($0.219$) & $0.529$ ($0.227$)& $0.504$ ($0.218$) \\ \hline
\end{tabular}
}
\caption{{Simulation results of the naive (uncorrected) estimator $G_{\beta\alpha}$ and corrected estimator $G_{\beta\alpha}^M$ on UKB genotype data. We perform simulation across a wide variety settings of heritability ($\h_{\beta}^2$, $\h_{\alpha}^2$), genetic correlation ($\varphi_{\beta\alpha}$), Population-I GWAS sample size ($n$), and genetic signal sparsity ($0.1$, $0.01$, $0.001$). 
We display the mean of estimates across $500$ simulation replications with corresponding standard errors in brackets. The ``mean'' column shows the average of the three signal sparsity levels.}}
\label{tab1}
\end{supptable}

\begin{supptable}[ht]
\centering
\begin{tabular}{rlrrrrrr}
  \hline
ID & Name & $G_{\alpha\beta}$ & P\_value &$G_{\alpha\beta}^M$ & $G_{\alpha\beta}$ & P\_value & $G_{\alpha\beta}^M$ \\ 
  \hline
1 & Heel\_bone\_mineral\_density & 0.197 & 8.98E-84 & 1.057 & 0.136 & 9.41E-14 & 0.656 \\ 
  2 & Hand\_grip\_strength\_left & 0.098 & 2.80E-85 & 0.782 & 0.079 & 3.30E-26 & 0.519 \\ 
  3 & Hand\_grip\_strength\_right & 0.100 & 5.82E-89 & 0.794 & 0.071 & 1.05E-21 & 0.468 \\ 
  4 & Waist & 0.195 & 3.19E-205 & 1.112 & 0.166 & 6.45E-78 & 0.990 \\ 
  5 & Hip & 0.155 & 1.07E-97 & 0.878 & 0.178 & 7.06E-69 & 0.758 \\ 
  6 & BMI (manual measure) & 0.195 & 1.25E-161 & 0.995 & 0.199 & 6.72E-87 & 0.799 \\ 
  7 & BMI (bioimpedance measure) & 0.190 & 1.49E-150 & 0.968 & 0.196 & 2.58E-84 & 0.788 \\ 
  8 & Weight (bioimpedance measure) & 0.160 & 3.03E-134 & 0.787 & 0.190 & 6.28E-98 & 0.754 \\ 
  9 & Weight (manual measure) & 0.156 & 2.26E-124 & 0.764 & 0.188 & 3.71E-94 & 0.745 \\ 
  10 & Body\_fat\_percentage & 0.167 & 2.54E-196 & 0.899 & 0.129 & 8.23E-73 & 0.610 \\ 
  11 & HDL & 0.198 & 1.83E-171 & 1.017 & 0.243 & 1.04E-131 & 1.289 \\ 
  12 & Platelet\_count & 0.200 & 1.30E-166 & 0.990 & 0.192 & 9.70E-78 & 1.284 \\ 
  13 & Platelet\_distribution\_width & 0.211 & 1.10E-178 & 1.094 & 0.208 & 9.04E-85 & 1.453 \\ 
  14 & Red\_blood\_cell\_count & 0.146 & 2.17E-113 & 0.827 & 0.135 & 3.72E-49 & 0.905 \\ 
  15 & Diastolic\_blood\_pressure & 0.106 & 1.79E-45 & 0.762 & 0.104 & 4.39E-21 & 0.970 \\ 
  16 & Systolic\_blood\_pressure & 0.159 & 2.02E-116 & 1.160 & 0.137 & 1.29E-40 & 1.196 \\ 
  17 & Pulse\_rate & 0.169 & 6.18E-111 & 1.226 & 0.175 & 6.59E-63 & 1.248 \\ 
  18 & FVC1 (zscore) & 0.133 & 1.01E-58 & 0.814 & NA & NA & NA \\ 
  19 & Vascular\_heart\_problems & 0.144 & 2.17E-90 & 1.210 & 0.129 & 5.78E-37 & 1.176 \\ 
  20 & Blood\_clot\_problems & 0.104 & 1.32E-45 & 1.111 & 0.100 & 1.88E-21 & 0.980 \\ 
  21 & Diabetes & 0.054 & 1.38E-13 & 1.000 & 0.062 & 1.57E-09 & 0.513 \\ 
  22 & Age\_at\_first\_live\_birth & 0.144 & 6.72E-32 & 1.236 & 0.052 & 2.77E-03 & 0.560 \\ 
  23 & Age\_menarche & 0.083 & 3.04E-17 & 0.950 & 0.043 & 6.48E-03 & 0.588 \\ 
  24 & Time\_to\_identify\_matches & 0.093 & 1.29E-39 & 1.230 & 0.042 & 7.50E-05 & 0.335 \\ 
  25 & Age\_completed\_education & 0.066 & 3.85E-11 & 1.153 & 0.071 & 1.82E-07 & 0.674 \\ 
  26 & Depression\_sum\_score & 0.075 & 2.89E-08 & 1.230 & 0.051 & 7.36E-02 & 0.834 \\ 
  27 & Neuroticism\_sum\_score & 0.109 & 4.79E-39 & 1.047 & 0.097 & 3.11E-13 & 0.956 \\ 
  28 & Anxiety\_sum\_score & 0.074 & 1.76E-08 & 1.243 & 0.046 & 1.01E-01 & 0.582 \\ 
  29 & Ever\_smoked & 0.083 & 2.38E-29 & 0.977 & 0.067 & 3.16E-11 & 0.816 \\ 
  30 & Alcohol\_drinker\_status & 0.033 & 7.57E-06 & 0.768 & 0.002 & 8.47E-01 & 0.021 \\ 
   \hline
\end{tabular}
\caption{{UK Biobank real data analysis results on $30$ complex traits.
$G_{\alpha\beta}$ is the uncorrected genetic correlation estimator, P\_value is the associated $T$-test $p$-value, and $G_{\alpha\beta}^M$ is the corrected genetic correlation estimator. 
The first three columns are results of the White non-British analysis, and the second three columns are results of the White Asian analysis.
}}
\label{tabs1}
\end{supptable}
\clearpage
\bibliographystyle{rss}
\bibliography{sample.bib}
